\documentclass[traditabstract]{aa}

\usepackage{graphicx}
\usepackage{txfonts}
\usepackage{natbib}
\usepackage{subfigure}
\usepackage{url}
\usepackage{longtable}
\usepackage{supertabular}
\usepackage{rotating}
\usepackage{multirow}
\usepackage{gensymb} %degrees
\usepackage{lscape}
\usepackage{color}
\usepackage{dblfloatfix}
\usepackage{rotfloat}
\usepackage{mwe,tikz}\usepackage[percent]{overpic}
\usepackage{caption}

\def\ujy{\,$\mu$Jy}

\def\mum{\,$\mu$m}
\def\msun{\,M$_{\odot}$}
\def\msunyr{\,M$_{\odot}$ yr$^{-1}$}
\def\lsun{\,L$_{\odot}$}
\def\nodata{...}

\def\herge{HeRG\'E}

\def\spitzer{{\it Spitzer}}
\def\herschel{{\it Herschel}}
\def\hst{{\it HST}}

\def\mips1{MIPS (24\,\mum)}

\def\lbolagn{$L^{\rm int}_{\rm AGN}$}

\def\lir{$L^{\rm IR}$}

\def\mstel{$M_{\rm stel}$}

\def\pegase{P\'EGASE}
\def\pegaseii{P\'EGASE.2}
\def\pegaseiii{P\'EGASE.3}
\def\tautorus{$\tau_{9.7\mu \rm m}$}
\def\relagn{$f_{\rm AGN}^{20\mu \rm m}$}
\def\d4000{D$_{\rm 4000}$}
\def\sfrdens{$\Sigma_{\rm SFR}$}

\def\l500{$L^{\rm 500 MHz}_{\rm ext}$}

\begin{document}

\title{Disentangling star formation and AGN activity in powerful infrared luminous radio galaxies at 1$<$$z$$<$4}

\author{G. Drouart,
          \inst{1,2,3}
            B. Rocca-Volmerange,
           \inst{2}
           C. De Breuck,
           \inst{4}
           M. Fioc
           \inst{2}
            M. Lehnert,
          \inst{2}
           N. Seymour,
           \inst{1}
           D. Stern,
           \inst{5}
           J. Vernet,
           \inst{4}
}

\institute{\inst{1}International Centre for Radio Astronomy Research, Curtin University, Perth, Australia\\
	      \inst{2}Institut d'Astrophysique de Paris, 98bis boulevard Arago, 75014 Paris, France \\
              \inst{3}Department of Earth and Space Science, Chalmers University of Technology, Onsala Space Observatory, 43992 Onsala, Sweden\\
              \inst{4}European Southern Observatory, Karl Schwarzschild Stra\ss e 2, 85748 Garching bei M\"unchen, Germany\\
              \inst{5}Jet Propulsion Laboratory, California Institute of Technology, Mail Stop 169-221, Pasadena, CA 91109, USA
}

\date{Version: 31 May 2016}

% 5 {} token are mandatory
\abstract{ {High-redshift radio galaxies present signs of both star formation and AGN activity, 
making them ideal candidates to investigate the connection and coevolution of AGN and star formation in the progenitors of present-day massive galaxies. We make use of a sample of 11 powerful radio galaxies spanning 1$<$$z$$<$4 which 
have complete coverage of their spectral energy distribution (SED) from UV to FIR wavelengths. Using \herschel\ data, 
we disentangle the relative contribution of the AGN and star formation by 
combining the galaxy evolution code \pegaseiii\ with an AGN torus model. We find that three components are necessary to reproduce the 
observed SEDs: an evolved and massive stellar component, a submm bright young starburst, and an
AGN torus. We find that powerful radio galaxies form at very high-redshift,
but experience episodic and important growth at 1$<$$z$$<$4 as the mass of the associated starburst varies 
from 5 to 50\% of the total mass of the system. The properties of star formation differ 
from source to source, indicating no general trend of the star formation properties in the most infrared luminous 
 high-redshift radio galaxies and no correlation with the AGN bolometric luminosity. Moreover, we find 
 that AGN scattered light have a very limited impact on broad-band SED fitting on our sample. 
Finally, our analysis also suggests a wide range in origins for the observed star formation,which we partially constrain for some sources.
 } {}{}{}{} }
%update keywords

\titlerunning{Disentangling SF and AGN in luminous HzRGs}
\authorrunning{G. Drouart et al.}

\maketitle

\section{Introduction}
\label{sec:intro}

%general radio galaxies, history
High-redshift radio galaxies (HzRG) are among the most luminous objects in the Universe and
are the target of a considerable number of investigations aiming to understand 
galaxy evolution \citep[][for a review]{Miley2008}. While powerful radio emission betrays the presence of a supermassive black hole, near-infrared (NIR) images reveal that their host galaxies are amongst the brightest galaxies in
the Universe at any redshift \citep[the $K$-$z$ relation, e.g.,][]{Lilly1984,Inskip2002}.
Modelling the $K$-band emission suggests that radio galaxies are extremely massive galaxies, 10$^{12}$\msun\ 
\citep{Rocca2004}. Optical images reveal that most radio galaxies present a surprising alignment
between the ultraviolet(UV)/optical emission and the radio jet axis \citep[e.g.,][]{Chambers1987, McCarthy1987, McCarthy1993, Pentericci1999,Pentericci2001}. More generally, the discovery that the mass of the central supermassive black hole is related to its host galaxy and dark matter halo \citep[][]{Magorrian1998,Gebhardt2000,Ferrarese2000, Neumayer2004}
implicates radio galaxies as important objects for understanding the formation and evolution of massive galaxies.

%spitzer
The \spitzer\ {\it Space Telescope} \citep{Werner2004} provided host galaxy masses of these objects, that place them among the 
most massive galaxies in the Universe \citep{Seymour2007}. \spitzer\ also allowed the 
first characterisation of their mid-infrared (MIR) emission, which showed hot dust close to the supermassive black hole \citep{Haas2008,DeBreuck2010}. 
In the orientation-based Unification scheme \citep[e.g.][]{Antonucci1993}, the hot dust near central active galactic nuclei (AGN) is completely obscured by material along the line of sight in type 2 AGN, and becomes gradually less obscured when transitioning to type 1 AGN \citep[e.g.,][]{Leipski2010,Drouart2012,Rawlings2013}.
\spitzer\ observations also revealed that high-redshift radio galaxies are preferentially found
in denser environments, consistent with galaxy clusters in formation \citep[e.g.,][]{Mayo2012,Galametz2012,Wylezalek2013b,Wylezalek2013a}.

%submm
The availability of sensitive submm bolometer arrays, such as SCUBA or LABOCA \citep[][]{Holland1999,Siringo2009} allowed 
the first detection of dusty star formation at high-redshift through the negative K-correction \citep{Blain1999}. 
Extensive surveys at $z$$>$3 showed that radio galaxies exhibit high submm flux, 
indicating vigorous, on-going star formation \citep[e.g.,][]{Archibald2001,Stevens2003,Reuland2004}, 
similar to submm galaxies \citep[SMGs, e.g.,][]{Borys2003,Coppin2006,Weiss2009} at similar redshift. 
However, single-dish observations are limited in spatial resolution (typical $20$\arcsec), and
only a handful of sources have been observed with interferometric facilities \citep{DeBreuck2005,Ivison2012}, 
which makes general conclusions on the entire sample of high-redshift radio galaxies difficult.

%infrared
The \herschel\ {\it Space Observatory} \citep{Pilbratt2010} covers the 60-700\mum\ range, allowing the first 
sensitive exploration of the entire infrared (IR) spectral energy distribution (SED) of galaxies in the distant Universe. This part of the SED is of utmost 
importance because most of the energy of star-forming galaxies is emitted at IR wavelengths. The IR comes
from the emission of dust that has absorbed a significant fraction of the UV and optical emission from young stars \citep[e.g.,][]{Chary2001,Sanders1996}. 
AGN emission by dust also contributes mainly in the IR, which can make it difficult to
determine the origin of the IR emission. While the submm emission is generally associated with cold dust and therefore with star formation,
the complete IR SED allows for a simultaneous characterisation of the AGN and star formation emission 
\citep[e.g.,][]{Tadhunter2007,Dicken2009,Mullaney2011,Feltre2012,Rosario2012,Drouart2014,Leipski2014}. 
The transition regime between dust heated by the AGN and dust heated by star formation is at 40-60\mum\ and is therefore important to observe
because it affects the total infrared emission and characterisation of each component.

% models
As the wavelength coverage of galaxies increased over the past decade, more advanced models were developed to
predict the emission of galaxies and AGN for given evolutionary scenarios and/or morphologies. The literature now offers
a large number of these models; we only mention here the models using stellar libraries and scenarios of evolution to
predict the emission of the integrated emission of the galaxy over a wide wavelength range \citep[e.g.,][]{Leitherer1995,Fioc1997,Maraston1998,Bruzual2003}. 
As the high-redshift infrared observations are becoming increasingly common, the models have extended their coverage
to take the dust emission into account \citep[e.g.][ for a recent review, see \citealt{Conroy2013}]{Leitherer1999,Burgarella2005,Maraston2005,daCunha2008,Groves2008,Noll2009}
The main characteristic of \pegaseiii\, we use here are to coherently predict attenuated metal-dependent SEDs extended to the 
dust emission in coherence with metal abundances. The code has been used in RV13 and will be detailed in \cite{Fioc2014}.
Simultaneously, numerous AGN torus models have been developed to characterise nearby AGN. These
models predict the UV to far-infrared (FIR) SED as a function of torus geometry, solving the radiative transfer equations. Different approaches have been used
to describe the torus properties, particularly to reproduce the 9.7\mum\ silicate feature. The two model types assume either
a continuous dust distribution or a clumpy distribution \citep[e.g.,][ for a recent review, see \citealt{Antonucci2012}]{Pier1992,Honig2006,Fritz2006,Nenkova2008a,Stalevski2012}. 

%outline
In this paper, we present the first in-depth analysis of the different emission components present
for a sample of powerful radio galaxies, handling data continuously covering 
UV to submm wavelengths. Disentangling the different components of a galaxy is particularly 
challenging in the UV-NIR domain because several components can contribute simultaneously to the 
emission and to highly variable proportions. The FIR brings an essential piece of 
information as, thanks to the energy balance between absorption and emission, constraints on each component are added. We aim in this paper 
to take advantage of the broad wavelength coverage of both the data and models to reliably disentangle the different components
of a sample of high-redshift radio galaxies. To investigate the different components, we use two models,
one to model the host galaxy and one to model the AGN. 

%attempts
It is difficult to simultaneously characterise the UV-to-submm SED with AGN and galaxy evolution codes: (i) the multiple components mean an increased number of free parameters, (ii) an efficient fitting procedure is challenging, (iii) a sample of sources needs to be constrained that includes the AGN and host contribution at similar 
levels, and (iv) homogeneous coverage in wavelength for these samples is necessary. For these reasons, only a few attempts exist
 in the literature (although the number is increasing  thanks to the availability of new techniques and data). Most of these 
 attempts suffer from simplifying one component to study the SED, either the host galaxy or the AGN, 
 depending on the sample or the scientific goals of the study \citep[e.g.,][]{Feltre2012,Trichas2012,Rocca2013,Banerji2014,Karouzos2014,Leipski2014,Ciesla2015}.

This paper is structured as follows. We present the 
sample and the data  in Sect. \ref{sec:sample}. 
Section~\ref{sec:fitting} describes the fitting procedure and the model used in this analysis. We
present the results in Sect. \ref{sec:results} and discuss their implications in Sect. \ref{sec:discussion}.
Throughout this paper, we assume the standard concordance cosmological model
\citep[$H_0= 70$\,kms$^{-1}$\,Mpc$^{-1}$, $\Omega_{\Lambda}=0.7$, $\Omega_{M}=0.3$,][]{Spergel2003}.

\section{Sample and data}
\label{sec:sample}

\begin{figure}[t]
    \centering
    \includegraphics[width=0.5\textwidth]{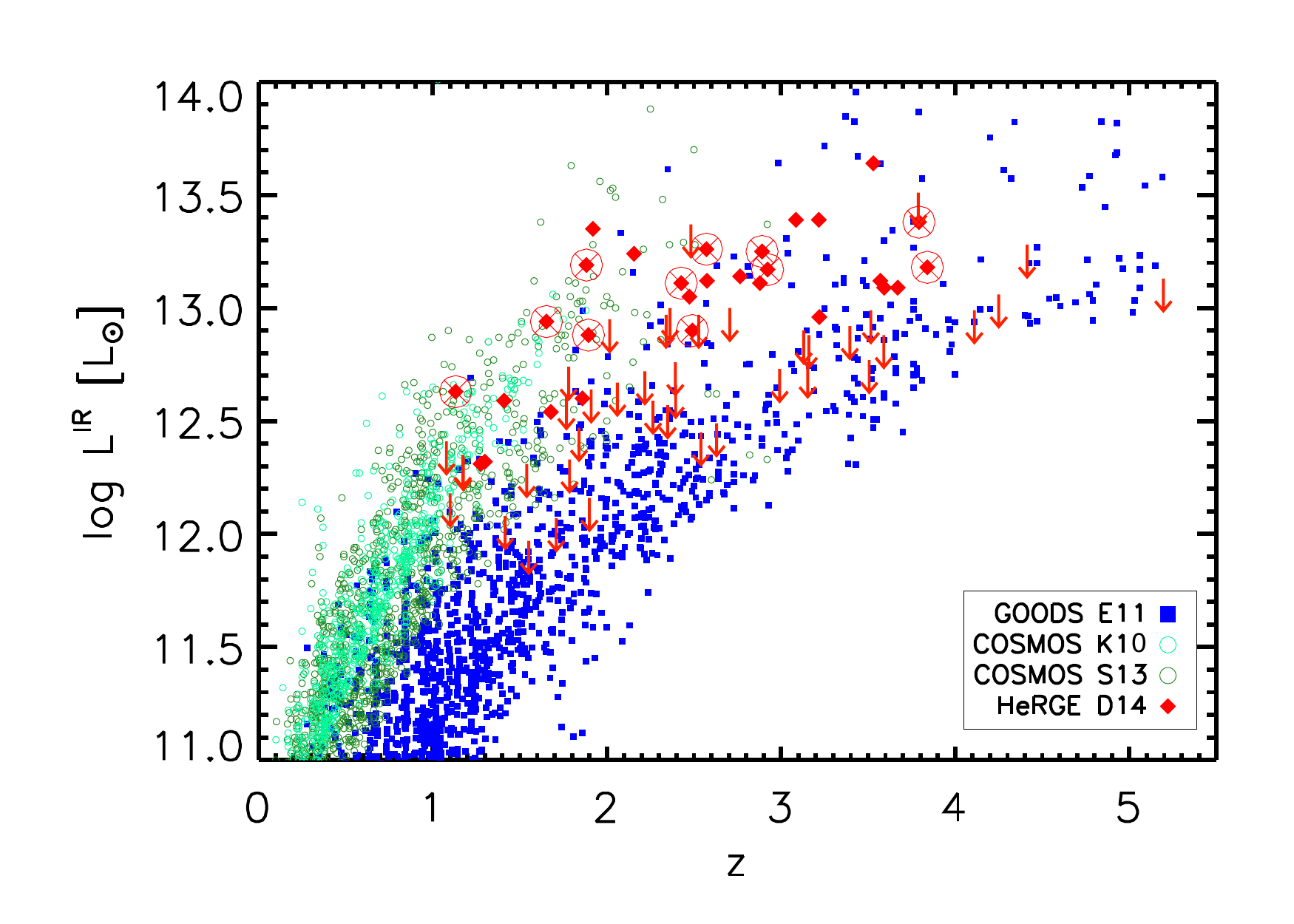}
    \caption{Total infrared luminosity versus redshift for several samples 
    available in the literature. The light green dots show the 
    COSMOS sample from \citet[][]{Kartaltepe2010}, making use of \spitzer\ data. 
    The blue squares denote the GOODS samples from \citet{Elbaz2011} 
    and the dark green dots represent the selection from \citet{Symeonidis2013}, both making use of \herschel\ data. The red circles indicate
    the subsample considered in this paper. The red diamonds show the sample of HzRGs presented in \cite{Drouart2014}. }
    \label{fig:arpege_sample}
\end{figure}

\begin{table*}[t] \centering
\caption{Summarised characteristics of our sample. Columns 2 and 3 indicate the 
spectroscopically confirmed redshifts and their references. 
Column 6 refers to morphological peculiarities at optical/NIR wavelengths: {\it companion:} source 
close to the object identified as the radio galaxy (few arcsecond) and {\it components:} potential substructures when detected in the images.
Column 7 indicates the alignment of the radio jet axis compared to the optical/NIR extension when detectable in the images. Column 8 lists the restframe 3GHz luminosity from \cite{DeBreuck2010}. The penultimate column indicates the polarisation and the band.}
\label{tab:arpege}
\makebox[0.8\textwidth]{
\begin{tabular}{l c c c c c c c c c}
\hline
Name & $z$& Ref. & RA & Dec & Morph. peculiarities & Radio jet axis & log $L_{\rm 3GHz}$ & Polarisation & Ref. \\
& spec. & & [J2000] & [J2000] & & & [W Hz$^{-1}$] & [\%] & \\
\hline \hline
  3C~368               &  1.132 &$a$ & 18:05:06.37  & $+$11:01:33.1 & two components      &  $aligned$    & 27.63 & $7.6 \pm 0.9$ (V)       & $g$ \\
   3C~470              &1.653   &$b$ & 23:58:36.00  & $+$44:04:45.0 & two components      &  $misaligned$ & 28.24 &      \nodata                 & \nodata \\   
 MRC~0324-228          & 1.894  &$c$ & 03:27:04.54  & $-$22:39:42.1 & one close companion & \nodata       & 28.49 & $6.5^{+3.8}_{-3.3}$ ($R$) & $h$ \\ 
PKS~1138-262           & 2.156  &$c$ & 11:40:48.38  & $-$26:29:08.8 & multiple companions &  $aligned$    & 28.14 & \nodata                 & \nodata \\ 
MRC~0406-244           &  2.427 &$d$ & 04:08:51.46  & $-$24:18:16.4 & multiple components &  $aligned$    & 28.11 & $2.5^{+2.3}_{-2.5}$ ($I$) & $h$ \\ 
MRC~2104-242           &  2.491 &$e$ & 21:06:58.28  & $-$24:05:09.1 & two components      &  $misaligned$ & 27.88 & \nodata                 & \nodata \\ 
USS~0828+193           &  2.572 &$c$ & 08:30:53.42  & $+$19:13:15.7 & multiple components &  $aligned$    & 27.47 & 10.0$\pm 2.0$ (spec)     & $i$ \\ 
 4C~28.58              &  2.891 &$c$ & 23:51:59.20  & $+$29:10:29.0 & multiple components &  $aligned$    & 27.89 & \nodata                 & \nodata \\ 
 USS 0943-242          & 2.923  &$c$ & 09:45:32.73  & $-$24:28:49.7 & one close companion & $aligned$     & 27.95 & $6.6 \pm 0.9$ (spec)    & $i$ \\
  4C~41.17$^a$         &  3.792 &$c$ & 06:50:52.23  & $+$41:30:30.1 & multiple components &  $aligned$    & 28.17 & $<$2.4 (spec)           & $j$ \\ 
 TN~J2007-1316$^a$     & 3.840  &$f$ & 20:07:53.26  & $-$13:16:43.6 & one close companion & \nodata       & 27.79 & $\sim3$ (spec)          & $k$ \\ 
\hline
\end{tabular} }
\tablebib{($a$) \cite{Meisenheimer1992}; ($b$) \cite{Hewitt1991}; ($c$) \cite{Rottgering1997}; ($d$) \cite{McCarthy1996}; ($e$) \cite{McCarthy1990}; ($f$) \cite{Bornancini2007}; ($g$) \cite{diSerego1989}; ($h$) \cite{Buchard2008}; ($i$) \cite{Vernet2001}; ($j$) \cite{Dey1995}; ($k$) \cite{Rocca2013}.}
\end{table*}

To reliably separate the different emission components of radio galaxies, a wide and 
 homogeneous coverage of the SED is of prime importance. The \herge\ 
 sample \citep[\textit{Herschel} Radio Galaxy Evolution sample][]{Drouart2014} 
 is ideally suited to estimate the relative contributions of 
 AGN and their host galaxies thanks to the  existing 
\spitzer, \herschel\ and submm data. We  briefly summarise here 
the criteria selected to build this sample. The radio galaxies were selected 
to cover homogeneously the radio luminosity-redshift plane, applying the criteria 
$L^{\rm{3\,GHz}}>10^{26}$\,WHz$^{-1}$, where $L^{\rm{3\,GHz}}$ is
the total luminosity at a rest-frame frequency of 3\,GHz \citep[Table 1;][]{Seymour2007}. 
To complete our SED coverage from UV to submm, we therefore defined a subsample with the following criteria. Each source (i) must have at least 
four \herschel\ detections, (ii) must have at least two broad-band photometric observations bracketing the discontinuity at 4000\AA\ 
restframe, and (iii) the filter response curves should avoid strong emission lines 
as much as possible.
We aim with this selection to obtain more than ten data points from UV to submm wavelengths as homogeneously as 
possible to allow for a reliable fit. In total, 11 sources remain out of 70 from the \herge\ sample (see Fig. \ref{fig:arpege_sample}).
In addition, a polarisation measurement at restframe 1500\AA\ is beneficial, if available, to measure the 
scattered AGN light. Of our eleven sources, seven have polarisation measurements (from broad-band photometry or spectroscopy, 
see Table \ref{tab:arpege} and the detailed discussion in Sect. \ref{sec:polarisation}).
The sources span 1$<$$z$$<$4, with a median redshift $\langle z \rangle$=2.5. The requirement of \herschel\ detections biases this 
sample towards the brightest IR emitters, with \lir$>$4$\times$10$^{12}$\lsun\ \citep{Drouart2014}. When \lir is taken into account, these radio galaxies cannot be distinguished from ULIRGs detected by other samples (see Fig. \ref{fig:arpege_sample}). 
The two highest redshift sources, 4C~41.17 and TN~J2007-1316 are extensively analysed in \citealt[][hereafter RV13]{Rocca2013}. 
They had been selected on the basis of their weak AGN contribution in the UV and optical to test the method, which is
further improved in this paper.  We now extend the sample to 11 sources with a wider range of AGN contributions.

\subsection{Building the SED from broad-band photometry}

% photometry 
We analysed galaxies with multi-wavelength data, therefore the consistency of the photometry throughout the entire spectral range
is essential. Specifically, the data from the optical/NIR and 
FIR domain have very different spatial resolutions. The optical data have sub-arcsec resolution 
(\hst), while the \herschel\ beam can reach $\sim$35\arcsec (FWHM). For this reason, we report the total flux when we compile photometry from the literature. The total
flux was calculated by applying corrections to aperture photometry based on the user manual or
primary reference of each corresponding instrument, or by selecting a large aperture, that is, $>$30\,kpc ($>$4\arcsec) or
64\,kpc ($\sim$8\arcsec). 

%emission lines
Based on the filter response and redshift of the source, 
strong emission lines (i.e., Ly$\alpha$, [OII], [OIII], H$\alpha$...) often fall in a 
band, increasing the measured flux relative to the pure continuum contribution (by up to $\sim$40\%). 
If spectroscopic observations are available for the contaminated bands, we subtracted the estimated flux of the line
based on the spectroscopy. All reported fluxes in the tables are corrected and therefore correspond to 
 pure continuum emission for the remainder of this paper (see Table \ref{tab:3C368_data}-\ref{tab:tnj2007_data}). 
 Only MRC~0406-244 does not have the information on line contamination, and fluxes are used as given.

\subsection{Notes on individual sources}

The data set for each source is mainly rich and of high quality. Especially, morphological 
peculiarities can be isolated and help to understand the overall picture in term of 
evolution in each source. We report the notes on each source in the 
Appendix \ref{sec:individual}. We describe the combined high-resolution 
image and radio contours (when electronically available) in 
Sect. \ref{sec:discussion} along with the discussion on the evolutionary status of the galaxy.

\section{Models, fitting procedure, and template libraries}
\label{sec:fitting}

\begin{table*}[ht] \centering
\caption{Fixed and fitted parameters for our library of evolved, SB, and AGN templates. We combined the AGN and 
SB libraries to create hybrid templates. Considered ranges and values  in 
the fitting are reported. In particular, the age of the starburst, $t_{\rm SB}$, must be younger than the age of the 
evolved component, $t_{\rm evolved}$, which in turn must be younger than the age of the Universe at the 
considered redshift. The single stellar population (SSP) is without infall, has no galactic winds, 
and is instantaneous.} 
\label{tab:param}
\begin{tabular}{cccc}
\cline{2-4}
& Evolved & SB & AGN \\
\hline \hline
\multirow{3}{*}{{Fixed parameters}} 
& Hubble type [Sa, S0, E, E2]  & SSP scenario & Opening angle $\theta$=[40\degree] \\
& IMF [Kroupa] & IMF [Kroupa] & Radial density profile [$\alpha$=0] \\
& & & Azimuthal density profile [$\gamma$=2] \\
\hline
\multirow{4}{*}{{Fitted parameters}} 
& Normalisation & Normalisation & log$_{10}$\relagn=[-3,-2,-1.7,-1,-0.7,0,0.7,1,2,3] \\
& Age, $t_{\rm evolved}$[$<$$t_{\rm Universe,z}$] & Age, $t_{\rm SB}$[$<$$t_{\rm evolved}$] & Inclination $i$=[40,50,60,70,80,90] \\
& & Column density factor $K$=[1,10]  & Size $Y$=[10,60,150] \\
& & Initial met. $Z_{\rm init}$=[0,0.0005,0.001,0.005,0.01] & Opacity $\tau_{9.7 \mu \rm m}$=[0.1,1,10] \\
\hline
\end{tabular}
\end{table*}

The broad wavelength coverage from UV to FIR allows us to probe multiple regimes of energy and therefore to
better isolate emission from specific components of the radio galaxy. 
The final goal is to disentangle the AGN and stellar emission through SED fitting.
We made use of two models, the AGN model of \citealt[][]{Fritz2006} and the stellar population model \pegaseiii\ for the host galaxy \citep[][]{Fioc2014}.
We present the fitting procedure, considered models and template libraries for each component (AGN and galaxy) 
in the following sections. We also explain in detail the creation of the new templates and the assumptions made
on both models to limit our total parameter space.
 
\subsection{Fitting procedure}
\label{sec:fitting_procedure}

The fitting procedure is presented in RV13. The code allows fitting of up to two stellar components
simultaneously based on $\chi^2$ minimisation by automatically exploring a library of templates. 
To adapt the fitting procedure from \pegaseii\ to the new code \pegaseiii, 
we made the following modifications: (i) we added an estimate of the 
uncertainties on the fitted quantities, (ii) modified the graphical 
displaying procedure to include the AGN component, (iii) extended 
the filter database, and (iv) updated the compatibility with the new outputs of \pegaseiii. 

Adding the AGN component was challenging in the context of the two-component fitting procedure without 
modifying a large part of the core procedure. We solved this difficulty by creating hybrid 
templates composed of the sum of the evolving starburst (SB) and AGN components (see Sect. \ref{sec:hybrid}). 
The principal problem in adding new components stems from the degeneracies induced
by the increase in the number of free parameters. We addressed this specific problem by 
combining a broad and homogeneous SED coverage (with 12-19 data points) and 
carefully choosing models and parameters: the self-consistent treatment of the 
optical and IR emission of the AGN and \pegaseiii\ models (calculated with radiative transfer codes), 
enabled us to leave a significant number of parameters free with only limited parameter 
degeneracies (see Sect. \ref{sec:discussion} for a discussion of the uncertainties).

Our approach uses grid-model fitting and requires brute-force calculation through the 
parameter space. We therefore set some parameters to physically justified values and took 
representative models to ensure that the code remained executable 
on a desktop machine in a reasonable computing time. More 
extensive exploration of the parameter space would require a new fitting algorithm
using a Bayesian approach and Monte Carlo sampling of the parameter space, but this
is beyond the scope of this paper because it requires a thourough redesign of the core procedure. Although not complete, the coverage of the parameter 
space is still satisfying, with $>$10$^7$ templates tested for each source. 

\subsubsection{\pegase\ model and library}
\label{sec:pegase_library}

We used the \pegaseiii\ code to generate a library of galaxies and starbursts.
The new code \pegaseiii\ is a coherent spectro-chemical evolution model
predicting simultaneously the metallicity enrichment, the corresponding SED, and 
the attenuation or emission by dust grain models calculated with radiative transfer
Monte Carlo simulations, taking into account scattering and stochastic heating of grains. 
The continuous synthetic SEDs that extend from the far-UV to submm wavelengths, 
were built by minimising the input parameter numbers to avoid large degeneracies in the solutions. We 
characterised each star formation scenario with only four free parameters: 
star formation rate, initial mass function (IMF), inflows and
outflows. We refer to \cite{Fioc2014} for a more detailed presentation of the code \pegaseiii\ and
its detailed documentation, which is available on the \pegase\ website\footnote{\url{www2.iap.fr/pegase}}. In the present
version, used hereafter as in RV13, the star formation parameters by spectral types are taken from \cite{LeBorgne2002},
fitting the Hubble sequence galaxy properties in the local Universe. 
These parameters are set to reproduce observed local galaxies from 
SDSS \citep{Tsalmantza2009,Tsalmantza2012}, including evolution 
with a formation at $z_{\rm form}$=10. We focused our library on four galaxy types, 
E, E2, S0, and Sa to test our sensitivity to the galaxy type when modelling the radio galaxy host. 
Despite their names, these templates refer to different star formation histories and not 
to specific morphologies, see \cite{LeBorgne2002}.

The starburst (SB) templates are defined differently than the evolved, main component. 
They are single stellar populations (SSPs), assuming a short formation in one step of time, that is, 
1Myr ($\delta$ function). SSPs do not take into account any infall or
galactic winds. We discuss the effect of this 1Myr formation assumption in detail in Sect. \ref{sec:starburst}.
We adopted a Kroupa IMF \citep{Kroupa2001} and a range of initial metallicities of the gas ($Z_{\rm init}$). 
RV13 showed that the column density plays a large role in the observed SED. 

We explored the same quantities for the evolved population with the SB component,
that is, the age and mass of the stellar populations, but we also added the initial metallicity and
 the column density factor $K$ (see RV13, Sect. 3). Table \ref{tab:param} reports the range of values considered in our 
 library. We here explore the age and mass of the starburst to determine whether the ten times higher
 column density preferentially found in radio galaxies by RV13 
is confirmed for a larger sample. We also note that the fitting procedure requires the evolved component to be younger than  the age of the Universe at a given redshift (t$_{\rm evolved}$$<$t$_{\rm Universe,z}$)  
and the young component to be younger than the evolved component (t$_{\rm SB}$$<$t$_{\rm evolved}$; see also Table \ref{tab:param})

\subsubsection{AGN model and library}
\label{sec:agn_library}

The AGN component was incorporated using the AGN model from \citealt[][]{Fritz2006}. To briefly describe this model:
a central, point-like source emits an SED composed of a sum of power laws.
The torus is defined in spherical coordinates by its geometry (size and opening angle), 
dust profile density (radial and azimuthal), opacity at 9.7\mum\ and inclination with respect to the 
observer. The torus is divided into cells assuming an axisymmetric geometry
and the radiative transfer equations are solved using the $\Lambda$-iteration 
technique (see \citealt{Fritz2006} for a complete description of the model).
\cite{Feltre2012} showed that the \cite{Fritz2006} model is directly 
comparable to the clumpy model from \cite{Nenkova2008a} for a certain set of parameters 
(only partially covered in our current library).

We adopted the \cite{Fritz2006} model for the present study, leaving the comparison with
clumpy models for further publications. Moreover, a precise characterisation of the torus properties is beyond the scope of this paper, 
because it requires spectroscopy around the silicate feature at $9.7$\mum\ restframe. 
We focus on the largest impact parameters, that is, the size, the opacity, and the inclination of the torus. 
The size, $Y$, is the ratio $R_{\max}$/$R_{\rm min}$, where $R_{\rm min}$ is set as the 
sublimation radius of the dust (which, in turn depends on the luminosity of the central source). 
The opacity, \tautorus, corresponds to the integrated opacity at 9.7\mum\ along 
the equatorial axis ($i$=90\degree). The dust density (radial and azimuthal) of the 
torus is calculated with respect to the size, opacity
and profile chosen by the user. These values were set to their default values here, radially 
constant and azimuthally decreasing as the square of the altitude (they also have a limited effect on the SED).
The inclination, $i$, is the orientation of the torus with respect to the observer. An
edge-on view corresponds to $i$=90\degree, and $i$=0\degree, corresponds to a 
face-on view. The opening angle, $\theta$, was set to $\theta$=40\degree, as suggested by previous studies 
\citep[e.g.,][]{Barthel1989,Drouart2012}. In practice, all inclinations $i$$<$50\degree present a line 
of sight free of dust and therefore similar SEDs. To save computing time but include the possibility of a type 1 AGN,
we only used $i$=40\degree as a type 1 AGN template. Table \ref{tab:param} 
presents all the assumed values for the fixed and fitted parameters. The objective of this selection is to 
explore typical torus configurations and evaluate how our method handles the addition of this third (AGN) component. 

\subsection{Hybrid AGN-SB component}
\label{sec:hybrid}

We created hybrid templates, corresponding to the sum of the SB templates produced by \pegaseiii\ 
and an AGN torus. For each evolving SB template, we normalised an AGN torus (for a given inclination $i$, size $Y$, opacity $\tau_{9.7 \mu m}$)
to a fraction of the total flux at 20\mum\ following the equation:

\begin{equation}
F_{\rm hybrid}(t) = F_{\rm SB}(t) + F_{\rm AGN}(i,Y,\tau_{9.7 \mu m}) \times f_{\rm AGN}^{20\mu m},
\label{eq:hybrid}
\end{equation}

\noindent{where} $F_{\rm hybrid}$ is the sum of the AGN and the SB templates,
$F_{\rm SB}$ is the SB template, $F_{\rm AGN}$ is the AGN template and $f_{\rm AGN}^{20\mu m}$ is 
the relative fraction of the AGN flux compared to the SB template at 20\mum\footnote{We evaluated \relagn\ with an 
idealised filter centred on 20\mum\ of 4\mum\ bandwidth to minimise 
effects from sharp gradients and/or emission and absorption lines (i.e. PAHs).} (see Table \ref{tab:param}). 
We performed a step-wise $\chi^2$-minimisation that permitting determining of intermediate 
solutions and checking for degeneracies.
In total, $>$10$^{7}$ templates were fitted for each source.

\subsection{Uncertainties of parameters, 68\% confidence intervals}
\label{sec:uncertainties}

Calculating uncertainties on the parameters can be challenging for a 
multi-dimensional analysis. However, as we performed brute-force calculation
through our parameter space, therefore each fit possesses a $\chi^2$ value, that indicates the
goodness of fit. We can therefore consider the $\chi^2$ distribution to gain insights into 
the confidence interval of the adopted free parameters. 
We used the description from Numerical recipes 
\citep[][Chapter 15]{NumericalRecipes} to estimate the uncertainties on the parameters,
using the complementary incomplete gamma function to calculate the corresponding $\Delta\chi^2$
corresponding to a 68.3\% confidence interval. For our ten free parameters, this corresponds to $\Delta\chi^2$=11.5. 
We defined the confidence interval as $\chi^2_{min}+\Delta\chi^2$ (see Figs. \ref{fig:results_3C368}---\ref{fig:results_TNJ2007} 
and Table \ref{tab:results}).

The uncertainties on stellar masses and luminosities were estimated with a different 
technique because they are directly expressed 
from the normalisation of the different components. We considered the extreme 
values of each mass within the allowed 68\% confidence interval
of all other parameters and defined this as the lower and upper uncertainties. 
For the AGN luminosity we added the uncertainty from the 
AGN fraction (\relagn) in quadrature and the normalisation of the hybrid component. 
Finally, we add a special caution of interpreting these uncertainties. 
They only represent the statistical uncertainties on the parameters induced from the noise in the observations 
and do not take into account potential systematics associated with the model 
choices and assumptions (see Appendix \ref{sec:systematics} for
a more extensive discussion of uncertainties and parameter calculation). 

\section{Results}
\label{sec:results}

\begin{figure*}[ht] \centering
\begin{overpic}[width=1.0\textwidth,angle=0,trim= 0 0 0 0,scale=1.0]{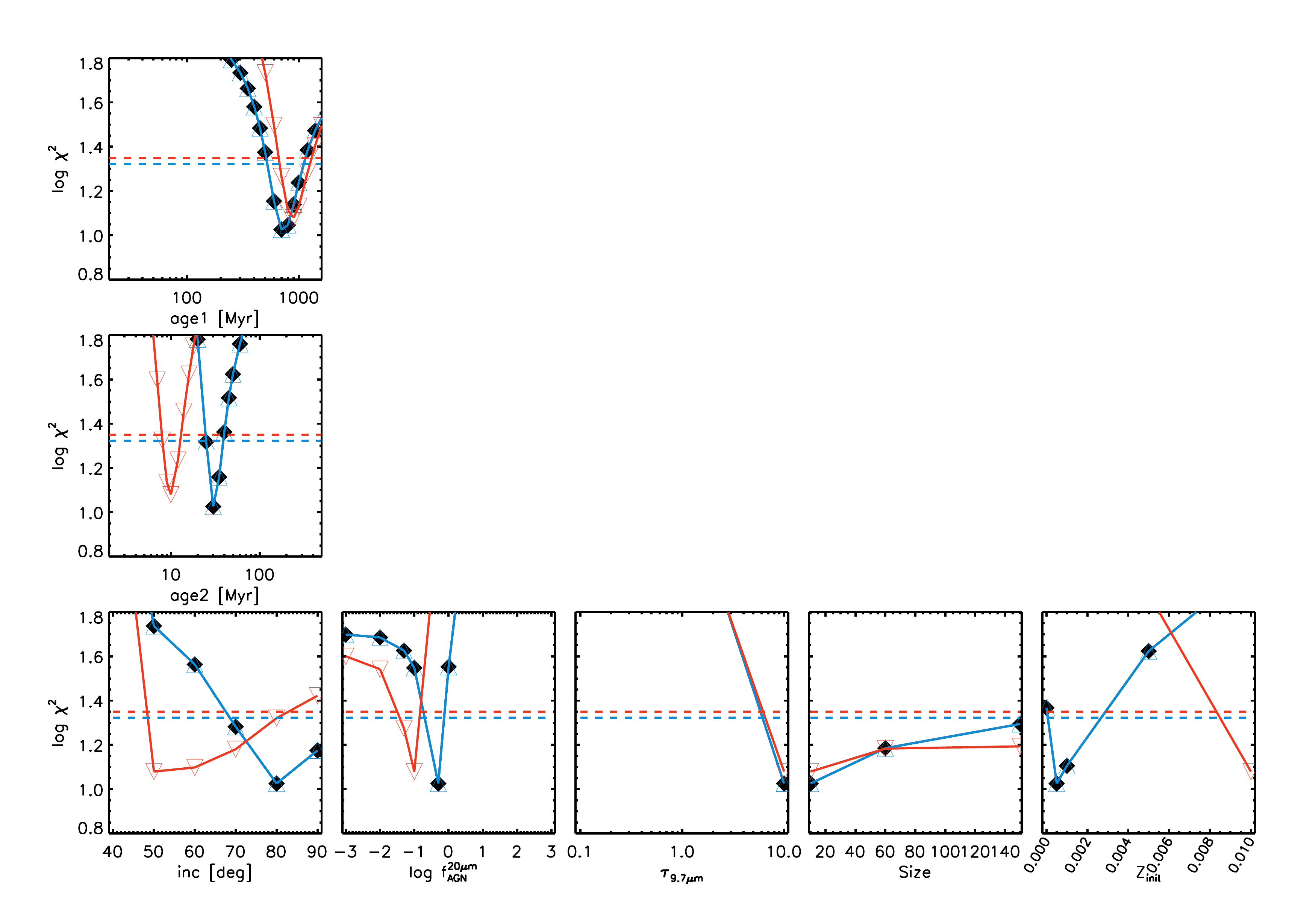}
\put(25,75){\includegraphics[height=0.7\textwidth,angle=270,trim= 0 0 0 0,scale=1.0]{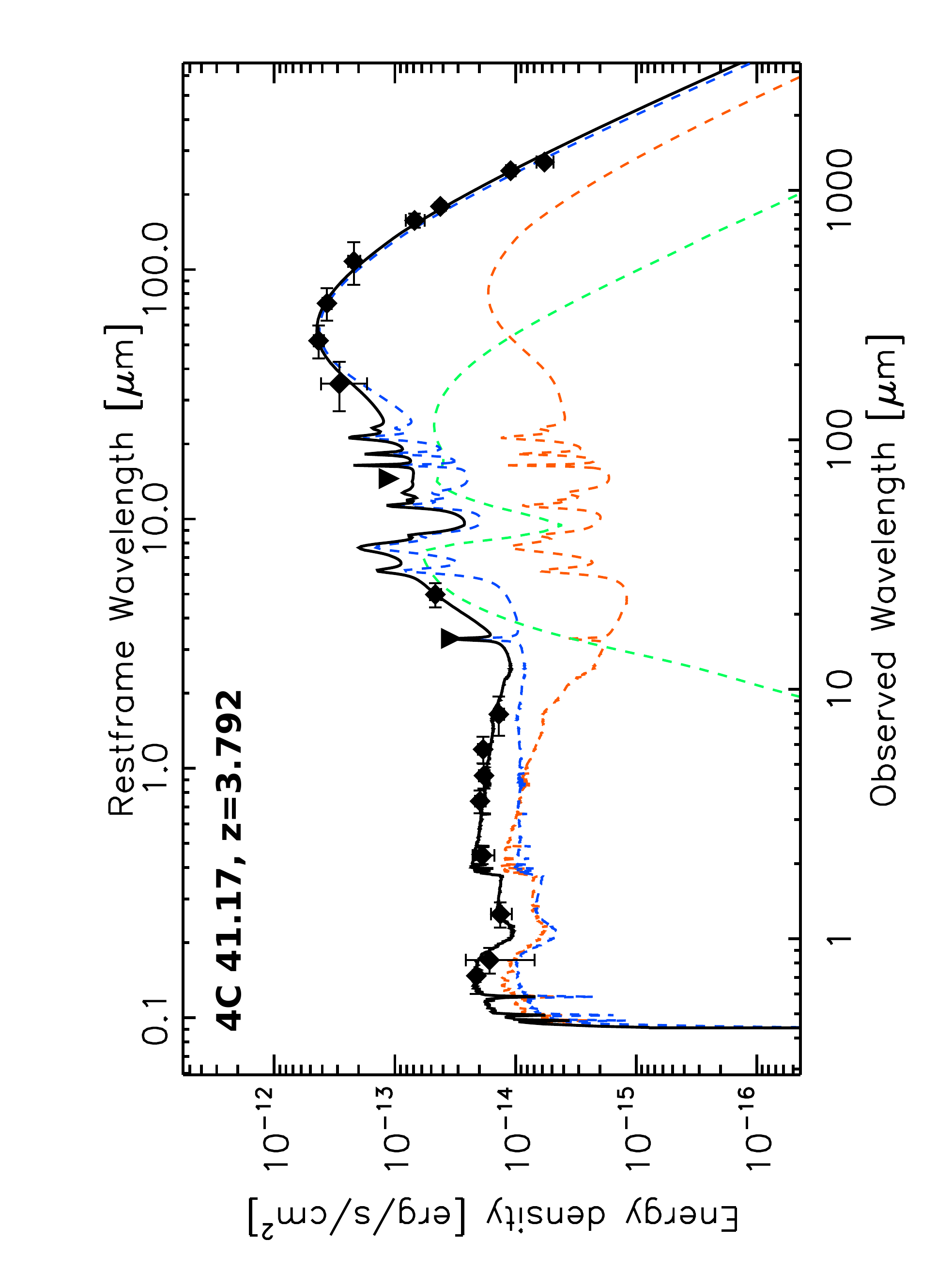}}
\end{overpic}
\caption{Best fit for 4C~41.17 ($z=3.792$). The orange, blue and green dashed lines symbolise the evolved, starburst, and AGN components, respectively. The sum of the components is the dark line, the black diamonds correspond to broad-band photometry, the vertical lines to the 1$\sigma$ uncertainties and horizontal lines to the FWHM of the filters. The downward triangles represent the 3$\sigma$ upper limits. The insets show the $\chi^2$ distribution for seven of the free parameters of the fit. From top left to bottom right: age of the evolved component (orange line in the SED), age of the starburst (blue line in the SED), inclination of the AGN torus (green line in the SED), fraction of the AGN at 20\mum, equatorial opacity of the torus, size of the torus and initial metallicity of the starburst. Black, blue and red indicate the three different approaches we used to estimate the polarisation effect, without, with the lowest and the highest contamination, respectively. We note that in this case the lowest polarisation contribution is zero, hence the black and blue lines are superposed. The horizontal coloured lines correspond to the 68\% confidence interval described in Sect. \ref{sec:uncertainties}.}
\label{fig:results_4C4117}
\end{figure*}

\begin{table} \centering
\caption{Best $\chi^2$ for one (evolved only), two (evolved and starburst), and three (evolved, starburst, and AGN) components. We also indicate the number of degrees of freedom $dof$ in brackets
$dof=N-p-1$ where $N$ is the number of data points and $p$ is the number of free parameters.}
\label{tab:chi2_comp}
\begin{tabular}{lccc}
\hline
Name & 1 comp. & 2 comp. & 3 comp. \\
\# free parameters & 2 & 6 & 10 \\
\hline \hline
       3C\,368            & 3615         [17] & 101      [13]   & 73   [9] \\
       3C\,470            & 4000         [12] & 71        [8]     & 32   [4] \\
     MRC\,0324-228 & 2152         [11] & 57         [7]    & 25    [3] \\
     PKS\,1138-262  &  14896      [14] & 834       [10]  & 79    [6] \\
     MRC\,0406-244 & 1208         [14] & 65         [10]  & 19    [6] \\
     MRC\,2104-242 & 167           [10] & 20         [6]    & 9      [2] \\
     USS\,0828+193 &  5235        [12] & 619       [8]   & 36     [4] \\
      4C\,28.58          &  247          [13] & 87         [9]   & 29     [5] \\
     USS\,0943-242  & 250           [13] & 16         [9]   &  5       [5] \\
      4C\,41.17          &  491          [16] & 39         [12]  & 11    [8] \\
     TN\,J2007-1316 & 148           [13] & 73         [9]   & 9       [5] \\
\hline
\end{tabular}
\end{table}

We performed the SED fitting described in the previous sections on the 11 sources in our sample, taking
one (evolved component only), two (evolved and starburst), and three (evolved, starburst, and AGN) components into account. 
Table \ref{tab:chi2_comp} shows the best $\chi^2$ depending on the number of components. The addition 
of extra components clearly dramatically increases the quality of the fit. This improvement 
is particularly strong from one to two components and is 
still significant from two to three components (amelioration by a factor superior to two). We therefore only consider
the three-component fit in the remainder of this paper.  Table \ref{tab:results} 
summarises the results of the fitting, including their associated (68\%) uncertainties.
Figure \ref{fig:results_4C4117} shows an example  of the best-fit model for 4C 41.17, while the results for the 
remaining the sample are presented in Figs. \ref{fig:results_3C368}--\ref{fig:results_TNJ2007} in the Appendix. 
We present here the overall trends on the different components
for the entire sample. Overall, the fittings are satisfying for all sources even if 
some discrepancies are still observed (see Sect. \ref{sec:limitations}).  It is important to 
note that three components are necessary to reproduce the SED of the powerful radio 
galaxies in this sample, and each appears to be dominating  at a particular wavelength range.

\begin{figure*}[t]
\begin{minipage}[c]{0.46\linewidth}
    \includegraphics[width=1\textwidth]{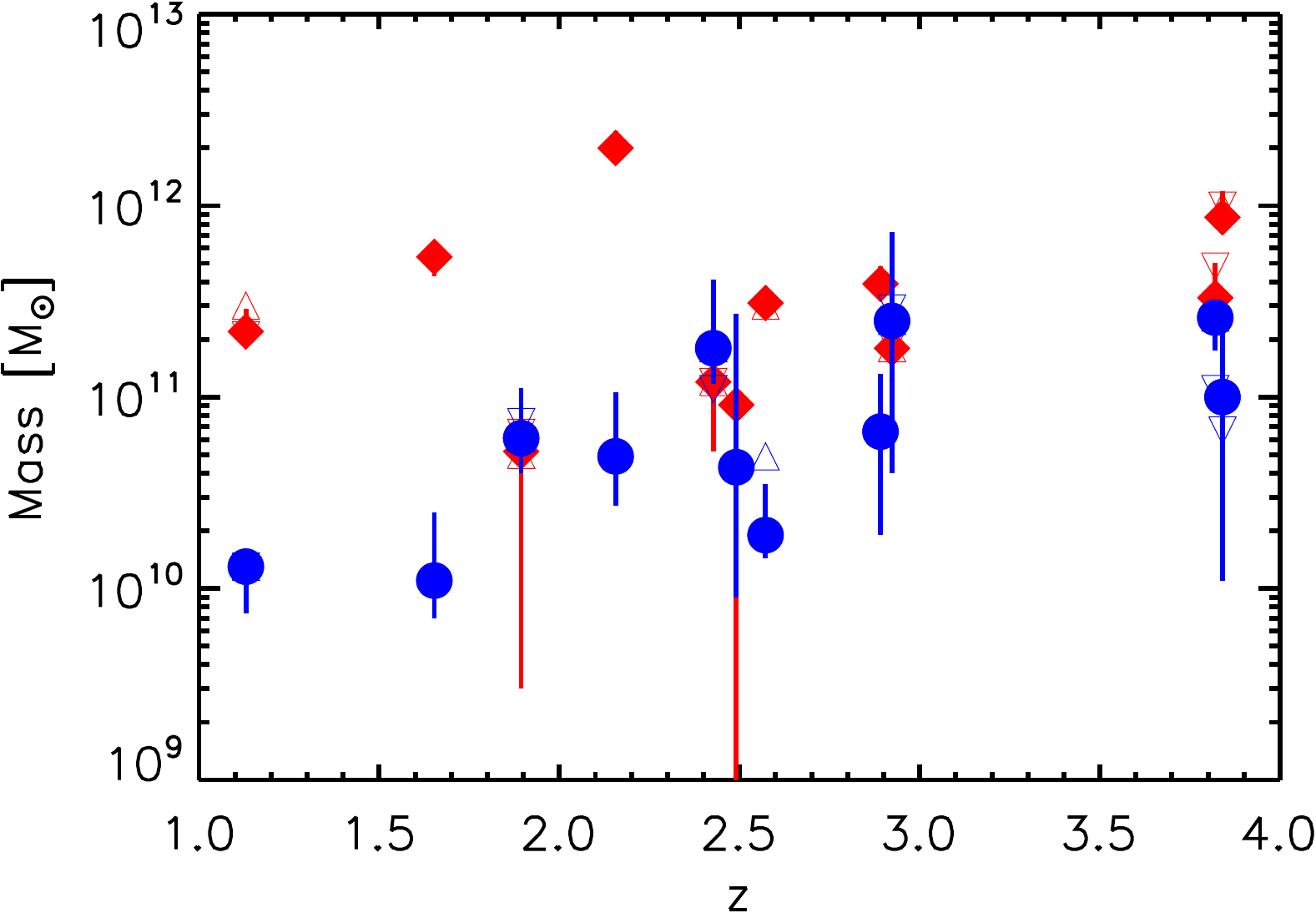}
\end{minipage}
\begin{minipage}[c]{0.46\linewidth}
    \includegraphics[width=1\textwidth]{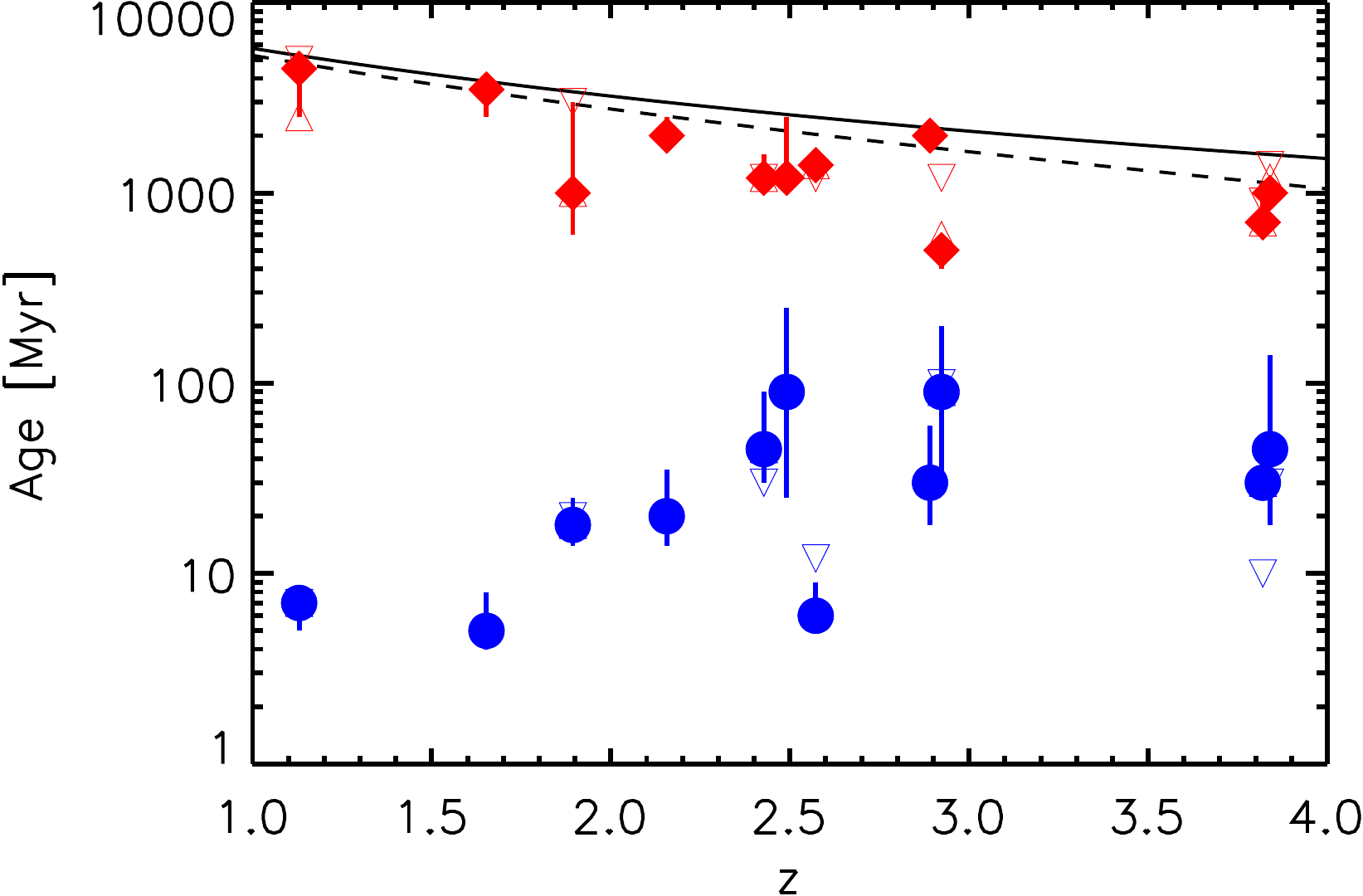}
\end{minipage}
    \caption{
    Two stellar populations, young and evolved, 
    shown as blue circles and red diamonds, respectively.
    The vertical lines indicate the 68\% confidence intervals from the fit. 
    When polarisation data are available, we
    report the fit with lowest and highest contribution from scattered AGN light as the
    downward and upward empty triangles respectively. For clarity, we do not 
    add the 68\% confidence intervals for the polarisation-subtracted fits. 
    {\it Left:} Masses of the two stellar populations versus redshift. 
    {\it Right:} Ages of the two stellar populations where
    the black line shows the oldest 
    universe age as a function of redshift assuming the standard cosmological concordance 
    model (Sect. \ref{sec:intro}). The dashed line represents the age of a galaxy for $z_{\rm form}$=10. 
    We note that the 68\% confidence interval can be smaller than point size and polarisation-subtracted fits are
    mainly contained in the 68\% confidence intervals when scattered AGN light is ignored.}
     \label{fig:z_mass}
\end{figure*}

\subsection{Evolved component}
\label{sec:evolved}

%mass
We call the oldest stellar component the evolved component ($>0.5$\,Gyr), which 
can be defined as the radio galaxy host. Figure \ref{fig:z_mass} 
shows the mass of the evolved stellar population as a function of redshift (red points).
The host galaxy appears massive at $z$=4, and this evolved massive component 
is present in every source throughout our redshift range (\mstel$>$10$^{11}$\msun, 
except for MRC~0324-228 with 10$^{10.7}$\msun). 
Previous studies have shown that the most powerful radio galaxies present exceptionally high 
masses \citep[e.g.][]{DeBreuck2002,Rocca2004,Seymour2007}.
This places powerful radio galaxies at the top of the galaxy mass distribution
seen in large field galaxy samples \citep[e.g.,][]{Marchesini2009,Ilbert2013}. 
Interestingly, these masses are also comparable to SMGs at similar redshift \citep[e.g.,][]{Borys2003}. 
This suggests that powerful radio galaxies are drawn from the same sample of massive galaxies,
formed at high-redshift ($z$$>$6) on a relatively short timescale. Similarly as in RV13, the 
evolved component clearly dominates the \spitzer\ photometry, consistent with the shift of the 
1\mum\ peak of the most evolved early-type galaxies.

%age
The right panel of Fig. \ref{fig:z_mass} confirms this result because all red points are clustered 
along the black lines (plain and dashed), consistent with an early formation epoch.
We also note that the four Hubble types used in our analysis provide similar $\chi^2$ values within the 68\% confidence ranges. 
We therefore conclude that low-resolution broad-band photometry alone cannot constrain the exact star formation history 
of the galaxies, but still favours a formation at early times in the Universe. We emphasise that the
ages determined by our method are well constrained, as illustrated in 
Figs. \ref{fig:results_3C368}-\ref{fig:results_TNJ2007} where the best $\chi^2$ solutions are located in deep minima. 

\subsection{Intense and massive starburst}
\label{sec:starburst}

Figure \ref{fig:z_mass} and Table \ref{tab:results} show two remarkable features of the young component
required to reproduce the strong \herschel\ and submm emission in our sample. Firstly, the starburst is extremely 
massive, \mstel$>$10$^{10}$\msun. In each case, it represents a significant fraction of the mass of the 
system (5\% to 50\%). These high-mass fractions indicate that even if the bulk of the stars has already been 
formed at high-redshift (Sect.  \ref{sec:evolved}), the radio galaxies in our sample still experience
vigorous star formation\footnote{We note that our sample selection requires bright IR emission 
($\ge$4 \herschel\ detections), implying the sample is likely biased towards strong star-forming systems.}.
Secondly, Fig. \ref{fig:z_mass} also shows that this starburst spans a wide range in age, 
5\,Myr$<t_{\rm SB}<$100\,Myr, indicating that star formation is on-going or in the aftermath of a violent event. 

%column density
It is also interesting to note that all but one (MRC~2104-242) of the galaxies have
a column density factor $K$=10. This indicates that starbursts associated with radio galaxies have an on average
ten times higher column density than the Milky Way\footnote{$N_{HI}$$\sim$6.8$\times$10$^{21}$ 
atoms cm$^{-2}$ is the standard galactic column density in \pegaseiii.}. This indicates that high-redshift 
radio galaxies, or at least the most IR-luminous radio galaxies, favour higher 
covering dust fractions, indicating a high obscuration, similar to SMGs in the same redshift range \citep[e.g.,][]{Chen2015}.
 
%metallicity
About half of the solutions favour a high initial metallicity (see Table \ref{tab:results}).
This could be expected for a starburst forming from IGM gas, which is
expected to be metal-rich at high-redshift \citep[e.g.,][]{Pieri2014}. Interestingly, the other half tends to have a low (or null)
initial metallicity, indicating a more pristine gas source. The evolution of the 
starburst implies ISM enrichment when SN explode and therefore the observed metallicity 
will be higher than the one reported here (initial metallicity of the gas).
Given that we used only broad-band photometry, the derived
initial metallicities should be treated with a grain of caution, even though they do provide some clues
about the overall metallicity of the initial gas of the starburst (null, sub-solar, or super-solar). The 
broad trend is well shown in Figs. \ref{fig:results_3C368}-\ref{fig:results_TNJ2007}, 
where the best-fit solutions are shown not to be located in 
deep and localised minima. Moderately high-resolution spectroscopic observations are necessary to further explore this part
of the parameter space because the emission lines are the result of the ISM photoionisation from both the young stars and the AGN. 

%SFR
We make a final note about the star formation rate related to the starburst component. By definition 
(in the model), all the star formation is taking place in a single time step (we discuss the duration of SB in Sect. \ref{sec:discussion}). When the fit converges
to an age $>$1\,Myr, the related star formation is therefore null. Moreover, the absence of infrared imaging at arcsec resolution and the
uncertainties about the duration of the burst make deriving a star formation rate from our fitting difficult. We therefore
used simple physical assumptions and empirical relations to estimate star formation rates from the starburst 
component (see Sect. \ref{sec:physical_properties}). 

\subsection{Bolometrically luminous AGN with an opaque torus}

%luminous, opaque, large
Our SED fitting allows us to derive some of the AGN and torus properties of HzRGs, such as the size, opacity and inclination.
With Eq. \ref{eq:hybrid} we calculated the 
corresponding intrinsic luminosity of the AGN (Table \ref{tab:results}, last column).
All our AGN are luminous, with \lbolagn$\ga$10$^{12}$\lsun, most with \lbolagn$\approx$10$^{13-14}$\lsun. Unfortunately, 
the small span of properties of our sample (in terms of luminosity or mass), usually only covering one order of magnitude, does not
allow us to identify any correlation between the AGN and host properties. 
Interestingly, these values are similar to optically luminous quasars observed at similar redshift \citep[e.g.,][]{Kollmeier2006}. This is 
consistent with radio galaxies experiencing a strong radiation-dominated accretion phase with copious accretion onto the
central supermassive black hole. In Figs. \ref{fig:results_4C4117} and \ref{fig:results_3C368}-\ref{fig:results_TNJ2007}, 
we show that the relative contribution at 20\mum\ and therefore the AGN luminosity is well constrained. 

Except for 3C\,368, the tori in our sample appear opaque, with \tautorus$\ge$1.0.
This opacity corresponds to $A_V$$>$18 using the \cite{Draine2003} conversion. The obscuration is also related to 
the inclination. The inclination represents the amount of dust on the line of sight. Most of our best fits
 converge to solutions with $i$$>$40\degree, indicating Type 2 AGN (edge-on view), as expected for radio galaxies. 
The one exception is 3C\,368, the source with the lowest redshift in our sample, with $i$=40\degree. 
The best solutions vary from $Y$=10 to $Y$=150 for the size.
As we calculated the intrinsic luminosity, we can estimate the
physical size of the torus \citep[see Eq. 1 of ][]{Fritz2006}. Taking the median of our sample, \lbolagn$\sim$2.5$\times$10$^{13}$\lsun\ 
translates into $R_{\rm min}$$\sim$4.1 pc and $R_{\rm max}$$\sim$41\,pc, $\sim$246\,pc, and $\sim$616\,pc
for $Y$=10, 60, and 150, respectively. We note the difficulties in constraining the torus properties are expected because even with dedicated samples in the nearby Universe, some parameters are only loosely
constrained \citep[e.g.,][]{AsensioRamos2009}. Interestingly, even after chosing the most relevant parameters 
for the AGN torus, the properties of the torus vary from source to source, with no clear trend on the entire sample
 (see Figs. \ref{fig:results_4C4117} and \ref{fig:results_3C368}-\ref{fig:results_TNJ2007}).

\begin{table} \centering
\caption{Summary of the fitting results. The $\chi^2$ is given in the last column of Table \ref{tab:chi2_comp}. The table is divided in three parts corresponding to each fitted component. For the evolved component: $Z_*$ refers to the average metallicity lock in stars for the given age. For the SB component: $Z_{init}$ refers to the initial metallicity of the gas, $K$ is the factor relating to the column density. For the AGN component: $Y$ refers to the size ratio between the outer and inner radius of the torus, $\tau_{\rm 9.7\mu\,m}$ the opacity at 9.7\mum, and $i$ to the inclination with respect to the observer. Asymmetric uncertainties of each parameter are provided as explained in Sect. \ref{sec:uncertainties}. We note that zero values of the uncertainties are due to the lack of coverage in the parameter space and are contained in the 68\% uncertainties, refer to Figs. \ref{fig:results_3C368}-\ref{fig:results_TNJ2007}.}
\label{tab:results}
\begin{tabular}{lc cccc}
\hline
&   \multicolumn{5}{c}{Evolved Component}  \\
\cline{2-6} 
Name    & Hubble  & Mass                & Age                       & $Z*$     & $L^{\rm tot}$                \\
                 &      type      & $log_{10}$[M$_{\odot}$]           & [Myr]                     &          & $log_{10}$[L$_{\odot}$]               \\
\hline \hline
       3C368 &   Sa &  $  11.3_{  -0.0}^{  +0.1}$ & $  4500_{-  2000}^{+   500}$ & 0.0112 & $  11.6_{  -1.3}^{  +0.2}$ \\
       3C470 &   Sa &  $  11.7_{  -0.1}^{  +0.1}$ & $  3500_{-  1000}^{+     0}$ & 0.0092 & $  12.0_{  -0.7}^{  +0.3}$ \\ 
     MRC0324 &  E2 &  $  10.7_{  -1.2}^{  +0.3}$ & $  1000_{-   400}^{+  2000}$ & 0.0264 & $  11.8_{  -0.0}^{  +0.0}$ \\ 
     PKS1138 &   S0 &  $  12.3_{  -0.0}^{  +0.1}$ & $  2000_{-   200}^{+   500}$ & 0.0142 & $  12.7_{  -1.0}^{  +0.2}$ \\ 
     MRC0406 &   E2 &  $  11.1_{  -0.4}^{  +0.1}$ & $  1200_{-     0}^{+   400}$ & 0.0259 & $  11.8_{  -0.3}^{  +0.3}$ \\ 
     MRC2104 &   E2 &  $  11.0_{  0.0}^{  +0.3}$ & $  1200_{-   200}^{+  1300}$ & 0.0259 & $  11.7_{   0.0}^{  +0.0}$ \\ 
     USS0828 &   E2 &  $  11.5_{   0.0}^{  +0.0}$ & $  1400_{-     0}^{+     0}$ & 0.0257 & $  12.2_{  -0.0}^{  +0.3}$ \\ 
      4C2858 &  S0 &  $  11.6_{  -0.0}^{  +0.1}$ & $  2000_{-   200}^{+     0}$ & 0.0142 & $  12.0_{  -1.1}^{  +0.2}$ \\ 
     USS0943 &   E &  $  11.3_{  -0.4}^{  +0.1}$ & $   500_{-   100}^{+   100}$ & 0.0114 & $  12.1_{  -0.2}^{  +0.2}$ \\ 
      4C4117 &   S0 &  $  11.5_{  -0.1}^{  +0.2}$ & $   700_{-   100}^{+   300}$ & 0.0057 & $  12.3_{  -0.6}^{  +0.2}$ \\ 
     TNJ2007 &   E &  $  11.9_{  -0.1}^{  +0.1}$ & $  1000_{-   100}^{+   200}$ & 0.0148 & $  12.4_{  -0.7}^{  +0.2}$ \\ 
\hline
 & & & & & \\
\end{tabular}

\begin{tabular}{lccccc}
\hline
&  \multicolumn{5}{c}{SB component$^{*}$}  \\
\cline{2-6} 
Name   & Mass                    & Age                       & $Z_{init}$ & $K$   & $L^{\rm tot}$             \\
         & $log_{10}$[M$_{\odot}$]               & [Myr]                     &           &       & $log_{10}$[L$_{\odot}$]  \\
\hline \hline
       3C368 & $  10.1_{  -0.2}^{  +0.1}$ & $     7_{-     2}^{+     1}$ &  0.01 &       10 & $   12.7_{  -0.4}^{  +0.3}$  \\ 
       3C470 & $  10.0_{  -0.2}^{  +0.4}$ & $     5_{-     1}^{+     3}$ &  0.01 &       10 & $   12.9_{  -0.5}^{  +0.1}$  \\ 
     MRC0324 & $  10.8_{  -0.2}^{  +0.3}$ & $    18_{-     4}^{+     7}$ &  0.01 &       10 & $   12.9_{  -0.5}^{  +0.1}$ \\ 
     PKS1138 & $  10.7_{  -0.3}^{  +0.3}$ & $    20_{-     6}^{+    15}$ &  0.0005 &     10 & $   13.2_{  -0.4}^{  +0.1}$  \\ 
     MRC0406 & $  11.3_{  -0.2}^{  +0.4}$ & $    45_{-    15}^{+    45}$ &  0.0 &        10 & $   13.1_{  -0.4}^{  +0.1}$  \\ 
     MRC2104 & $  10.6_{  -0.7}^{  +0.8}$ & $    90_{-    65}^{+   160}$ &  0.0 &        1 & $   12.9_{  -0.1}^{  +0.7}$  \\  
     USS0828 & $  10.3_{  -0.1}^{  +0.3}$ & $     6_{-     1}^{+     3}$ &  0.01 &       10 & $   13.2_{  -0.6}^{  +0.1}$ \\ 
      4C2858 & $  10.8_{  -0.5}^{  +0.3}$ & $    30_{-    12}^{+    30}$ &  0.005 &      10 & $   13.3_{  -0.1}^{  +0.0}$  \\ 
     USS0943 & $  11.4_{  -0.8}^{  +0.5}$ & $    90_{-    60}^{+   110}$ &  0.01 &       10 & $   13.0_{  -0.1}^{  +0.3}$  \\ 
      4C4117 & $  11.4_{  -0.2}^{  +0.0}$ & $    30_{-     5}^{+     5}$ &  0.001 &      10 & $   13.4_{  -0.5}^{  +0.3}$  \\ 
     TNJ2007 & $  11.0_{  -1.0}^{  +0.4}$ & $    45_{-    27}^{+    95}$ &  0.001 &      10 & $   13.0_{  -0.1}^{  +0.2}$  \\ 
\hline
 & & & & & \\
\end{tabular}

\begin{tabular}{lcccc}
\hline
& \multicolumn{4}{c}{AGN component} \\
\cline{2-5}
Name   & $Y$                          & $\tau_{\rm 9.7\mu\,m}$           & $i$                        & $L^{\rm tot}$ \\
             &                             &                            & [degrees]                  & $log_{10}$[L$_{\odot}$] \\ 
\hline \hline
       3C368 & $   150_{-     0}^{+     0} $ & $   0.1_{-   0.0}^{+   0.0} $ & $    40_{-     0}^{+     0} $ & $  11.7_{  -0.4}^{  +0.3}$ \\ 
       3C470 & $    60_{-     0}^{+    90} $ & $  10.0_{-   0.0}^{+   0.0} $ & $    50_{-     0}^{+    10} $ & $  13.2_{  -0.4}^{  +0.5}$ \\ 
     MRC0324 & $    60_{-     0}^{+    90} $ & $   1.0_{-   0.9}^{+   0.0} $ & $    90_{-    10}^{+     0} $ & $  12.3_{  -0.2}^{  +0.4}$ \\ 
     PKS1138 & $    10_{-     0}^{+     0} $ & $   1.0_{-   0.0}^{+   0.0} $ & $    50_{-     0}^{+    10} $ & $  13.5_{  -0.3}^{  +0.5}$ \\ 
     MRC0406 & $    10_{-     0}^{+     0} $ & $  10.0_{-   0.0}^{+   0.0} $ & $    50_{-     0}^{+    20} $ & $  13.3_{  -0.4}^{  +0.5}$ \\ 
     MRC2104 & $   150_{-    90}^{+     0} $ & $  10.0_{-   0.0}^{+   0.0} $ & $    80_{-    30}^{+    10} $ & $  13.7_{  -0.1}^{  +0.9}$ \\ 
     USS0828 & $    10_{-     0}^{+     0} $ & $   1.0_{-   0.0}^{+   0.0} $ & $    70_{-     0}^{+    10} $ & $  13.5_{  -0.6}^{  +0.5}$ \\ 
      4C2858 & $    60_{-     0}^{+     0} $ & $  10.0_{-   0.0}^{+   0.0} $ & $    90_{-     0}^{+     0} $ & $  14.3_{  -0.1}^{  +0.7}$ \\ 
     USS0943 & $   150_{-   140}^{+     0} $ & $  10.0_{-   0.0}^{+   0.0} $ & $    70_{-    20}^{+    10} $ & $  13.4_{  -0.0}^{  +0.6}$ \\ 
      4C4117 & $    10_{-     0}^{+   140} $ & $  10.0_{-   0.0}^{+   0.0} $ & $    80_{-    10}^{+    10} $ & $  13.9_{  -0.5}^{  +0.3}$ \\ 
     TNJ2007 & $    10_{-     0}^{+     0} $ & $   1.0_{-   0.0}^{+   0.0} $ & $    50_{-     0}^{+    20} $ & $  13.0_{  -0.1}^{  +0.8}$ \\ 
\hline

\end{tabular}
\end{table}

\subsection{Moderate effect of AGN-scattered light in broad-band SED fitting}
\label{sec:polarisation}

\begin{figure}[t] \centering
\includegraphics[width=0.5\textwidth]{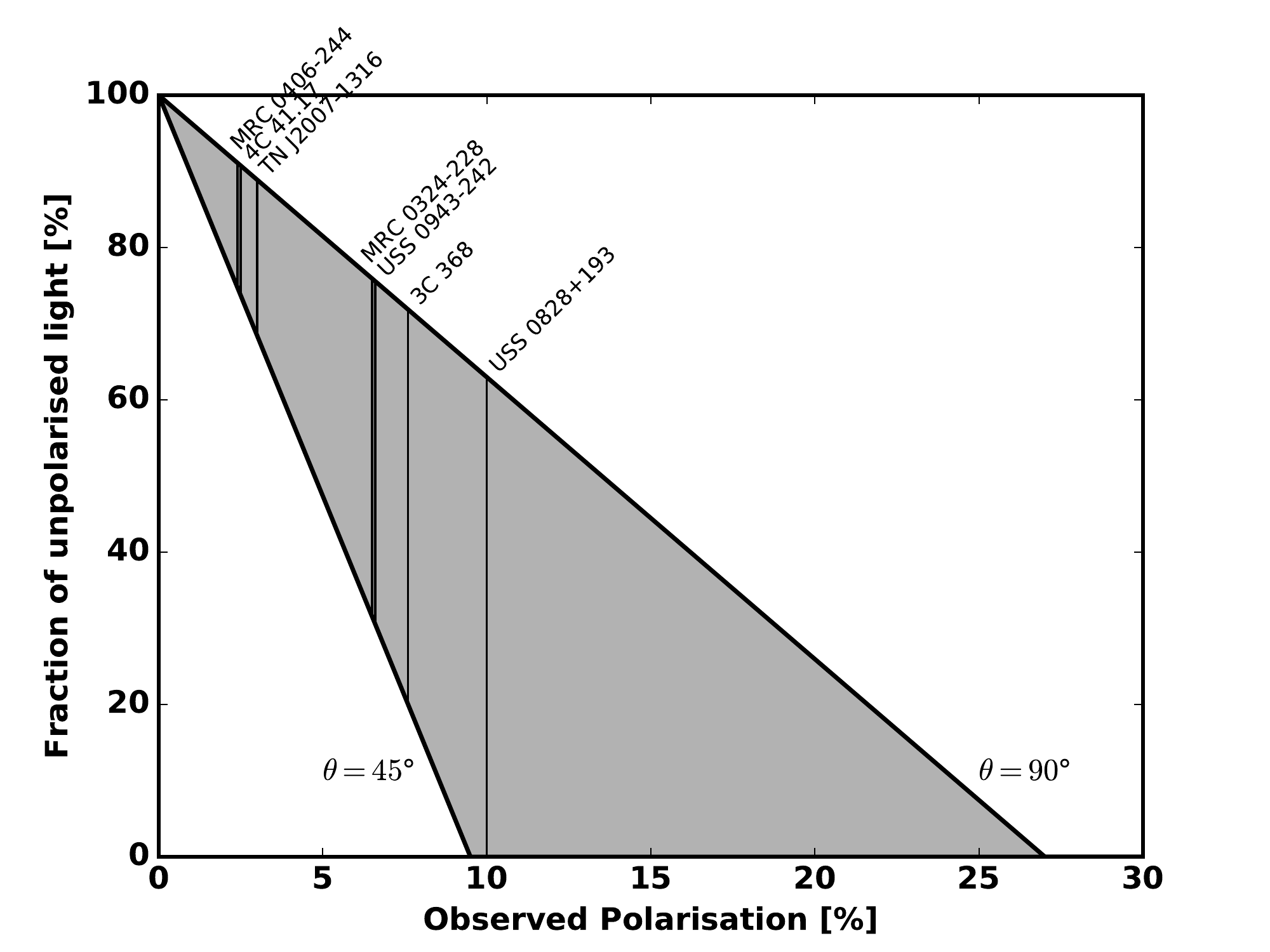}
\caption{Highest and lowest polarisation diagram at 1500\AA. The grey shaded area represents the allowed fraction of unpolarised light depending on the inclination (see text). Polarisation measurements of our sources are reported as vertical lines {\it} without the uncertainties for visibility (i.e. the uncertainties increase the allowed range of unpolarised light accordingly).}
\label{fig:pol_diagram}
\end{figure}

The AGN emission is partly absorbed by the dusty torus or escapes along the open part of the torus. Escaping photons can be scattered by
dust present in the ionisation cone toward the observer, and thereby contaminate the overall UV and optical light.
This contribution especially affects type 2 AGN (seen edge-on), and can be strong in the 
UV and optical rest-frame \citep[up to 100\% ---][]{diSerego1996,Cimatti1998,Vernet2001}. It is important to quantify this effect in our SED fitting. We used the
technique from \cite{Vernet2001} to estimate the range of possible contributions
of the optical AGN0scattered light, as illustrated in Fig.\ref{fig:pol_diagram}. We stress that this range relies on several assumptions 
regarding the geometry and inclination of the obscuring dust distribution, which 
are consistent with the adopted torus models ($\theta$=40\degree, inclination $i$).

One fundamental property assumed here is the grey behaviour of polarisation, that is, the fact that
polarisation from dust scattering is approximately independent of wavelength \citep{Vernet2001}. This 
allows us to estimate the range of AGN contribution in the UV and optical domain from a 
single polarisation measurement at 1500\,\AA\. When a polarisation measurement was available, we therefore considered the following three cases: 
(i) without polarisation, (ii) with lowest possible AGN contamination, and (iii) with the highest possible AGN contamination. We applied the following correction to the UV and optical photometry before the fitting: 
We scaled the 1500\AA\ contribution from Fig. \ref{fig:pol_diagram} and subtracted the corresponding fraction from our broad-band photometry\footnote{We are interested only in the continuum emission, therefore we removed the 
strong emission lines from the templates by linearly interpolating 
the continuum below the line. In addition, line polarisation is typically close to null because the physical processes 
involved in line emission are different from the continuum} with a typical UV and optical continuum quasar spectrum from \citet{Cristiani1990}. The latter represents a typical UV and optical SED of a type 1 AGN (i.e. without obscuration). We recall that these are the boundary cases and that the true contamination is located between these two extremes.

\begin{figure}[t] \centering
\includegraphics[height=0.5\textwidth,angle=270,trim= 80  0  0  0]{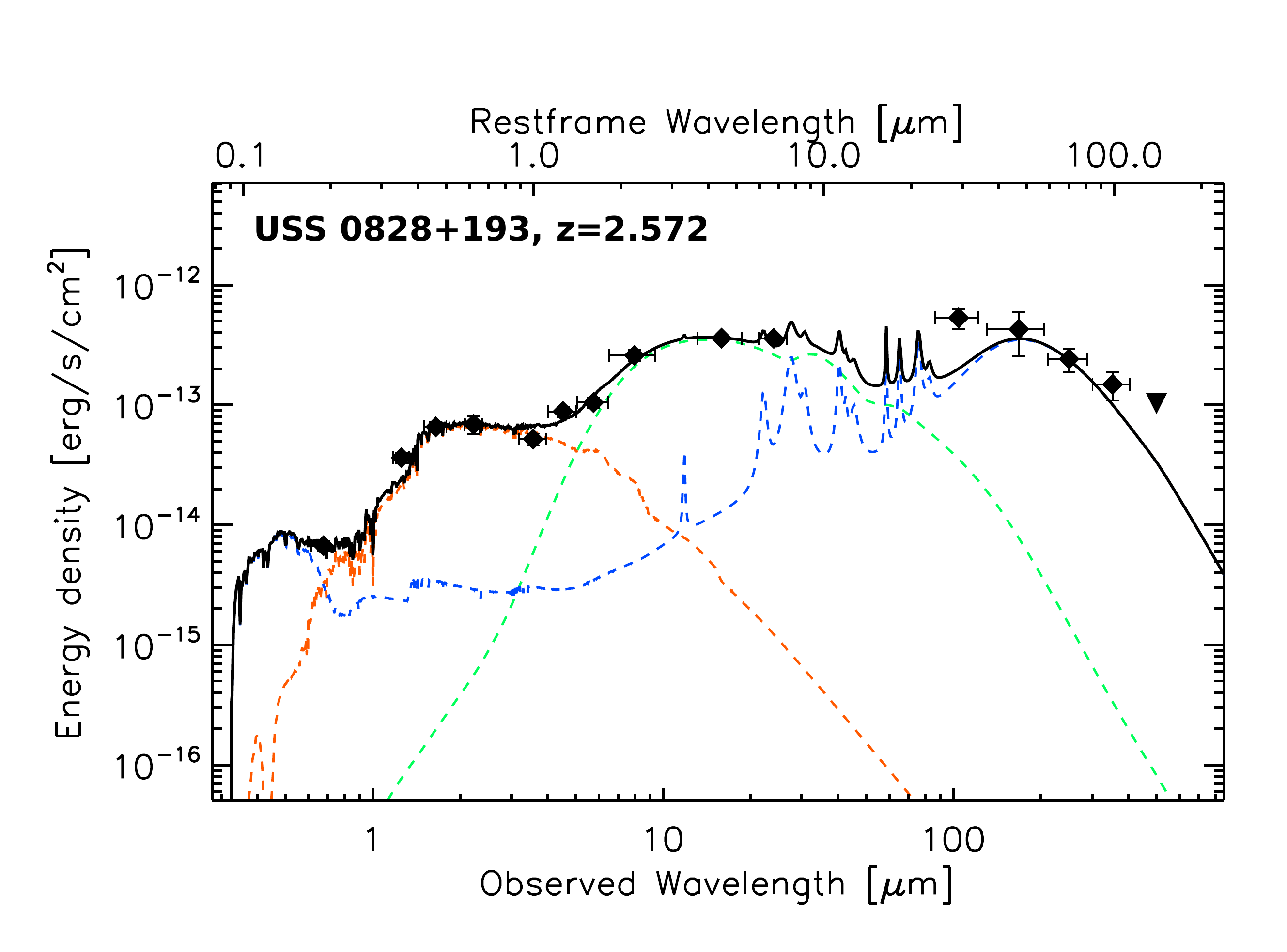}
\includegraphics[height=0.5\textwidth,angle=270,trim= 80  0  0  0]{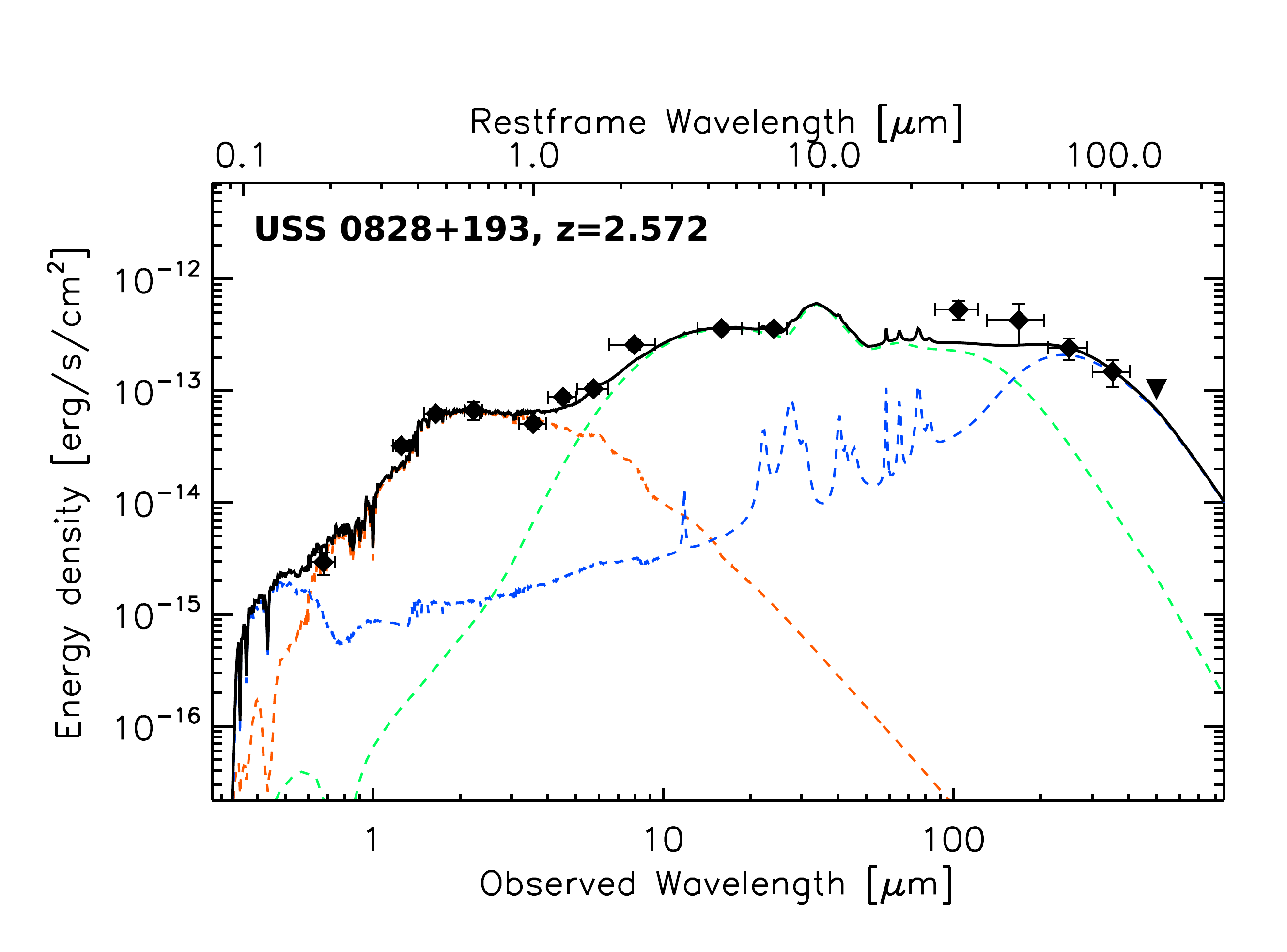}
\caption{SEDs of USS\,0828+193 showing the impact of the polarisation on our fitting. The colour coding is the same than Fig. \ref{fig:results_4C4117}. The top figure is without polarisation, while the bottom figure is the minimal subtraction case. The maximal case of a 100\% subtraction is not reported (see text).}
\label{fig:pol_minmax}
\end{figure}

The most polarised source of our sample, 
USS~0828+193, with $P$=10.0$\pm$2.0\% is predicted to have a contribution from unpolarised light of between 0\% and $\sim$65\%. The remaining AGN contribution is thus between $\sim$35 and 100\% (illustrated in Fig. \ref{fig:pol_minmax}).
Subtracting the lowest (35\%) AGN contribution has almost no effect on the fit ($\chi^2=35.7$ or $\chi^2=37.1$, respectively). This suggests either that (i) the addition
 of this component is not particularly justified, or that (ii) the true contamination is not close to the lowest 35\% contribution of the AGN. As the highest possible AGN contribution for this source is 100\% in the UV, this would leave no room for any stellar contribution. Such an exceptional case (only one in our sample) cannot be properly taken into account in our fitting procedure. We therefore consider the lowest possible AGN contribution as an upper limit on the effect on USS~0828+193. 
For this lowest contribution, the effect on the fitting is limited to a factor of $\sim$2 for the mass and $\sim$2 for the age of the stellar components (see Fig. \ref{fig:z_mass}). 
For the remaining six sources with polarisation measurements, the difference induced by polarisation is typically contained in our 68\% confidence intervals 
(see empty symbols in Fig. \ref{fig:z_mass}). The effect on the other parameters is presented
in the insets in the SEDs in Figs. \ref{fig:results_4C4117} and \ref{fig:results_3C368}-\ref{fig:results_TNJ2007} (blue 
and red points for the lowest and highest contamination, respectively).

We conclude that the effect of polarisation for our broad-band SED fitting is weaker than our global uncertainties on each parameter. Polarisation therefore has a negligible effect on our main conclusion, and we ignore this in the remainder of this paper. 

\section{Discussion}
\label{sec:discussion}

% summary, outline

Our fitting procedure enables us to carefully disentangle the three main spectral components of our  
sources: an evolved stellar component, an SB component and an AGN. We focus on the new information provided 
by the \herschel\ data in the far-infrared (FIR), allowing us to disentangle the SB from the AGN contribution. We find that the two stellar components 
are massive ($>10^{10}$\msun) and that both the AGN and the SB are very luminous ($>10^{10}$\lsun). The properties of the evolved 
stellar population confirm a very high formation redshift of the galaxy host, leading to local massive early-type galaxies as previously proposed in literature \citep[e.g.][]{Lilly1984}.
Interestingly, our sources have similar properties as quasars and SMGs in the same redshift range\citep[e.g.,][see  Sect. \ref{sec:results}]{Alexander2005,Wang2011,Leipski2014}. 
We discuss the limitations of our modelling in a first step and then explore the properties of star formation
making use of physical arguments and empirical laws from SED fitting. Finally, we classify and discuss 
the morphology of our sources making use of a combination of multi-wavelength data.

\subsection{Modelling uncertainties, limitations, and multi-wavelength data set}
\label{sec:limitations}

The self-consistency in our models is a key point of our modelling: 
for instance, the computed FIR emission is dependent on the dust properties defined by the 
evolutionary scenario and produced from the stellar population evolution (ISM enrichment, see Sect. \ref{sec:pegase_library}). 
We discuss here the limitations and degeneracies within our libraries and from our observations on the fitting procedure.
We refer to Sect. \ref{sec:fitting} and Appendix \ref{sec:systematics} for the discussion of the model assumptions and systematics introduced by our choice of code and libraries. 

%pegase
The \pegaseiii\ model only explores a limited number of star formation histories. For the evolved component, the star formation history 
is set by the Hubble type, although the data presented here are unable 
to prefer a single type within the confidence intervals  (Sect. \ref{sec:uncertainties}).
However, the effect on the age and mass of stellar populations is limited, within a factor of 2
(see RV13 for better details). For the starburst, the star formation is considered instantaneous (in one time step). 
Forming  $\sim$10$^{10}$\msun\ in such a short time appears highly unlikely. Nevertheless, we note that only a short burst can reproduce 
the strong submm emission in our sample. We stress that star formation is not necessarily 
spatially concentrated, but is most probably temporally connected. Therefore, the SB component
may be interpreted as a collection of smaller clumps, formed in a short period.
In addition, RV13 have discussed the effect of this $\delta$ function, showing that the age from the fitting is measured within a factor of $\sim$2 for constant  star formation over a longer period of time.

%agn
The \cite{Fritz2006} AGN model is one of several models available in the literature. Direct comparisons between
different models are difficult because of the variation in the number of free parameters and because of the assumptions on 
the torus properties itself (smooth, clumpy, bi-phased). More investigations are reserved for further papers, taking different 
AGN models into account \citep[e.g.][]{Podigachoski2016}.
We therefore do not further discuss the implications of the smooth torus model on our analysis, but we 
warn that the results provided here are dependent on this choice. We note that part of our best fit has difficulties 
to reproduce the 30\mum\ restframe data (e.g. PKS~1138-262 and USS~0828+193). The most likely explanation
is the lack of exploration in the parameter space, especially on the size parameter. A larger torus, having 
dust at larger radii, would better reproduce this cold emission.

% observations
Finally, observations themselves can be affected by systematics. This is particularly true for our set of multi-wavelength 
observations that were obtained with a range of instruments and telescopes, each presenting their own calibration uncertainties, especially at moderate
signal-to-noise ratio. The FIR and submm domain are particularly challenging observations and the photometry estimation method can have a significant effect on the
final measurement, especially in the $<$5$\sigma$ range \citep[by a factor of $\sim$20\%][]{Popesso2012}. A good illustration is the 160\mum\
photometry in 4C~28.58 (Fig \ref{fig:results_4C2858}), where the 160\mum\ flux might be expected to be higher. Similarly, sky-subtraction strategy 
(on-off versus raster) may affect the 850\mum\ photometry, possibly explaining the differences in the cold dust part of the SED. 
These uncertainties are mitigated by the good sampling of our SEDs, however.

\subsection{Star formation properties from the SED fitting}
\label{sec:physical_properties}

In the previous sections we estimated several crucial properties 
related to the star formation in our sample, such as the age, average column density, mass and, to some extent, initial metallicity. 
Even if the star formation most likely occurs simultaneously in different clumps,
we treat this component as a single idealised starburst (completely localised and isolated). This provided us with information useful for comparing our sample to other galaxies. Assuming the starburst has a projected 
circular geometry (at constant density), we estimate the size of the starburst region as follows: 

\begin{equation}
r_{\rm SB} \propto \left(\frac{M_{\rm SB}}{\pi K N_{\rm HI}}\right)^{\frac{1}{2}},
\end{equation}

\noindent where $r_{\rm SB}$ is the projected radius, $M_{\rm SB}$ is the mass of the starburst, $K$ is the column density factor and 
$N_{HI}$ the column density (here $N_{HI}$=6.8$\times$10$^{21}$atoms cm$^{-2}$). These values are reported in the third column of Table \ref{tab:sf_properties}. 
With an estimated size, we can define a typical velocity dispersion assuming equilibrium:

\begin{equation}
\sigma_{\rm SB} \sim \sqrt{\frac{G M_{\rm SB}}{r_{\rm SB}}},
\end{equation}

\noindent where $\sigma_{\rm SB}$ is the velocity dispersion of the gas, and $G$ is the gravitational constant. We recall that when the system  is collapsing, this value is an upper limit, while for an expanding system, this value becomes a lower limit.
Assuming the relation $V_{\rm circ}$=$\sigma$/0.6 \citep[e.g.,][]{Rix1997}, we calculate the dynamical time as $t_{\rm dyn}$=2$r_{\rm SB}$/$\sigma_{\rm SB}$. 

The projected star formation per surface unit is a powerful proxy to estimate the physical conditions of 
the starburst \citep[e.g.,][]{Lehnert1996}. Based on the physical sizes and timescales derived above,  
we calculate the star formation density, \sfrdens, with the following formula \sfrdens=SFR/(2$\pi$r$_{\rm SB}^{2}$) with SFR=$M_{\rm SB}$/$t_{\rm dyn}$.

It is interesting to note that the overall properties from these empirical calculations indicate (i) a high efficiency of forming stars with an sSFR $>$20Gyr$^{-1}$ in
all cases, consistent with measurements from \cite{Drouart2014}, (ii) a dense star-forming region \sfrdens$>$5\msunyr kpc$^{-2}$, 
similar to the local starburst \citep[e.g.][]{Bigiel2008} and largely exceeding the limit for starburst driven winds \citep[e.g.][]{Lehnert1996}, and (iii) half (5/11 sources) of our sample 
presents an SB mass of the same order as the host component (Table \ref{tab:sf_properties}), strongly suggesting merging activity. 
These estimates have important implications in terms of star formation properties because
it suggests that the star formation is very efficient in very confined regions. However, the large projected radius ($>$1\,kpc), 
suggests that the star formation arises in several locations within the same galaxy.

\begin{table*}[t] \centering
\caption{Physical properties of the star formation, assuming empirical relations and simple physical assumptions 
for an {\it isolated} starburst (columns 2-6, see also Sect. \ref{sec:physical_properties}). All calculated values only take into account the SB component (see Sect. \ref{sec:physical_properties} for details). None of the indicated properties are fully constrained (except for 4C~41.17(compact) 
where we constrain the size; See Sect. \ref{sec:jet_induced_sf} for details). Columns 7-9 refer to the likelihood from visual inspection on images described in Sect. \ref{sec:loc_sf}. Question marks refer to ambiguous case (see respective figures).
(\nodata) indicates a lack of data to assess any visual association.}
\label{tab:sf_properties}
\begin{tabular}{lcc ccc ccc}
\hline
 & &  \multicolumn{4}{c}{Empirical properties from SED} & \multicolumn{3}{c}{Visual association of from images}  \\
Radio galaxy &   $\frac{M_{\rm SB}}{M_{\rm tot (SB+evol.)}}$ & $r_{\rm SB}$ &  $t_{\rm dyn}$ & sSFR & \sfrdens & ex situ($>$5kpc) & in situ($<$5kpc) & AGN-driven \\
              &           [\%] & [kpc]        &  [Myr]       & [Gyr$^{-1}$]  & [\msunyr kpc$^{-2}$] &  \multicolumn{3}{c}{likelihood} \\
\hline \hline
3C~368          & 6 & 2.7 &  22 &  44 &   12.0 & low & high & medium \\
3C~470           & 2 & 2.4 &  21 &  46 &   12.8 & medium? & medium? & none \\
MRC~0324-228     & 56 & 5.4 &  31 &  31 &    8.5 & \nodata & \nodata & \nodata \\
PKS~1138-262     & 2 & 6.8 &  35 &  27 &    7.6 & high & low & none \\
MRC~0406-244    & 61 & 7.7 &  37 &  26 &    7.2 & none & high & low \\
MRC~2104-242    & 28 & 4.8 &  30 &  33 &    9.0 & high & medium? & low? \\
USS~0828+193   & 6 & 3.4 &  25 &  39 &   10.7 & none & high & medium? \\
4C~28.58      &   14 & 6.1 &  33 &  29 &    8.1 & medium? & medium? & medium? \\
USS 0943-242  &  56 & 12.1 &  47 &  20 &    5.7 & high & low & none \\
4C~41.17 (all) &   44 & 12.1 &  47 &  20 &    5.7 & high & low & medium  \\
TN~J2007-1316 &  11 & 6.1 &  33 &  29 &    8.1 & high & \nodata & \nodata \\ [1ex]
\hline
4C~41.17 (compact) & \nodata & {\bf 1}  & 2.5 & 400 & 3200 & \nodata & \nodata & \nodata \\  
\hline
\end{tabular}
\end{table*}

\subsection{Localisation of the star formation from the images}
\label{sec:loc_sf}

The limited spatial resolution of \herschel\ makes it difficult to answer the key questions 
of the location of star formation and of the process that leads to the observed 
star formation. Complementary to the spectral information, spatial information can be used to 
understand the localisation of star formation and the overall evolutionary status of the radio galaxies in our sample. 
We made use of the high-resolution \hst\ and radio imaging, along with the moderate resolution of
the NIR data, to investigate the relative location of the emission at different 
wavelengths\footnote{The \hst\ data are downloaded from the HLA. The \herschel\ 
data are unresolved when compared to 
optical or NIR images and are not reported for simplicity \cite[but available in ][]{Drouart2014}.}.
 We visually classified our sources into three non-exclusive categories 
 as follows (and report their likelihood in Table \ref{tab:sf_properties} as none, low, medium, and high): 
\begin{enumerate}
\item ``ex situ star formation'', 
where star formation occurs outside the galaxy, at $>$5 kpc of the radio galaxy (late-stage mergers and star-forming companions 
fall into this category);
\item ``in situ star formation'', which refers to star formation that occurs within the radio galaxy, 
at $<$5 kpc (rotating structure and nuclear star formation fall into this category);
\item ``AGN-driven star formation''
corresponding to star formation linked to AGN activity, independently of spatial scale (both jet-induced and shock-induced star formation fall into this category).
\end{enumerate}
The limit of 5\,kpc is driven by the resolution obtained in the optical 
and NIR\footnote{We recall that at $z>1$, 1'' corresponds roughly to 8\,kpc, assuming the standard cosmological model.} domains 
and the typical size of a massive galaxy at $z>1$ \citep[][]{Fan2010}. 

\subsubsection{Ex situ star formation}

\begin{figure}[!h]
\includegraphics[width=0.5\textwidth]{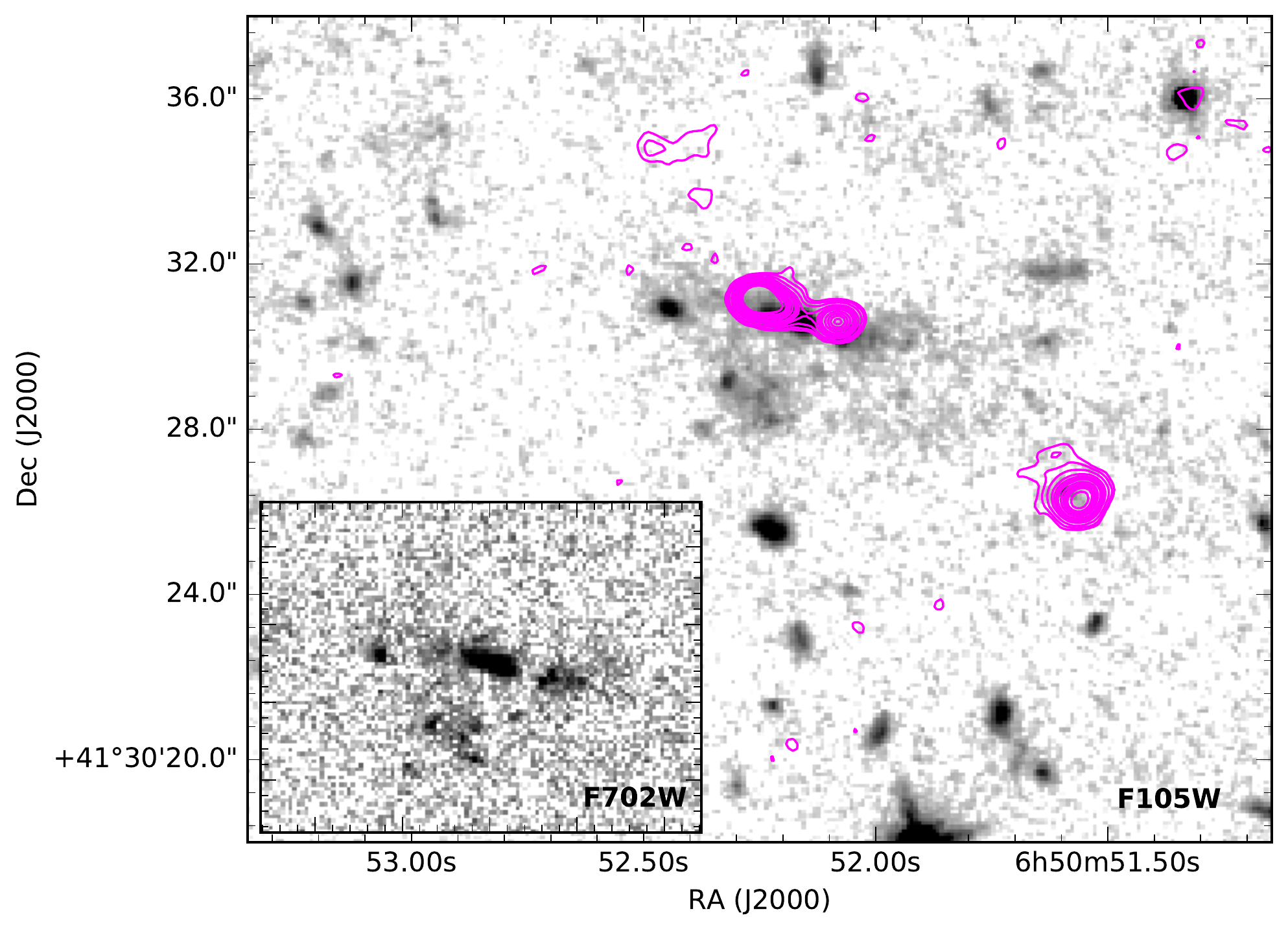}
\caption{Composite view of 4C~41.17 ($z$=3.792). The main frame is a 1.05\mum\ \hst\ image while the inset corresponds to the $R$ band. The magenta contours show the 4.8GHz data overlaid at 3, 5, 10, 20, 30, 50, 75, 100, 125, 150, and 200$\sigma$, with an {\it rms}$=$25$\mu$Jy}
\label{fig:optical_4C4117}
\end{figure}

\begin{figure}[!h]
\includegraphics[width=0.5\textwidth]{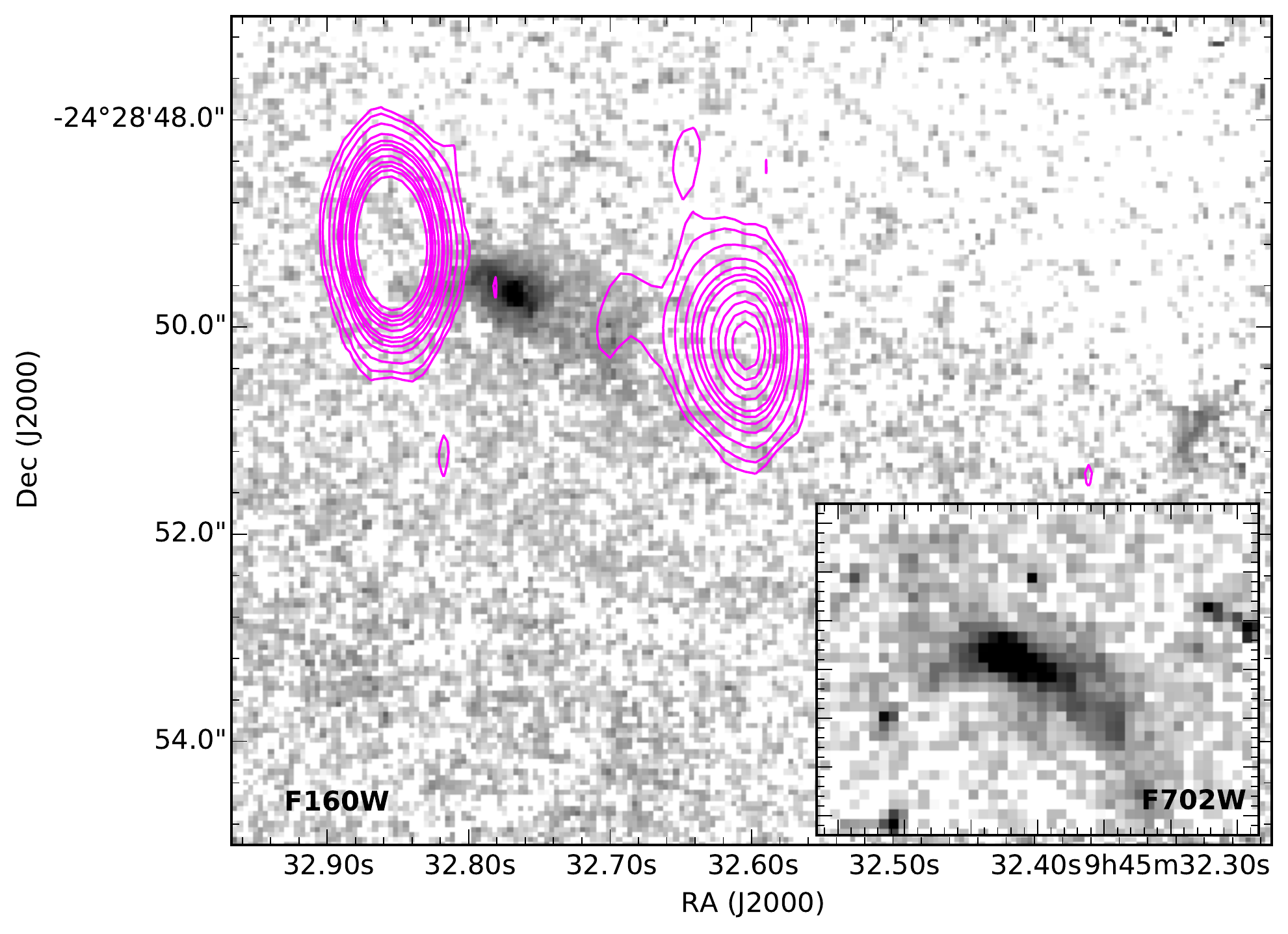}
\caption{Composite view of USS~0943$+$193 ($z$=2.923). The main frame is a 1.6\mum\ \hst\  image while the inset corresponds to the $R$ band. The magenta contours are the 4.8GHz data overlaid at 3, 5, 10, 20, 30, 50, 75, 100, 125, 150, and 200$\sigma$, with an {\it rms}$=$50$\mu$Jy.}
\label{fig:optical_USS0943}
\end{figure}

\begin{figure}[!h]
\includegraphics[width=0.5\textwidth]{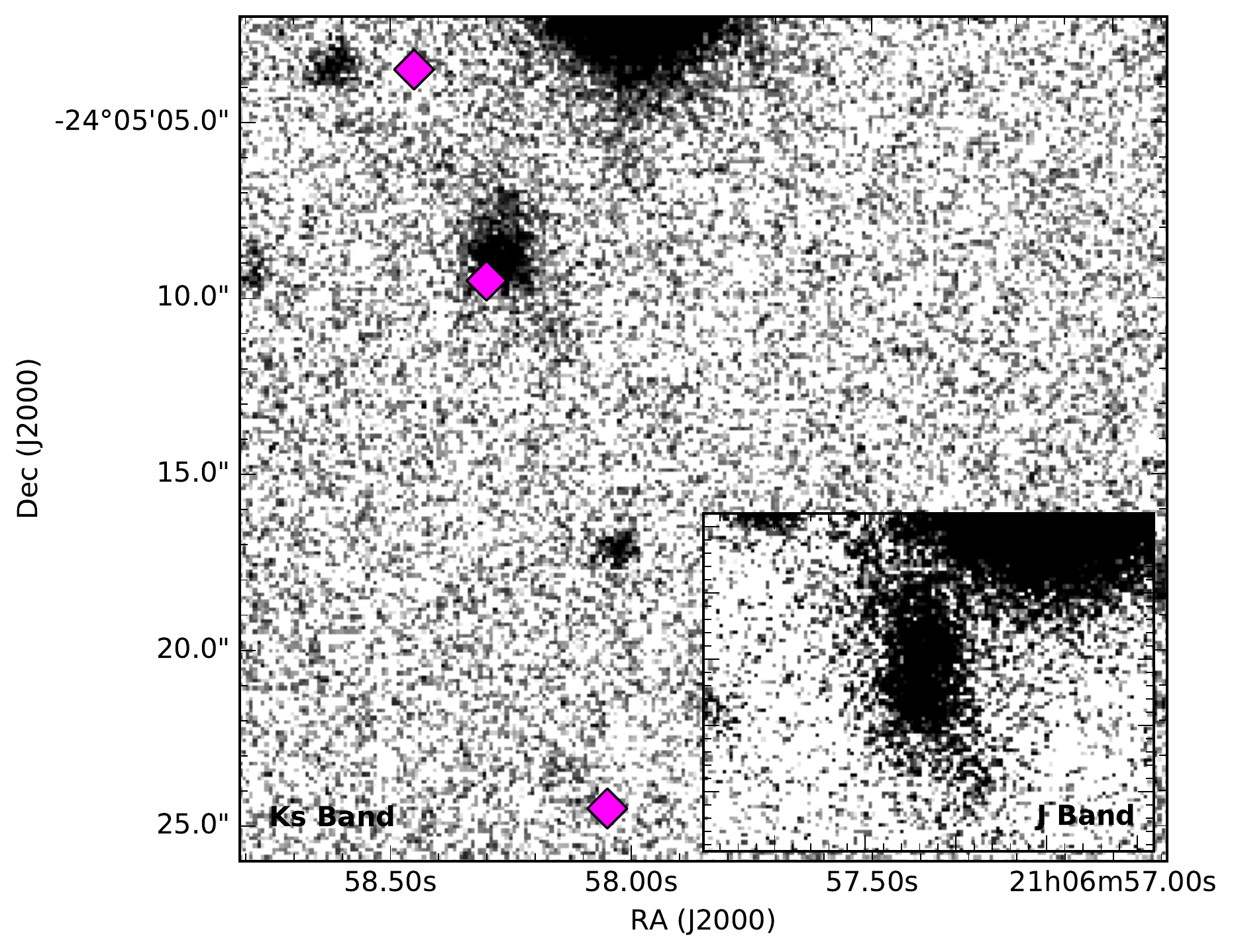}
\caption{Composite view of MRC~2104-242 ($z$=2.491). The main frame is a 2.2\mum\ image while the inset corresponds to the $I$ band. The magenta diamond shows the approximate position of the radio emission peak at 8.2\,GHz \citep{Pentericci2001}.}
\label{fig:optical_MRC2104}
\end{figure}

\begin{figure}[!h]
\includegraphics[width=0.5\textwidth]{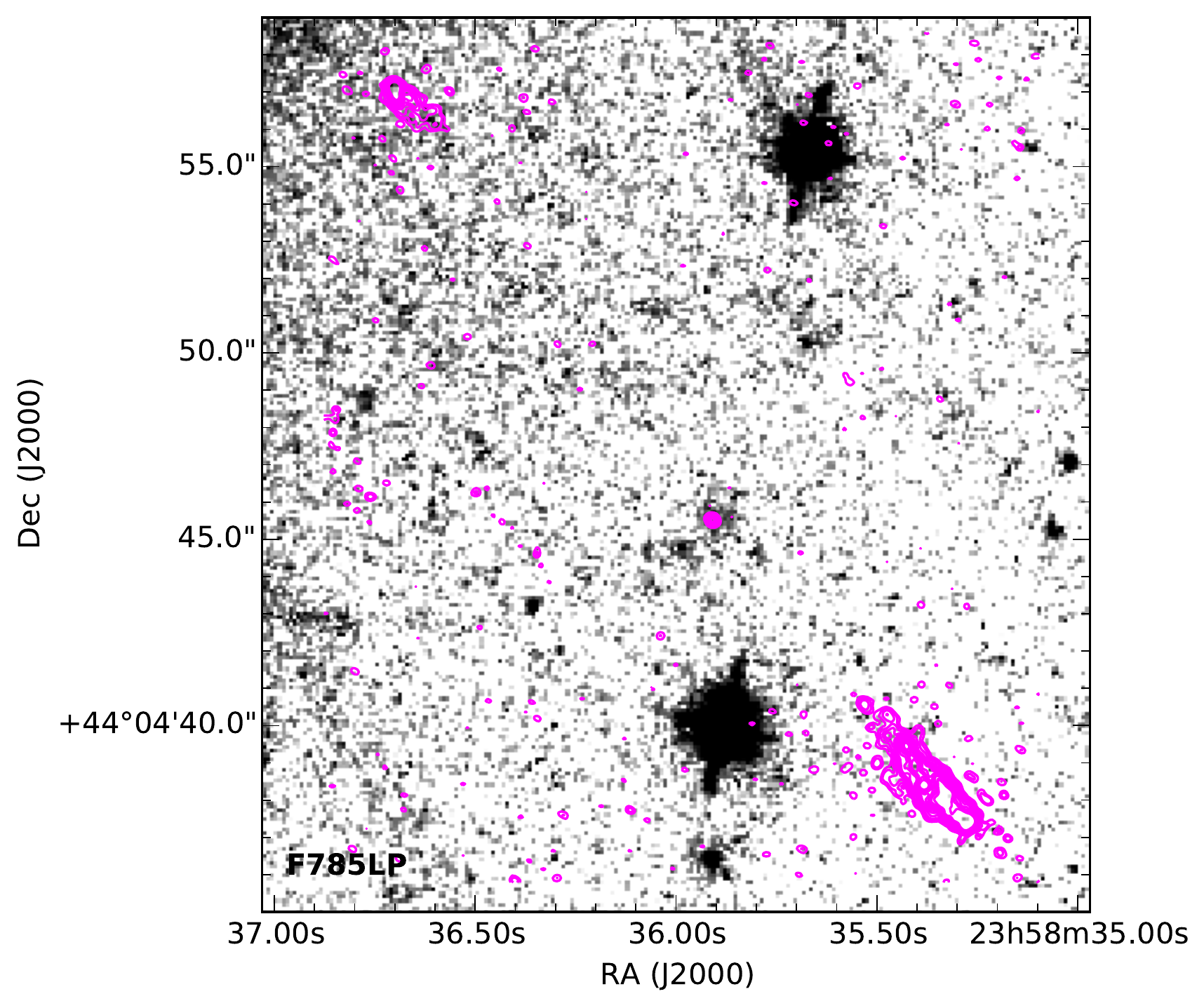}
\caption{Composite view of 3C~470 ($z$=1.653). The main frame is a $z$-band \hst\ image. The magenta contours show the 8.4GHz data overlaid at 3, 5, 10, 20, 30, 50, 75, 100, 125, 150, and 200$\sigma$, with an {\it rms}$=$25$\mu$Jy.}
\label{fig:optical_3C470}
\end{figure}

\begin{figure}[!h]
\includegraphics[width=0.5\textwidth]{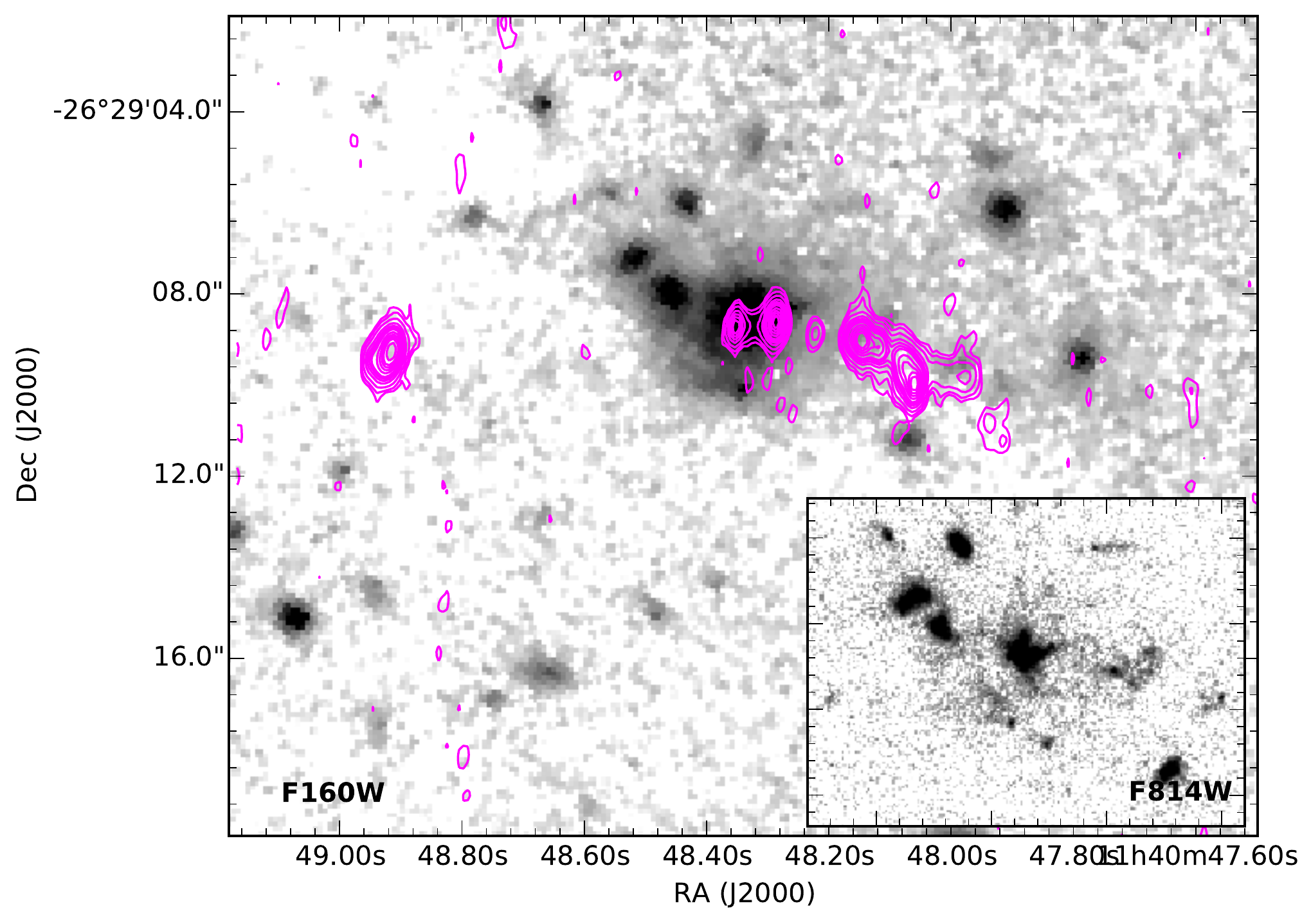}
\caption{Composite view of PKS\,1138-262 ($z$=2.43). The main frame is at 1.6\mum\ \hst\ image while the inset corresponds to the $I$ band. The magenta contours show the 8.4GHz data overlaid at 3, 5, 10, 20, 30, 50, 75, 100, 125, 150, and 200$\sigma$, with an {\it rms}$=$30$\mu$Jy. Note that most of the sources around the radio galaxy are at the same redshift \citep{Hatch2011b}.}
\label{fig:optical_PKS1138}
\end{figure}

\begin{figure}[!h]
\includegraphics[width=0.5\textwidth]{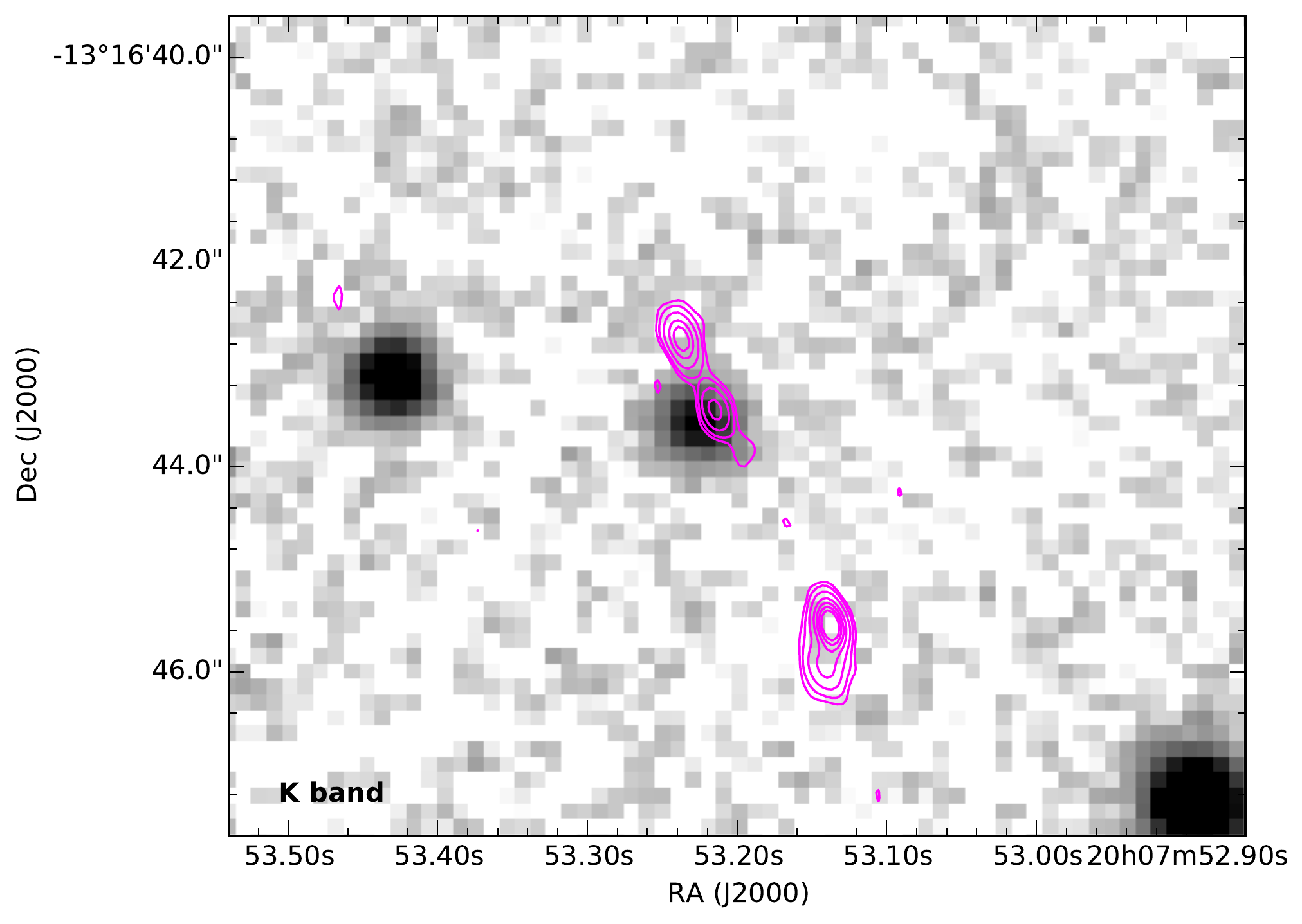}
\caption{$Ks$-band image of TN\,J2007-1316 ($z$=3.840). The magenta contours are the 8.4GHz data overlaid at 3, 5, 10, 20, 30, 50, 75, 100, 125, 150, and 200$\sigma$, with an {\it rms}$=$40$\mu$Jy}
\label{fig:optical_TNJ2007}
\end{figure}

\begin{figure}[!h]
\includegraphics[width=0.5\textwidth]{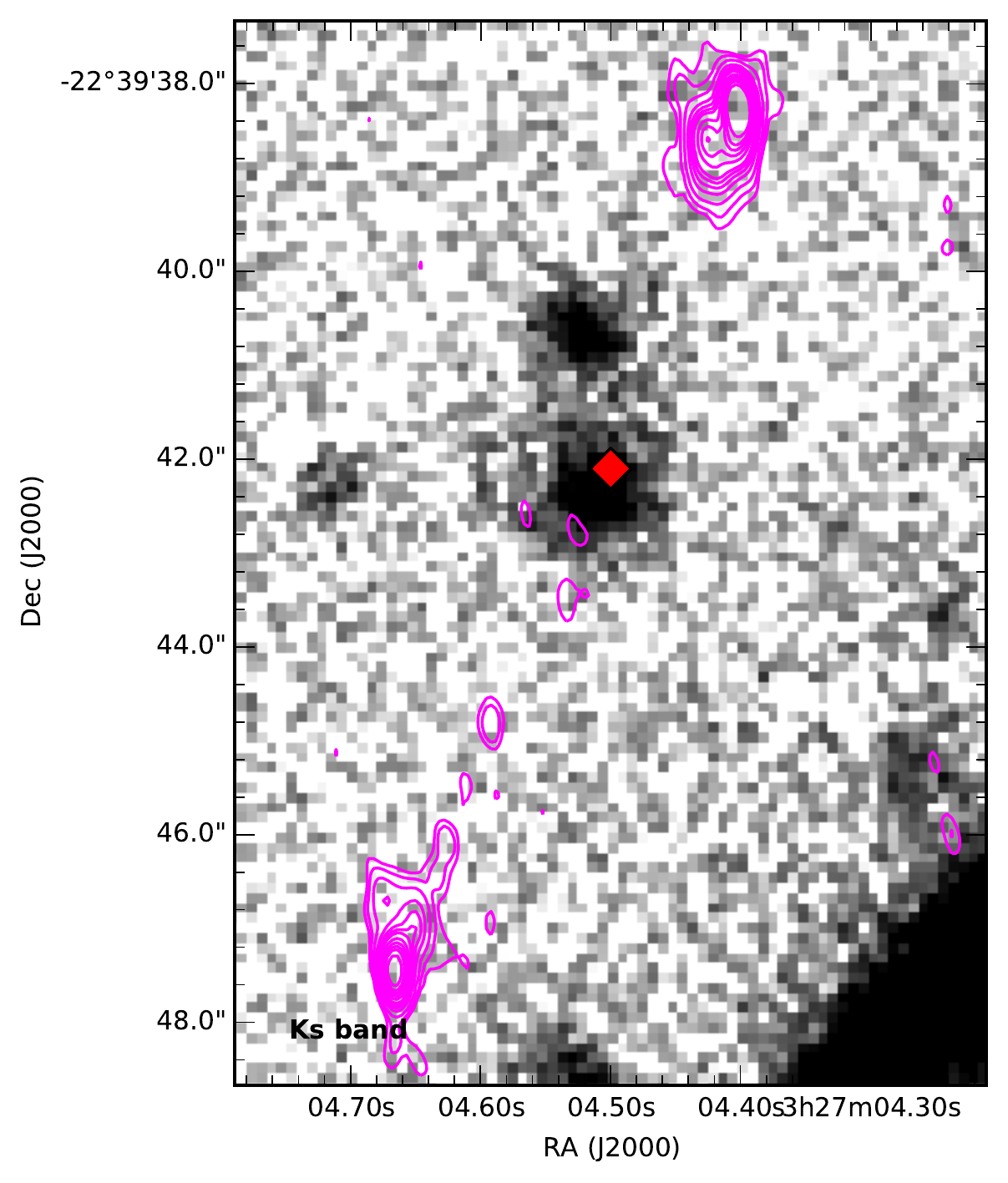}
\caption{Composite view of MRC~0324-228 ($z$=1.874). The main frame is at 2.2\mum\ image. The magenta contours are the 8.4GHz data overlaid at 3, 5, 10, 20, 30, 50, 75, 100, 125, 150, and 200$\sigma$, with an {\it rms}$=$30$\mu$Jy.}
\label{fig:optical_MRC0324}
\end{figure}

\begin{figure}[!h]
\includegraphics[width=0.5\textwidth]{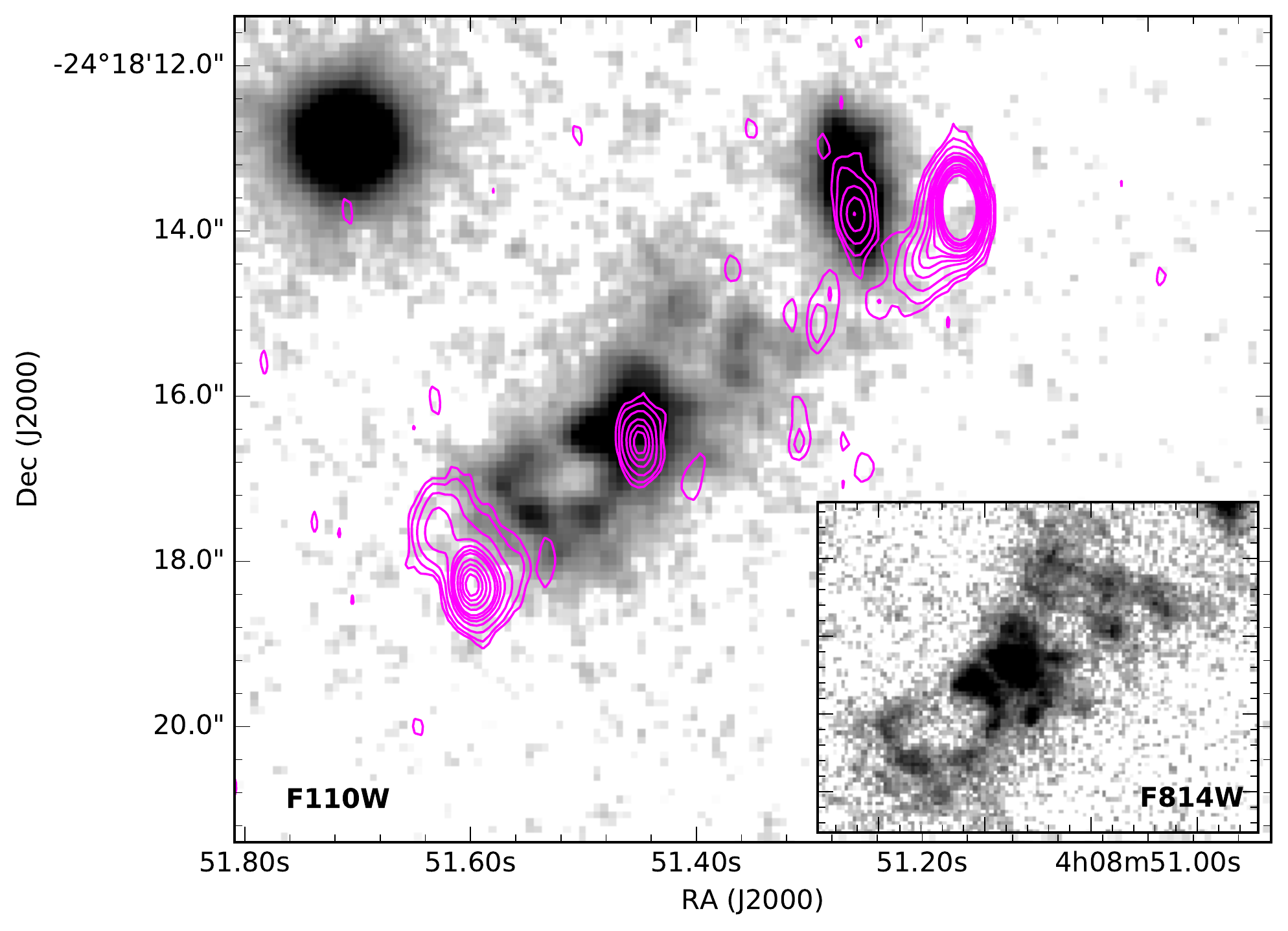}
\caption{Composite view of MRC\,0406-244 ($z$=2.43). The main frame is at 1.1\mum\ \hst\ image while the inset corresponds to the $I$ band \citep{Hatch2013}. The magenta contours are the 8.4GHz data overlaid at 3, 5, 10, 20, 30, 50, 75, 100, 125, 150, and 200$\sigma$, with an {\it rms}$=$30$\mu$Jy. The source North-West of the radio galaxy is a foreground source \citep{Rush1997a}.}
\label{fig:optical_MRC0406}
\end{figure}

\begin{figure}[!h]
\includegraphics[width=0.5\textwidth]{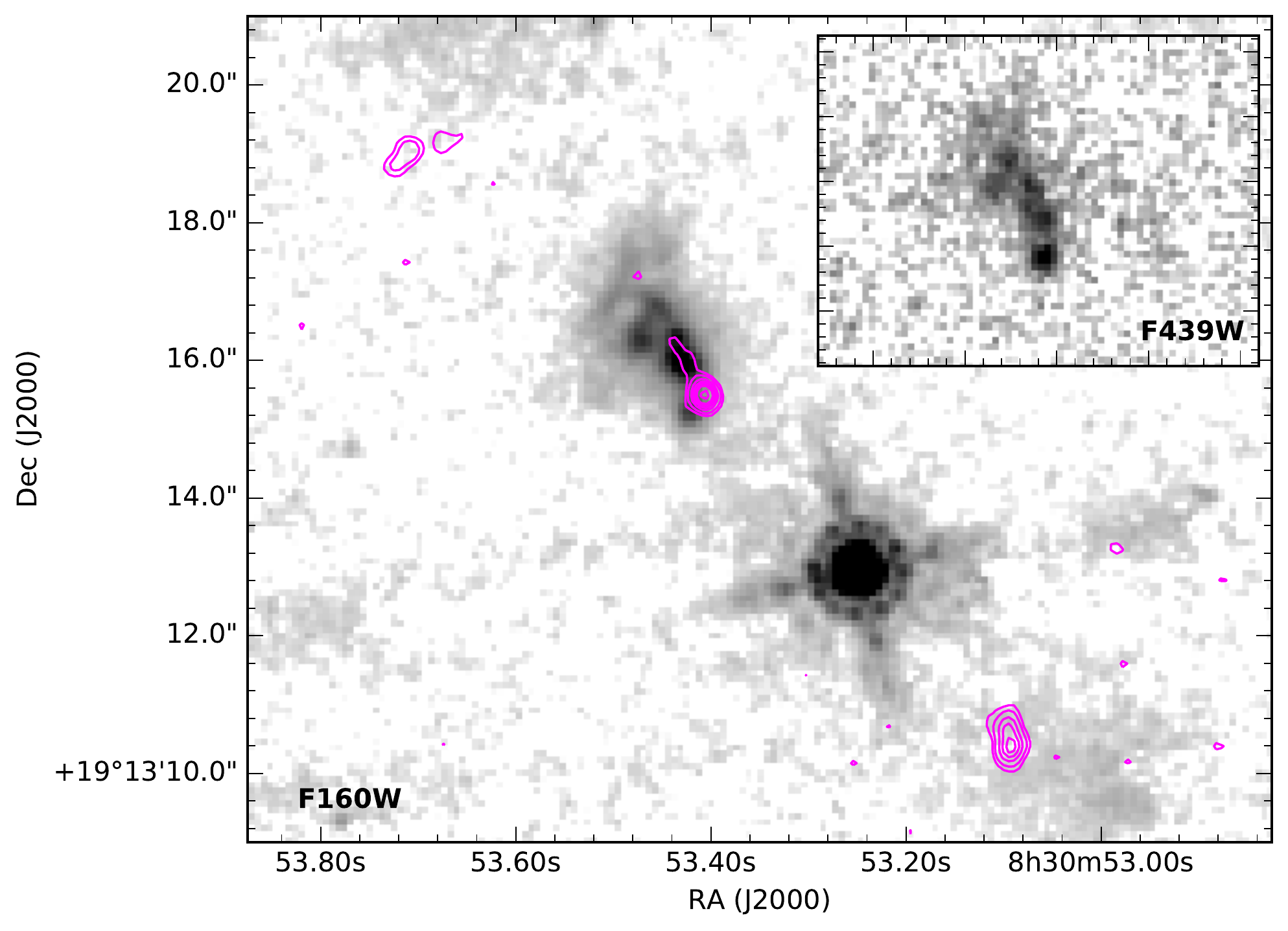}
\caption{Composite view of USS~0828$+$193 ($z$=2.572). The main frame is at 1.6 \hst\ \mum\ image while the inset corresponds to the $u$ band. The magenta contours are the 8.4GHz data overlaid at 3, 5, 10, 20, 30, 50, 75, 100, 125, 150, and 200$\sigma$, with an {\it rms}$=$40$\mu$Jy. The source located South-West of the radio galaxy.}
\label{fig:optical_USS0828}
\end{figure}

Ex situ star formation can be divided into two subcategories, a massive star-forming companion or a multiple system of less massive sub-units, which can
be interpreted as a signature of a major merger or multiple minor mergers, respectively.
Based on our visual classification (Col. 7 in Table \ref{tab:sf_properties}), seven of the eleven sources have a medium or high likelihood of ex situ star formation. 

Among the ``high'' likelihood ex situ star forming sources, three sources have a SB mass fraction $>$28\% (4C\,41.17 Fig. \ref{fig:optical_4C4117}, USS\,0943-242, Fig.\ref{fig:optical_USS0943} 
and MRC\,2104-242, Fig. \ref{fig:optical_MRC2104}), strongly suggesting major-merger events\footnote{Recently, USS\,0943-242 was 
observed at submm wavelength at arcsec resolution, indicating that a significant fraction of star formation occurs outside the host galaxy \citep{Gullberg2015}, which confirms the ex situ star formation scenario.}. 
The four remaining sources (3C\,470 Fig. \ref{fig:optical_3C470}, PKS\,1138-262 Fig. \ref{fig:optical_PKS1138}, 4C\,28.58 and TN\,J2007-1316 Fig. \ref{fig:optical_TNJ2007}) present a 
mass fraction for the SB in the 2-14\% range, favouring a minor-merger scenario or orbiting companions. This is particularly clear in PKS\,1138-262
where the \hst\ images reveal many companions that are organised in a complex structure around the 
central massive galaxy that is located at the position of the radio core \citep[][see also Fig. \ref{fig:optical_PKS1138}]{Pentericci1998}. Almost all 
these companions are confirmed to be at the same redshift as PKS~1138-262 \citep{Hatch2009,Doherty2010}.

A note should be made on MRC\,0324-228 (Fig. \ref{fig:optical_MRC0324}), where the SB mass fraction is very high (56\%), but where the lack of high-quality data prevents any clear association with star formation. The $Ks$-band image suggests that the star formation originates from a close companion or the radio galaxy itself. 

When this classification is compared with the empirical properties of star formation, all seven sources but one (3C\,470 Fig. \ref{fig:optical_3C470}) present a 
projected radius r$_{\rm SB}$$>$4.8\,kpc (see Table \ref{tab:sf_properties}) for the idealised starburst. This size is comparable to the size of a galaxy itself at these redshift, which additionally support the multiple galaxies and merging scenarios. 

In our sample, 65\% of our sources show evidence of merging activities (both major and minor). While the size of our sample is limited, 
this result agrees with those of other studies. Several galaxies from our parent powerful radio galaxy sample
are known to be gas-rich major mergers (e.g., TXS J1809+7220, B3 J2330+3927, and MRC0152-209), but this is based on observations made outside the IR 
domain, on molecular gas mapping \citep{DeBreuck2005,Ivison2008,Ivison2012,Emonts2014,Emonts2015}.

\subsubsection{In situ star formation}

In situ star formation refers to star formation that occurs in the radio galaxy host ($<5$\,kpc). Once again, two subcategories can be defined:
a rotating star-forming structure from accreted gas, or a nuclear starburst embedded in the central part of the galaxy. 

Three of our sources (3C\,368 in Fig. \ref{fig:optical_3C368}, MRC\,0406-244 in Fig. \ref{fig:optical_MRC0406} 
and USS\,0828+193 in Fig. \ref{fig:optical_USS0828}) show a high likelihood for in situ star 
formation (Table \ref{tab:sf_properties}), representing 30\% of our sample (55\% when medium likelihood is included). 
MRC\,0406-244 suggests an SB mass fraction of up to 61\%, while in 3C\,368 and USS\,0828+193 the SB fraction is more moderate at 6\% of the total mass. 
The case of MRC 0406-244 appears related to a rotating star-forming structure. \cite{Hatch2013} 
showed that the light profile is consistent with a disc, and the galaxy forming stars up to 1000\msunyr (albeit with large uncertainties). 
Moreover, the system appears well isolated with no merger remnants seen in the \hst\ images (Fig. \ref{fig:optical_MRC0406}). 
According to the SFR-M$_*$ relation at $z$$\sim$2 \citep{Rodighiero2011}, a main-sequence galaxy 
in the 10$^{11-12}$\msun\ range will have a corresponding SFR in the
200-1000 \msunyr\ range, which appears to be consistent with the specific case of MRC 0406-244.

The nuclear starburst (star formation within the central 1 kpc) might be expected and related to the AGN activity as they both require gas. To trigger a bolometrically luminous phase, an SMBH needs to accrete at rates of the order of 10\msunyr \citep[e.g.][]{Drouart2014}.  This implies that cold gas needs to be present in the central parts of radio galaxies. Because most of the gas is not expected to directly fall onto the black hole (the gas has to lose $>$99\% of its angular momentum, \citealt{Jogee2006}), large reservoirs are expected to accumulate close to the centre of the galaxy. Collapse of the gas clouds is therefore likely to occur, producing central star formation. This element seems to agree with the correlation found between AGN activity and radial star formation dependency \citep[e.g.,][]{Asmus2011,Videla2013,Esquej2014}. However this relation between SB and AGN luminosities is still highly debated  and might not hold at higher redshift \citep[e.g.][]{Dicken2012,Drouart2014}. In this sample, our limited range of luminosity both for the SB and the AGN prevents us from finiding any correlation, which suggests a weak correlation if it exists at all. 

Several arguments in favour of a central nuclear starburst for these three source come from the SED fitting: the $K$=10 (higher column density), the short timescale ($\delta$-like function as used here), and the massive starburst required to reproduce the large submm emission. In such constrained volume, a fast depletion of the gas is expected and required to create high star formation rates. Our discussion on the origin of the nuclear starburst overlaps with Sect. \ref{sec:jet_induced_sf} because an AGN could trigger a nuclear starburst through intense activity \citep[e.g.][]{Silk2013}. In this case, the difference between AGN-driven star formation and secular star formation (not triggered by the AGN) is difficult to assess because higher resolution observation are crucial to distinguish between these two cases. 

\subsubsection{AGN-driven star formation}
\label{sec:jet_induced_sf}

\begin{figure}[t]
\includegraphics[width=0.5\textwidth]{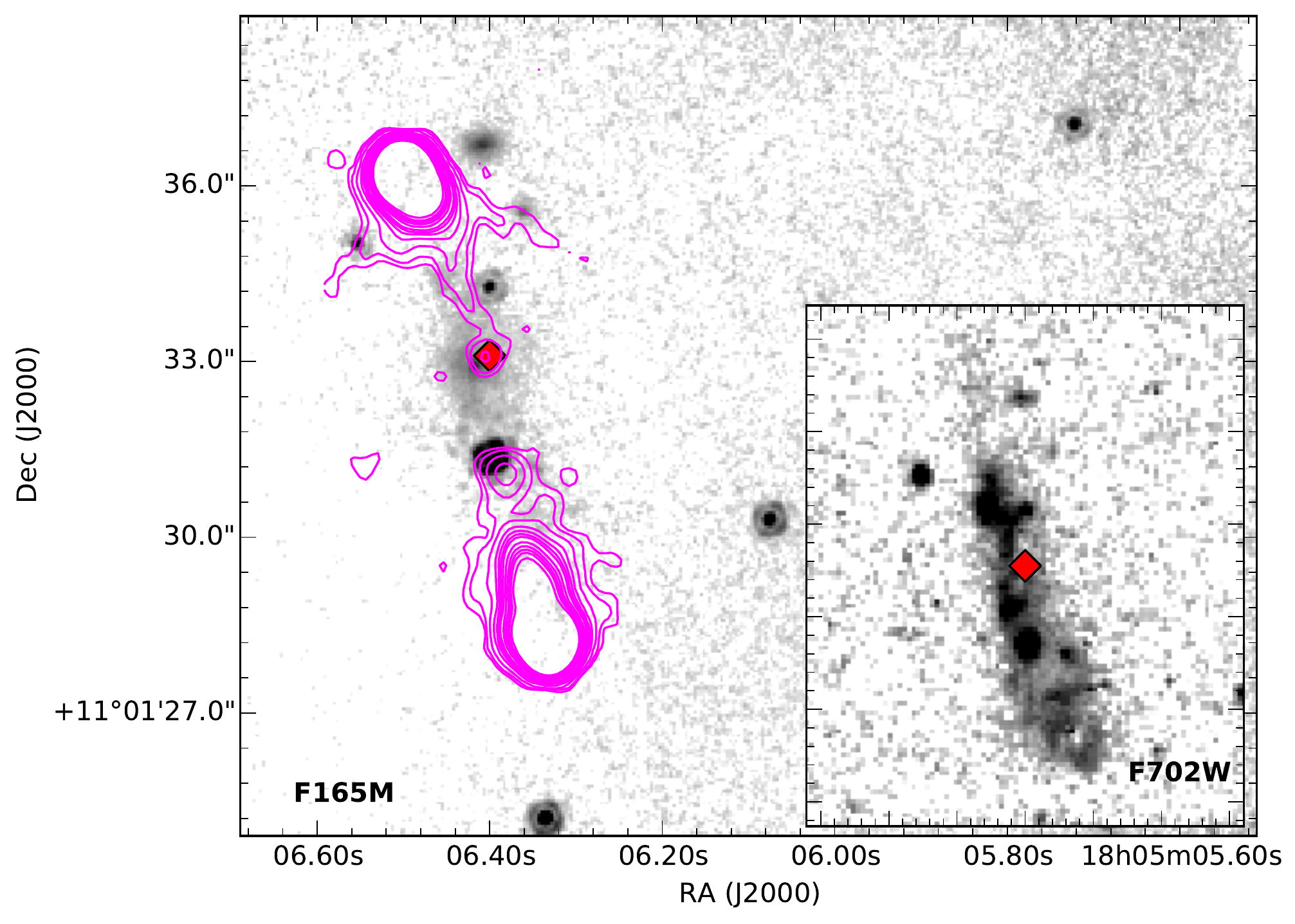}
\caption{Composite view of 3C~368 ($z$=1.132). The main frame is a 1.6\mum\ \hst\ image while the inset corresponds to the $R$ band. Two stars are located at $\sim$2\arcsec. The magenta contours show the 4.8GHz data overlaid at 3, 5, 10, 20, 30, 50, 75, 100, 125, 150, and 200$\sigma$, with an {\it rms}$=$20$\mu$Jy. }
\label{fig:optical_3C368}
\end{figure}

We include in this category star formation triggered from AGN activity, would it be mechanical and/or radiative (radio jets, UV, optical, and infrared radiation). The so-called alignment 
effect in radio galaxies was observed more than two decades ago, and jet-induced star formation was 
proposed as a possible explanation \citep[e.g.,][]{McCarthy1987,Miley1992,Chambers1996b,vanBreugel1998,Pentericci1999,Zirm2003}. 
We need to be careful, however, in associating elongated UV emission with star formation because other mechanisms can produce similar clumpy structures in high-resolution observations \citep[e.g.,][]{Dickson1995,Chambers1996b,Cimatti1998,vanBreugel1998,Pentericci1999}. 
The other likely candidates for the origin of such emission are either scattered emission from the AGN \citep{Cimatti1998} and direct nebular emission \citep[excited either from radiative or mechanical shocks; ][]{Humphrey2006,VillarMartin2007}. 
One simple test to distinguish between AGN-scattered emission and nebular emission are polarimetric measurements. Strongly polarised sources
will be AGN dominated, whereas weakly polarised sources will be dominated by nebular and/or young stars\citep[e.g.,][Sect. \ref{sec:polarisation}]{Vernet2001}. 

In our sample, we focused on the three sources with the richest data sets, 3C\,368 (Fig. \ref{fig:optical_3C368}), 4C\,41.17 (Fig. \ref{fig:optical_4C4117}) and MRC\,0406-244 (Fig. \ref{fig:optical_MRC0406}), where the AGN may have a limited effect on star formation. 
3C 368 presents striking elongated features aligned with the radio emission \citep{Best1998b}. The source is also
moderately polarised (7.6\%), with a wide range of contributions from 20\% to 80\% from the AGN-scattered light. This makes it particularly difficult to assess 
the origin of the emission, but given the low mass of the starburst compared to the total mass of the system, star formation along the jet axis
may be possible. For 4C\,41.17, observations provide strong evidence that a massive star-forming region is aligned with the radio jet axis \citep{Miley1992,Dey1997,Bicknell2000,Steinbring2014}, and 
polarisation for this source is weak $<$2.4\%. The source 4C 41.17 appears to be an exceptional case even among radio galaxies. A large molecular cloud is aligned with the radio jet axis \citep{DeBreuck2005}. 
However, recent mm observations suggest that this fortuitous alignment, although rare, may not be unique \citep{Emonts2014}.
Finally, in MRC\,0406-244 (weakly polarised $P$=2.5\%), \cite{Hatch2013} found structures along the jet axis, implying an SFR of  $\sim$75\msunyr\ assuming it is due to star formation and not scattered AGN emission.

%observational evidences
It seems that AGN-driven star formation, even if not dominant in our sample, does contribute to some extent. This is in line with a 
a comparison of radio-loud and radio-quiet AGN that found higher average SFRs for radio-loud AGN \citep{Zinn2013} which does suggest an effect of the jet on star formation \citep[though also see][for an opposite conclusion]{Karouzos2014}. When we use the information
available from the SED fitting and observational constraints, we are able to predict some of the properties of an SB triggered by a jet. As seen in the 
last line of Table \ref{tab:sf_properties}, when half of the mass of the SB component is contained in a region of  1\,kpc, the dynamical time is 
of the order of 10$^6$yr and the sSFR is higher by a factor of $\sim$20 compared to the rest of the sample. This short and very efficient star 
formation makes sense in such an extremely contained volume, where as the gas will be expelled extremely quickly and efficiently, as soon as the first SN explode, which will terminate the episode of star formation.

\section{Conclusion}

We presented 11 objects from the \herge\ sample, selected to obtain a homogeneous coverage over 
the UV to submm range. We performed SED fitting with the \pegaseiii\ galaxy evolution and the \cite{Fritz2006} AGN torus models, both 
of which use radiative transfer code to predict their respective SEDs. We showed that the SEDs of high-redshift radio galaxies 
require three spectral components: two stellar components, one evolved and one young, plus an AGN torus. 

All evolved components are very massive, $>$10$^{11}$\msun, and are consistent with formation at high-redshift. The young component represents up to 61\% of the total mass of the system and is much younger than the evolved component, indicating on-going or 
recently completed star formation. The physical properties of these starbursts broadly agree with SMGs 
at similar redshift and present a large variation from source to source. The radio galaxies are generally in a luminous phase with a 
preferentially extended, opaque torus, consistent with a type 2 AGN orientation. The bolometric luminosities 
are similar to quasars at similar redshift but the AGN activity observed at radio and optical wavelengths does not appear to be strongly correlated
with the star formation within our sample. 
We also considered reflected AGN contamination in our SED fitting based on polarisation measurements for seven sources in this 
sample and found that for broad-band photometry, fitting the differences induced by prior subtraction of the reflected AGN falls 
within the confidence intervals of the fitting and does not strongly affect the results. 

Finally, our sample contains evidence of ex situ star formation, with 65\% of the sources presenting evidence of merging activity. In addition, 30-55\% of the
sources also present in situ star formation, indicating large amounts of gas accreted directly by the galaxy. Some weak evidence for jet-induced star formation 
(although rather limited in terms of mass and relative bolometric contribution) appears in 35\% of our sample. 
An additional analysis, including high-resolution submm observations and optical/NIR
integral field maps, will allow characterisation of emission and absorption lines and will most likely bring new insights. 

\begin{acknowledgements} 
GD would like to warmly thank Alessandro Romeo, Kirsten Knudsen,  
and Clive Tadhunter for the useful discussions that contributed to improve this
paper. The authors also thank the referee for detailed suggestions and a thorough report that helped to clarify this paper. 
GD also thanks Nina Hatch for providing \hst\ fluxes for the Spiderweb galaxy and A. Galametz for providing images for part of the sample. GD thanks Philip Best for providing the 3C~368 and 3C~470 data. NS is the recipient of an ARC Future Fellowship. The work of DS was carried out at Jet Propulsion Laboratory, California Institute of Technology, under a contract with NASA. Based on observations made with the NASA/ESA Hubble Space Telescope, and obtained from the Hubble Legacy Archive, which is a collaboration between the Space Telescope Science Institute (STScI/NASA), the Space Telescope European Coordinating Facility (ST-ECF/ESA) and the Canadian Astronomy Data Centre (CADC/NRC/CSA). GD acknowledges the support from the ESO scientific visitor programme.
\end{acknowledgements} 

\bibliographystyle{aa.bst}
\bibliography{Library}

\appendix

\section{Systematics}
\label{sec:systematics}

A great part of the systematics in our analysis is inherent to the code and the various approximations and assumptions made
to build the stellar evolution tracks, metal enrichment, SN rates, etc. We discuss here only the internal systematics in \pegase.
For a comparison between models, we refer to \citealt[][]{Conroy2013}. 
In the case of \pegase, the choice of IMF and star formation law certainly produce systematics that can affect our results.

The IMF has been subject to considerable discussion, particularly concerning the potential universality
of the IMF \citep[see][for a recent review]{Bastian2010}. 
The difference in mass between IMF between a \cite{Salpeter1955} IMF and a \cite{Kroupa2001} IMF is contained 
in a factor of $\sim$2, as noted in RV13 and other studies \citep[e.g.][]{Marchesini2009}. 

As for the star formation law, the \pegase\ code sets the star formation history and star formation efficiency based on
the Hubble type galaxies at $z$=0. To test how sensitive our results are to this choice, we used four scenarios for the
host galaxy (evolved component): E, E2, S0, and Sa templates \citep{Fioc2014}. While the results show 
variations (up to 4 in mass and age, $\sim$2 in most cases), it is impossible to select a preferred scenario based on the $\chi^2$ values. The star 
formation history of the evolved component is therefore our largest systematic. That said, even if mass and age present variations,
the fact that two stellar components are needed to reproduce the SED (see Sect. \ref{sec:results}) remains unchanged and does not affect the primary conclusions of this paper. 

The main systematics for the AGN component are the continuous approach and
 radial and azimuthal dust density profiles. For the clumpy versus smooth tori, \cite{Dullemond2005} and \cite{Feltre2012}
showed that they can be equivalent for a given set of parameters. \cite{Fritz2006} showed that the
dust density profiles have a limited effect on the continuum properties in our considered range (at most 10\%, see their Figs. 7 and 8). 

\section{Domain, interval, and weight of the parameters}
\label{sec:interval}

Each parameter was defined onto a grid, presented in Table~\ref{tab:param}. Only the
stellar masses and the AGN luminosity were calculated directly from the normalisation of the components during the fitting
and were an output of our SED fitting. However, each parameter has a different effect on the final SED. They were also
constrained mostly in the wavelength range they dominate. For instance, a type 2 AGN consists of an accreting SMBH obscured 
by a dusty torus, with mostly high dust temperature ($>$200K), and it mostly dominates in the mid-IR domain. Therefore good mid-IR 
data will allow constraining better the parameters of the AGN model better. We discuss this weight of the parameters and their final effect on the broad-band SED. 

\subsection{AGN component}

As mentioned previously, we focused on typical configurations of the AGN torus. The size is important because the farther out the central part, the cooler the dust, causing the SED to peak
at longer wavelength. Opacity also plays a role in the overall AGN SED in the IR. In particular, the opacity affects
both the peak of the dust emission and its width \citep[see Fig. 7 and 8 of ][]{Fritz2006}. These parameters mostly affect the transition 
regime from dust heated by the AGN to dust heated by the SB, in the 40-60\mum\ range. 
We note that some of our results show a discrepancy in this wavelength range, suggesting that the torus size is not 
sufficiently explored in our fitting procedure (see Sect. \ref{sec:limitations}).
The AGN inclination will have the strongest effect at shorter
wavelength, affecting the peak of the dust emission (typically $\sim$4-10\mum\ restframe), where the fraction
of the view of the innermost part of the AGN (i.e. close the sublimation radius) is visible to the observer increases. 
This wavelength range is particularly critical because it concerns the transition between the emission of intermediate-to-low-mass 
stars, evolved stars (e.g. AGB) and the host dust of the AGN torus. We therefore selected three torus sizes, with 
$Y$=10, $Y$=60 and $Y$=150, which represent different extensions
of the outer part of the torus, to test the effect of the torus size on the AGN-SB transition ($\sim$40-60\mum\ restframe interval).

\subsection{P\'EGASE components}

For the SB component, mass and age are among the most important 
parameters (see Fig. 2 of RV13) as both parameters strongly affect the luminosity and 
shape of the SED. A young stellar population produces UV and strong dust heating, the mass
of the system and a rapid metal enrichment when the most massive stars eventually explode. 
The column density is also of great importance, as a higher column density 
implies more absorption by dust. The direct effect of this is
will be a shift of the SB dust peak to higher temperature (i.e. shorter wavelength) and 
decreased UV and optical emission. The two highest redshift sources of our sample, presented in RV13, show a 
clear preference for a higher column density factor to reproduce the SED ($K$=10), confirmed in this analysis.
The initial metallicity is the parameter with the most limited effect on the SED in this analysis. However, the difference between a
null metallicity and a highly enriched initial gas implies a strong change on the stellar evolutionary tracks, 
and therefore an effect on the other parameters and on the final SED. Spectroscopic data are required to fully assess this parameter,
while broad-band photometry provides very limited constraints on this parameter. However, we kept this parameter
in our fitting to determine the final effect on our best fit and whether a complete fitting from 
UV to submm can provide useful information on the initial metallicity. 

For the evolved component, we recall that \pegase\ predicts the integrated
SED for a given scenario, and the Hubble type refers, for convenience, to the combination of parameters necessary to reproduce the
observed colours or SED of the classical Hubble type galaxies at $z$$=$0. The model assumes a closed-box evolution (with a galaxy plus reservoir), where progressive enrichment
of the ISM translates into more dust. The main parameters are $z_{form}$, the astration, the star formation rate, the infall timescale and the 
galactic winds. In addition, each galaxy type corresponds to a different set of these parameters. By considering several types, we explore different
star formation histories that will reproduce observed galaxies in the local Universe \citep{Tsalmantza2009,Tsalmantza2012}. 
The initial metallicity is null as galaxies are assumed to form from primordial, metal-free gas as formed at high redshift $z_{form}$=10. A slight 
change of $z_{form}$ will not have a dramatic effect because the fraction of age at high redshift is not significant compared to the age at $z$=0. 
The star formation rate is proportional to the gas available in the galaxy, the gas supply being driven by the infall timescale. A longer infall timescale, such as the spiral scenario, allows more moderate star formation over longer periods, reproducing the star-forming disc, hence the galaxy appears bluer at optical wavelengths. The astration rate is the efficiency of converting gas into stars, increasing from spiral to elliptical. 
The galactic winds expel all the ISM from the galaxy, and thus prevent any further star formation. This parameter is only used in elliptical scenarios
to reproduce the red colours of elliptical galaxies in the local Universe. Complete details of each parameter are 
described in \cite{Fioc1997} for \pegaseii\ and \cite{Fioc2014} for \pegaseiii.

\section{Notes on individual sources}
\label{sec:individual}

We briefly summarise for each object
the morphological peculiarities and interesting aspects from a multi-wavelength point of view. Each of the objects is
extensively described in their respective papers, to which we refer for more details (see Tables \ref{tab:3C368_data}-\ref{tab:tnj2007_data}).
{\bf We present the the NIR and optical data, along with radio contours when data are available in Sect. \ref{sec:discussion} from Fig. 
\ref{fig:optical_4C4117} to Fig. \ref{fig:optical_3C368}.}

\paragraph{3C~368, in Fig. \ref{fig:optical_3C368} ---}
This galaxy is located a $z$=1.13 \citep{Djorgovski1987}, 
and has been extensively studied from X-ray to radio energies \citep[e.g.,][]
{Best1997,Zirm2003,Reuland2004,Leipski2010,Stacey2010}. Two stars visible 
in high-resolution imaging \citep{Best1997} could affect the final 
photometry presented in Table \ref{tab:3C368_data}. \cite{Best1997} identified two components 
assumed to be from the same galaxy. Between the optical and NIR images, 
the emission peak shifts to the north, implying that the northern component
is likely heavily obscured \citep{Stockton1996}. The contour of the radio emission is aligned
between the optical morphology and the jet axis.

\paragraph{3C~470, in Fig. \ref{fig:optical_3C470} ---}
This galaxy is located at $z$=1.65 \citep{McCarthy1988}, and shows a potential companion in the high-resolution optical image 
\citep{Best1997} that also affects the photometry. For simplicity, we use photometry including the two components
since the components blend at longer wavelength (i.e., the \herschel\ data). We note that the 
two components are not aligned with the radio jet axis, in contrast to most HzRGs \citep{Chambers1996b,vanBreugel1998}. 

\paragraph{MRC~0324-228, in Fig. \ref{fig:optical_MRC0324} ---}
This galaxy is at $z$=1.89 \citep{McCarthy1991} and has a companion in high-resolution 
NIR images \citep{Pentericci2001}. However, nothing indicates that this companion 
is part of the system.

\paragraph{PKS~1138-262, in Fig. \ref{fig:optical_PKS1138} ---}
Also named the ``Spiderweb galaxy'', this galaxy is at $z$=2.16 \citep{vanOjik1995}. Its unusual morphology consists of a central massive galaxy 
surrounded by fainter companions \citep{Pentericci1998,Miley2006,Kuiper2010,Hatch2011b,Seymour2012}. The 
galaxy has an extension that is aligned with the radio jet axis.

\paragraph{MRC~0406-244, in Fig. \ref{fig:optical_MRC0406} ---}
This galaxy is at $z$=2.43 \citep{McCarthy1991} and has an extensive data set
\citep{Rush1997a}. Some features, aligned with the radio jet axis in  
high-resolution optical images \citep{Pentericci2001}, suggest a complex system. New 
\hst\ high-resolution data are also available \citep{Hatch2013}, enabling a more 
precise multi-wavelength morphological characterisation. No companion is evident in
the direct vicinity ($<$3\arcsec) of the radio galaxy, and the ionisation cones are aligned with the radio jet axis.

\paragraph{MRC~2104-242, in Fig. \ref{fig:optical_MRC2104} ---}
This galaxy is at $z$=2.49 \citep{McCarthy1991} and shows several components in 
the optical and NIR images \citep{Pentericci2001}. The images and spectroscopy 
exhibit a large Ly$\alpha$  halo extended along the radio lobes \citep{Pentericci2001}. 
A companion is present $\sim$2\arcsec north of the galaxy. 

\paragraph{USS~0828+193, in Fig. \ref{fig:optical_USS0828} ---}
This source is at $z$=2.57 \citep{Knopp1997b} and has a star 3\arcsec\ 
south of the galaxy, complicating the photometry. The radio galaxy 
itself seems to be split into multiple components in the high-resolution NIR
images \citep{Steinbring2002}. The galaxy has polarimetric observations, that have been reported in 
 \cite{Knopp1997b}.

\paragraph{USS~0943-242, in Fig. \ref{fig:optical_USS0943} ---}
This source is at $z$=2.92 \citep{Rottgering1995} and has an elongation in the high-resolution \hst\ image
that is broadly aligned  with the radio jet axis. While the optical \hst\ image does not show a close companion, 
the MIR IRAC image reveals a close potential companion 6\arcsec\ south-west of the radio galaxy that appears
at redder wavelengths. The two sources blend in the SPIRE images, $\lambda$$>$250\mum. 

\paragraph{4C~28.58 ---}
This source is at $z$=2.89 \citep{Chambers1996a} and presents a very disturbed 
morphology in \hst\ images \citep{Chambers1996b}. Multiple components appear, 
the main flux coming from two components close to the radio core. 
Given astrometric uncertainties, we cannot determine which component is the primary radio galaxy. 
The elongation observed in the \hst\ image is roughly aligned with the radio jet axis.

\paragraph{4C~41.17 and TN~J2007-1316, in Fig. \ref{fig:optical_4C4117} and \ref{fig:optical_TNJ2007} ---}
These sources, at $z$=3.79 and $z$=3.84, respectively, have been presented extensively in RV13. 
We briefly report some unusual properties. 4C~41.17 is weakly polarised \citep{Dey1997}, and shows a low contribution
of the AGN in the IR (RV13). In addition, a star-forming 
companion is evident in the high-resolution optical images. TN~J2007-1316 is also weakly polarised, 
but has a stronger AGN contribution in the MIR (RV13).

\section{Additional material: Photometry and SED fit for the sample}

All \herschel\ flux are taken from \cite{Drouart2014}. Some of these flux present a signal-to-noise ratio $<$3 and are explained by the
method used to calculate photometry on our sources, from the 24\mum\ prior and the calibration uncertainties. 
We refer to Sect. 2 in \cite{Drouart2014} for better details.

\begin{table}[h!] \centering
\caption{Data for 3C~368 ($z$=1.132). $\lambda_0$ is the central wavelength, line 
contribution (in \%) in the broad-band, corrected in the final flux, flux in mJy 
(for a non-detection, the 3$\sigma$ upper limit 
is quoted).}
%\resizebox{17cm}{!}{
\label{tab:3C368_data}
%\makebox[0.8\textwidth]{
\begin{tabular}{lcccc}
\hline
Filter & $\lambda_0$ & Line & Flux & Ref.\\
 & [\mum] & [\%] & [\ujy]  & \\
\hline \hline
B(CFHT)      &   0.45   &       -      &        11.5      $\pm$     1.6   & A  \\
R(CFHT)      &   0.65   &       -      &        17.2      $\pm$     2.5   & A  \\
F702W        &   0.68   &       24     &        21.7      $\pm$     2.1   & B  \\
F791W        &   0.78   &       38     &        28.0      $\pm$     2.7   & B  \\
J(UKIRT)     &   1.27   &       3      &        36.9      $\pm$     4.0   & C  \\
H(UKIRT)     &   1.67   &       2      &        71.6      $\pm$     9.1   & C  \\
K(UKIRT)     &   2.2    &       4      &        85.0      $\pm$     8.2   & C  \\
IRAC1        &  3.6     &      -       &       126        $\pm$    13.0   & D  \\
IRAC2        &  4.5     &      -       &       112        $\pm$    11.0   & D  \\
IRAC3        &  5.8     &      -       &       112        $\pm$    11.0   & D  \\
IRAC4        &  8.0     &      -       &       210        $\pm$    21.0   & D  \\
IRS          &  16.0    &      -       &       1620       $\pm$    180    & D  \\
MIPS1        &  24.0    &      -       &       3350       $\pm$    50.0   & D  \\
PACS 70      &  70      &      -       &       32300      $\pm$    3500   & E  \\
PACS 160     &  160     &      -       &       52900      $\pm$    9100   & E  \\
SPIRE 250    &  250     &      -       &       35600      $\pm$    6200   & E  \\
SPIRE 350    &  350     &      -       &       19600      $\pm$    6500   &  E \\
SPIRE 500    &  500     &      -       &       $<$19000                   &  E \\
SCUBA 850    &  850     &      -       &       4080       $\pm$    1080   &  F \\
IRAM 1300    &  1300    &      -       &       $<$600.0                   &  G \\
\hline
\hline
\end{tabular} 
\tablebib{[A] \cite{Djorgovski1987};
[B] \cite{Best1997}; [C] \cite{Inskip2006}; [D] \cite{DeBreuck2010}; [E] \cite{Drouart2014}; 
[F] \cite{Archibald2001}; [G] \cite{Meisenheimer2001}.}
\end{table}

\begin{table} \centering
\caption{Data for 3C~470 ($z$=1.65). Columns are the same as in Table \ref{tab:3C368_data}.}
\label{tab:3C470_data}
\begin{tabular}{lcccc}
\hline
Filter & $\lambda_0$ & Line  & Flux  & Ref. \\
 & [\mum] & [\%] & [\ujy] & \\
\hline \hline
F785LP     &   0.87    &     15    &   4.71   $\pm$       1.5      & A   \\             
K(UKIRT)   &   2.2     &     -     &   38.9   $\pm$       5.8      & A   \\             
K(UKIRT)   &   2.2     &     -     &   35.8   $\pm$       4.2      & B   \\         
IRAC1      &  3.6      &    -      &  49.5    $\pm$      10.4      & C   \\      
IRAC2      &  4.5      &    -      &  75.2    $\pm$      11.8      & C   \\  
IRAC3      &  5.8      &    -      &  70.9    $\pm$      10.4      & C   \\           
IRAC4      &  8.0      &    -      &  266     $\pm$    30.0        & C   \\           
IRS        &  16.0     &    -      &  1540    $\pm$    180         & C   \\           
MIPS1      &  24.0     &    -      &  2750    $\pm$    40          & C   \\           
PACS 70    &  70       &    -      &  21600   $\pm$    3700        & D   \\            
PACS 160   &  160      &    -      &  22500   $\pm$    7200        & D   \\            
SPIRE 250  &  250      &    -      &  39200   $\pm$    5700        & D   \\            
SPIRE 350  &  350      &    -      &  33500   $\pm$    6000        & D   \\            
SPIRE 500  &  500      &    -      &  24900   $\pm$    6300        & D   \\            
SCUBA 850  &  850   &      -      &  $<$9300                      & E   \\
\hline                                    
\end{tabular}
\tablebib{[A] \cite{Best1997}; [B] \cite{Targett2011}; [C] \cite{DeBreuck2010}; [D] \cite{Drouart2014}; [E] \cite{Archibald2001}.}
\end{table}

\begin{table} \centering
\caption{Data for  MRC~0324-228 ($z$=1.894). Columns are the same as in Table \ref{tab:3C368_data}.}
\label{tab:mrc0324_data}
\begin{tabular}{lcccc}
\hline
Filter & $\lambda_0$ & Line  & Flux  & Ref. \\
 & [\mum] & [\%] & [\ujy] & \\
\hline \hline
R(Bessel)  &  0.65  &     -       &     1.5    $\pm$      0.25       & A    \\     
F165M      &  1.64  &     -       &     10.5   $\pm$      1.1        & B    \\
IRAC1      &  3.6   &     -       &     39.4   $\pm$      4.2        & C    \\  
IRAC2      &  4.5   &     -       &     39.7   $\pm$      4.3        & C    \\  
IRAC3      &  5.8   &     -       &     61.1   $\pm$      8.6        & C    \\  
IRAC4      &  8.0   &     -       &     89.9   $\pm$      9.9        & C    \\  
IRS        &  16.0  &     -       &     530.0  $\pm$      54.0       & C    \\  
MIPS1      &  24.0  &     -       &     1880.0 $\pm$      35.0       & C    \\  
PACS 70    &  70    &     -       &     $<$9100                      & D    \\   
PACS 160   &  160   &     -       &     27900  $\pm$      5400       & D    \\   
SPIRE 250  &  250   &     -       &     61800  $\pm$      6700       & D    \\   
SPIRE 350  &  350   &     -       &     35500  $\pm$      5900       & D    \\   
SPIRE 500  &  500   &     -       &     17500  $\pm$      7400       & D    \\   
LABOCA     &  870   &     -       &     9000   $\pm$      2000       & D    \\   
\hline                        
\end{tabular}
\tablebib{[A] \cite{Buchard2008}; [B] \cite{Pentericci2001}; [C] \cite{DeBreuck2010}; [D] \cite{Drouart2014}.}
\end{table}

\begin{table} \centering
\caption{Data for PKS~1138-262 ($z$=2.156). Columns are the same as in Table \ref{tab:3C368_data}.
 {\it (Priv.)} for "private communication".}
\label{tab:pks1138_data}
\begin{tabular}{lcccc}
\hline
Filter & $\lambda_0$ & Line  & Flux  & Ref. \\
 & [\mum] & [\%] & [\ujy] & \\
\hline \hline
F475W       &  0.48    &       -        &     7.66   $\pm$      0.8       &   N. Hatch (Priv.) \\                 
F814W       &  0.8     &       -        &     14.5   $\pm$      1.5       &   N. Hatch (Priv.) \\                 
F110W       &  1.1     &       -        &     29.4   $\pm$      3.0       &   N. Hatch (Priv.) \\                 
F160W       &  1.6     &       -        &     87.9   $\pm$      9.0       &   N. Hatch (Priv.) \\                 
Ks(ISAAC)   &  2.16    &       -        &     207    $\pm$      21.0      &   N. Hatch (Priv.) \\                 
IRAC1       &  3.6     &       -        &     318    $\pm$      32.0      &   A  \\           
IRAC2       &  4.5     &       -        &     497    $\pm$      50.0      &   A  \\           
IRAC3       &  5.8     &       -        &     887    $\pm$      89.0      &   A  \\           
IRAC4       &  8.0     &       -        &     1500   $\pm$      150       &   A  \\           
IRS         &  16.0    &       -        &     3020   $\pm$      100       &   A  \\           
MIPS1       &  24.0    &       -        &     3890   $\pm$      20        &   A  \\           
PACS 100    &  100     &       -        &     25200  $\pm$      2200      &   B  \\            
PACS 160    &  160     &       -        &     40200  $\pm$      10200     &   B  \\            
SPIRE 250   &  250     &       -        &     40400  $\pm$      5900      &   B  \\            
SPIRE 350   &  350     &       -        &     33000  $\pm$      6100      &   B  \\            
SPIRE 500   &  500     &       -        &     28900  $\pm$      6700      &   B  \\            
SCUBA 850   &  850     &       -        &     12800  $\pm$      3600      &   C  \\             
\hline                                    
\end{tabular}
\tablebib{[A] \cite{DeBreuck2010}; [B] \cite{Drouart2014}; [C] \cite{Reuland2004}.}
\end{table}

\begin{table} \centering
\caption{Data for  MRC~0406-244 ($z$=2.427).  Columns are the same as in Table \ref{tab:3C368_data}. 
No estimates available for the line contamination; so fluxes are not corrected. $^{*}$A galaxy NW might contaminate the photometry.}
\label{tab:mrc0406_data}
\begin{tabular}{lcccc}
\hline
Filter & $\lambda_0$ & Line  & Flux  & Ref. \\
 & [\mum] & [\%] & [\ujy] & \\
\hline \hline
F555W       &  0.52   &       -     &        2.11     $\pm$     0.31     & A   \\
r(SDSS)     &  0.62   &       -     &        2.73     $\pm$     0.4      & A   \\      
I(BESS)$^{*}$     &  0.79   &       -    &        6.6      $\pm$     1.0      & A   \\
J(NIRC)     &  1.25   &       yes  &        17.3     $\pm$     2.6      & A   \\
K(NIRC)     &  2.2    &       yes   &        32.0     $\pm$     4.7      & A   \\
IRAC1       &  3.6    &       -     &        40.4     $\pm$     4.3      & B   \\
IRAC2       &  4.5    &       -     &        43.3     $\pm$     4.6      & B   \\
IRAC3       &  5.8    &       -     &        $<$77.4                     & B   \\
IRAC4       &  8.0    &       -     &        63.5     $\pm$     14.5     & B   \\
IRS         &  16.0   &       -     &        637      $\pm$     86.0     & B   \\
MIPS1       &  24.0   &       -     &        1540     $\pm$     40.0     & B   \\
PACS 100    &  100    &       -     &        $<$12300                    & C   \\
PACS 160    &  160    &       -     &        21500    $\pm$     7900     & C   \\ 
SPIRE 250   &  250    &       -     &        47600    $\pm$     5600     & C   \\
SPIRE 350   &  350    &       -     &        38700    $\pm$     5300     & C   \\
SPIRE 500   &  500    &       -     &        22800    $\pm$     5900     & C   \\
LABOCA      &  870.   &       -     &        $<$17800                    & C   \\
 \hline
\end{tabular}
\tablebib{[A] \cite{Rush1997a}; [B] \cite{DeBreuck2010}; [C] \cite{Drouart2014}.}
\end{table}

\begin{table} \centering
\caption{Data for MRC~2104-242 ($z$=2.49). Columns are the same as in Table \ref{tab:3C368_data}.}
\label{tab:mrc2104_data}
\begin{tabular}{lcccc}
\hline
Filter & $\lambda_0$ & Line  & Flux  & Ref. \\
 & [\mum] & [\%] & [\ujy] & \\
\hline \hline
F606W       & 0.59   &     -      &      3.28      $\pm$   0.3       & A \\          
F160W       & 1.6    &     26     &      29.5      $\pm$   3.8       & B \\          
IRAC1       & 3.6    &     -      &      28.1      $\pm$   3.3       & C \\            
IRAC2       & 4.5    &     -      &      29.7      $\pm$   3.5       & C \\            
IRAC3       & 5.8    &     -      &      32.8      $\pm$   10.0      & C \\            
IRAC4       & 8.0    &     -      &      $<$54.4                     & C \\            
IRS         & 16.0   &     -      &      $<$325.5                    & C \\            
MIPS1       & 24     &     -      &      709       $\pm$   48.0      & C \\            
PACS 100    & 100    &     -      &      14400     $\pm$   3500      & D \\             
PACS 160    & 160    &     -      &      22000     $\pm$   8400      & D \\             
SPIRE 250   & 250    &     -      &      14200     $\pm$   5100      & D \\             
SPIRE 350   & 350    &     -      &      21100     $\pm$   6600      & D \\     
SPIRE 500   & 500    &     -      &      $<$15800                    & D \\             
\hline                                              
\end{tabular} 
\tablebib{[A] \cite{Pentericci1999}; [B] \cite{Pentericci2001}; [C] \cite{DeBreuck2010}; [D] \cite{Drouart2014}.}                                      
\end{table}                                        
                                                    
\begin{table} \centering
\caption{Data for USS~0828+193 ($z$=2.57). Columns are the same as in Table \ref{tab:3C368_data}. }
\label{tab:uss0828}
\begin{tabular}{lcccc}
\hline
Filter & $\lambda_0$ & Line  & Flux  & Ref. \\
 & [\mum] & [\%] & [\ujy] & \\
\hline \hline
F675W      &  0.59   &      -      &      1.51      $\pm$   0.15   & A  \\         
J(KIR)     &  1.25   &      -      &      15.2      $\pm$   1.5    & A  \\         
H(KIR)     &  1.63   &      -      &      36.1      $\pm$   3.5    & A  \\         
K(KIR)     &  2.2    &      -      &      51.2      $\pm$   8.9    & A  \\         
IRAC1      &  3.6    &      -      &      61.7      $\pm$   6.9    & B  \\            
IRAC2      &  4.5    &      -      &      133.0     $\pm$   13.0   & B  \\            
IRAC3      &  5.8    &      -      &      201.0     $\pm$   21.0   & B  \\            
IRAC4      &  8.0    &      -      &      687.0     $\pm$   74.0   & B  \\            
IRS        &  16.0   &      -      &      1910      $\pm$   130    & B  \\            
MIPS1      &  24     &      -      &      2880      $\pm$   40.0   & B  \\            
PACS 100   &  100    &      -      &      18500     $\pm$   3500   & C  \\             
PACS 160   &  160    &      -      &      24000     $\pm$   9600   & C  \\             
SPIRE 250  &  250    &      -      &      20200     $\pm$   4500   & C  \\             
SPIRE 350  &  350    &      -      &      17500     $\pm$   4700   & C  \\  
SPIRE 500  &  500    &      -      &      $<$17200                 & C  \\            
\hline                                              
\end{tabular}                                       
\tablebib{[A] \cite{Steinbring2002}; [B] \cite{DeBreuck2010}; [C] \cite{Drouart2014}.}
\end{table}                                                                                
                                            
\begin{table} \centering
\caption{Data for 4C~28.58 ($z$=2.89). Columns are the same as in Table \ref{tab:3C368_data}.}
\label{tab:4c2858_data}
\begin{tabular}{lcccc}
\hline
Filter & $\lambda_0$ & Line  & Flux  & Ref. \\
 & [\mum] & [\%] & [\ujy] & \\
\hline \hline
F439W      &  0.43    &      -       &   $<$0.84                        & A     \\    
F675W      &  0.67    &     -        &    1.23     $\pm$   0.1          & A     \\ 
F814W      &  0.83    &     -        &    0.92     $\pm$   0.1          & A     \\ 
K(NIRC)    &  2.2     &     -        &    13.1     $\pm$   1.3          & B     \\  
IRAC1      &  3.6     &     -        &    31.6     $\pm$   3.5          & C     \\ 
IRAC2      &  4.5     &     -        &    36.0     $\pm$   3.9          & C     \\ 
IRAC3      &  5.8     &     -        &    $<$62.6                       & C     \\
IRAC4      &  8.0     &     -        &    40.9     $\pm$   4.4          & C     \\ 
IRS        &  16.0    &     -        &    430.0    $\pm$   85.0         & C     \\ 
MIPS1      &  24.0    &     -        &    866.0    $\pm$   155.0        & C     \\ 
PACS 100   &  100     &     -        &    22800    $\pm$   2900         & D     \\  
PACS 160   &  160     &     -        &    23600    $\pm$   8700         & D     \\  
SPIRE 250  &  250     &     -        &    42500    $\pm$   4900         & D     \\  
SPIRE 350  &  350     &     -        &    29700    $\pm$   5400         & D     \\  
SPIRE 500  &  500     &     -        &    15500    $\pm$   4500         & D     \\  
SCUBA 850  &  850.0   &     -        &    3930     $\pm$   950.0        & E     \\
\hline                                             
\end{tabular}
\tablebib{[A] \cite{Chambers1996a}; [B] \cite{Akiyama2008}; [C] \cite{DeBreuck2010}; [D] \cite{Drouart2014}; [E] \cite{Archibald2001}. }
\end{table}

\begin{table} \centering
\caption{Data for USS~0943-242 ($z$=2.92). Columns are the same as in Table \ref{tab:3C368_data}.}
\label{tab:uss0943_data}
\begin{tabular}{lcccc}
\hline
Filter & $\lambda_0$ & Line  & Flux  & Ref. \\
 & [\mum] & [\%] & [\ujy] & \\
 \hline \hline
F702W      &   0.7    &     11.0    &      3.58  $\pm$      0.3       & A    \\
F160W      &   1.6    &     -       &      13.3  $\pm$      1.0       & B    \\    
IRAC1      &   3.6    &     -       &      21.5  $\pm$      2.6       & C    \\
IRAC2      &   4.5    &     -       &      28.4  $\pm$      3.2       & C    \\
IRAC3      &  5.8     &     -       &    $<$30.9                      & C    \\
IRAC4      &   8.0    &     -       &    25.8    $\pm$      11.7      & C    \\
IRS        &   16.0   &     -       &    170.0   $\pm$      48.0      & C    \\
MIPS1      &   24.0   &     -       &    518.0   $\pm$      40.0      & C    \\
MIPS2      &  70.0    &     -       &   $<$3390                       & C    \\
MIPS3      &  160.0   &     -       &    $<$50900                     & C    \\
PACS 100   &   100    &     -       &   $<$27600                      & D    \\
PACS 160   &   170    &     -       &     23600  $\pm$     7700       & D    \\ 
SPIRE 250  &   250    &     -       &    25700   $\pm$      5200      & D    \\
SPIRE 350  &   350    &     -       &     31700  $\pm$      5500      & D    \\
SPIRE 500  &   500    &     -       &     35200  $\pm$      7300      & D    \\
1.3mm(SEST)&  1300    &     -       &  $<$14200                       & E    \\   
\hline                                                                                                    
\end{tabular}
\tablebib{[A] \cite{Pentericci1999}; [B] \cite{Pentericci2001}; [C] \cite{DeBreuck2010}; [D] \cite{Drouart2014}; [E] \cite{Cimatti1998}.}
\end{table}

\begin{table} \centering
\caption{Data for 4C~41.17 ($z$=3.79). Columns are the same as in Table \ref{tab:3C368_data}.}
\label{tab:4c4117_data}
\begin{tabular}{lcccc}
\hline
Filter & $\lambda_0$ & Line  & Flux  & Ref. \\
 & [\mum] & [\%] & [\ujy] & \\
\hline \hline
F702W        & 0.7    &    -         &    5.01      $\pm$    0.36    & A  \\              
I(KPNO)      & 0.9    &    -         &    4.5       $\pm$    1.5     & B  \\         
J(NIRC)      & 1.25   &    -         &    5.6       $\pm$    1.1     & C  \\       
Ks(NIRC)     & 2.155  &    -         &    13.6      $\pm$    2.8     & C  \\       
IRAC1        & 3.6    &    -         &    23.4      $\pm$    2.40    & D  \\         
IRAC2        & 4.5    &    -         &    27.5      $\pm$    2.80    & D  \\         
IRAC3        & 5.8    &    -         &    35.6      $\pm$    3.70    & D  \\         
IRAC4        & 8.0    &    -         &    36.5      $\pm$    3.50    & D  \\         
IRS          & 16.0   &    -         &    $<$181.0                   & D  \\         
MIPS1        & 24     &    -         &    370.0     $\pm$    40.0    & D  \\         
PACS 70      & 70     &    -         &    $<$2600                    & E  \\           
PACS 160     & 170    &    -         &    16200     $\pm$    6700    & E  \\           
SPIRE 250    & 250    &    -         &    35800     $\pm$    3500    & E  \\           
SPIRE 350    & 350    &    -         &    43100     $\pm$    3700    & E  \\           
SPIRE 500    & 500    &    -         &    38000     $\pm$    4500    & E  \\           
UKT14 800    & 800    &    -         &    17400     $\pm$    3100    & F  \\          
SCUBA 850    & 850    &    -         &    12100     $\pm$    880     & G  \\        
IRAM 1200    & 1200   &    -         &    4400      $\pm$    400     & H  \\            
IRAM 1300    & 1300   &    -         &    2500      $\pm$    400     & I  \\            
\hline                                    
\end{tabular}
\tablebib{[A] \cite{Miley1992}; [B] \cite{Chambers1990}; [C] \cite{vanBreugel1998}; [D] \cite{DeBreuck2010},
[E] \cite{Wylezalek2013a}; [F] \cite{Dunlop1994n}; [G] \cite{Archibald2001}; [H] \cite{Greve2007}; [I] \cite{Chini1994}.}
\end{table}

\begin{table} \centering
\caption{Data for TN~J2007-1316 ($z$=3.84).Columns are the same as in Table \ref{tab:3C368_data}.}
\label{tab:tnj2007_data}
\begin{tabular}{lcccc}
\hline
Filter & $\lambda_0$ & Line  & Flux  & Ref. \\
 & [\mum] & [\%] & [\ujy] & \\
\hline \hline
R          &   0.65   &      -     &       2.06    $\pm$      0.4    & A  \\                
I(CFHT)    &   0.9    &     -      &       2.60    $\pm$      0.15   & A  \\                
H(ISAAC)   &   1.65   &     -      &       9.6     $\pm$      1.0    & A  \\                
K(UKIRT)   &   2.2    &     -      &       28.4    $\pm$      1.9    & B  \\          
IRAC1      &   3.6    &     -      &       46.6    $\pm$      4.8    & A  \\              
IRAC2      &   4.5    &     -      &       52.7    $\pm$      5.7    & A  \\              
IRAC3      &   5.8    &     -      &     $<$146.0                    & A  \\              
IRAC4      &   8.0    &     -      &       135.1   $\pm$      16.9   & A  \\              
IRS        &   16.0   &     -      &       378.0   $\pm$      113.0  & C  \\           
MIPS1      &   24     &     -      &       385.0   $\pm$      40.0   & C  \\           
PACS 100   &   105    &     -      &     $<$35025                    & A  \\             
PACS 160   &   170    &     -      &     $<$62034                    & A  \\             
SPIRE 250  &   250    &     -      &       13850   $\pm$      6100   & A  \\             
SPIRE 350  &   350    &     -      &       16500   $\pm$      6376   & A  \\             
SPIRE 500  &   500    &     -      &       7615    $\pm$      3322   & A  \\             
SCUBA 850  &   850    &     -      &       5800    $\pm$      1500   & D  \\            
\hline                                    
\end{tabular}
\tablebib{[A] \cite{Rocca2013}; [B] \cite{Bornancini2007}; [C] \cite{DeBreuck2010}; [D] \cite{Reuland2004}.}
\end{table}

\clearpage

\begin{landscape}
\clearpage
\begin{figure} \centering
\begin{overpic}[width=1.0\textwidth,angle=0,trim= 0 0 0 0,scale=1.2]{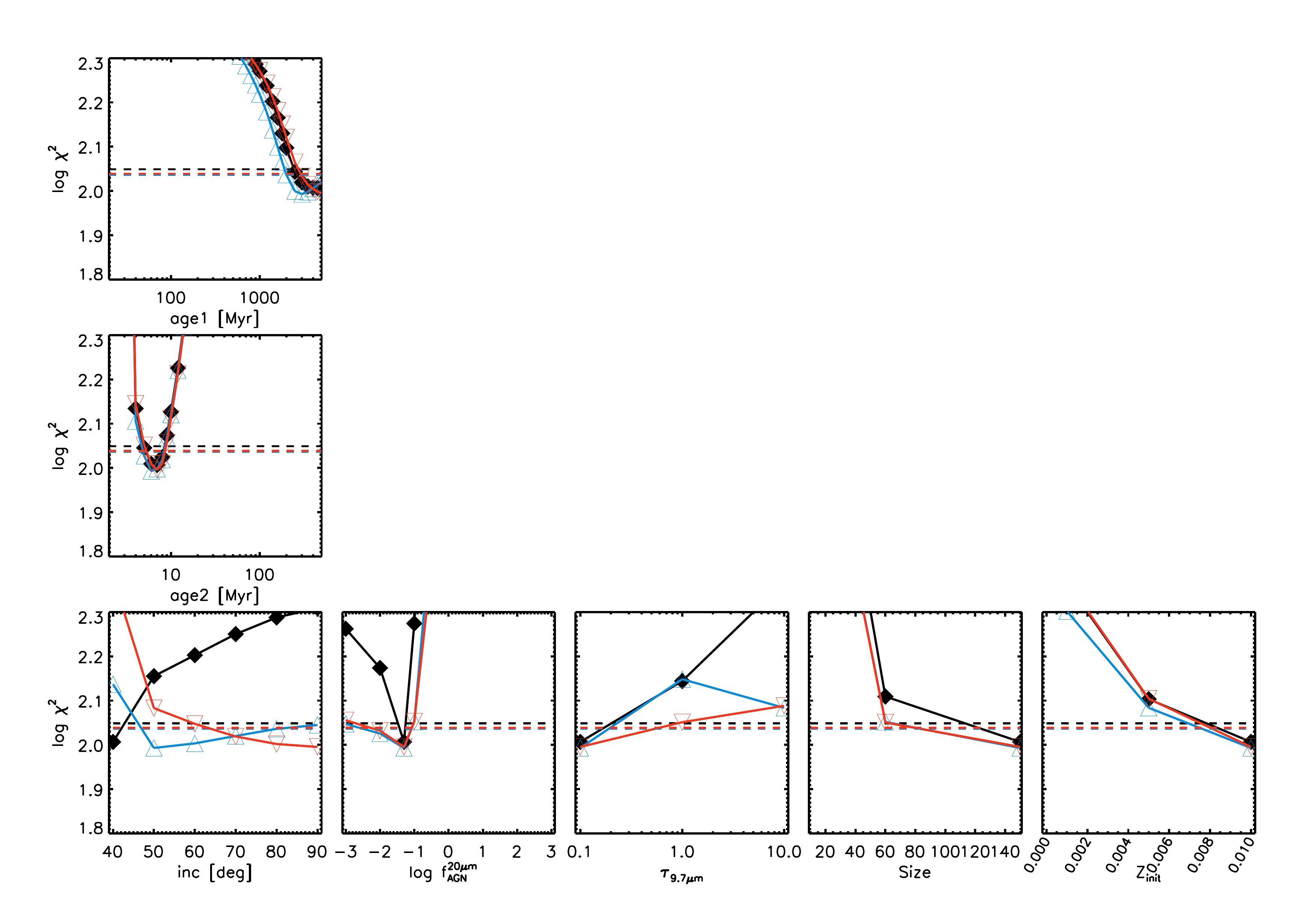}
\put(25,75){\includegraphics[height=0.7\textwidth,angle=270,trim= 0 0 0 0,scale=1.2]{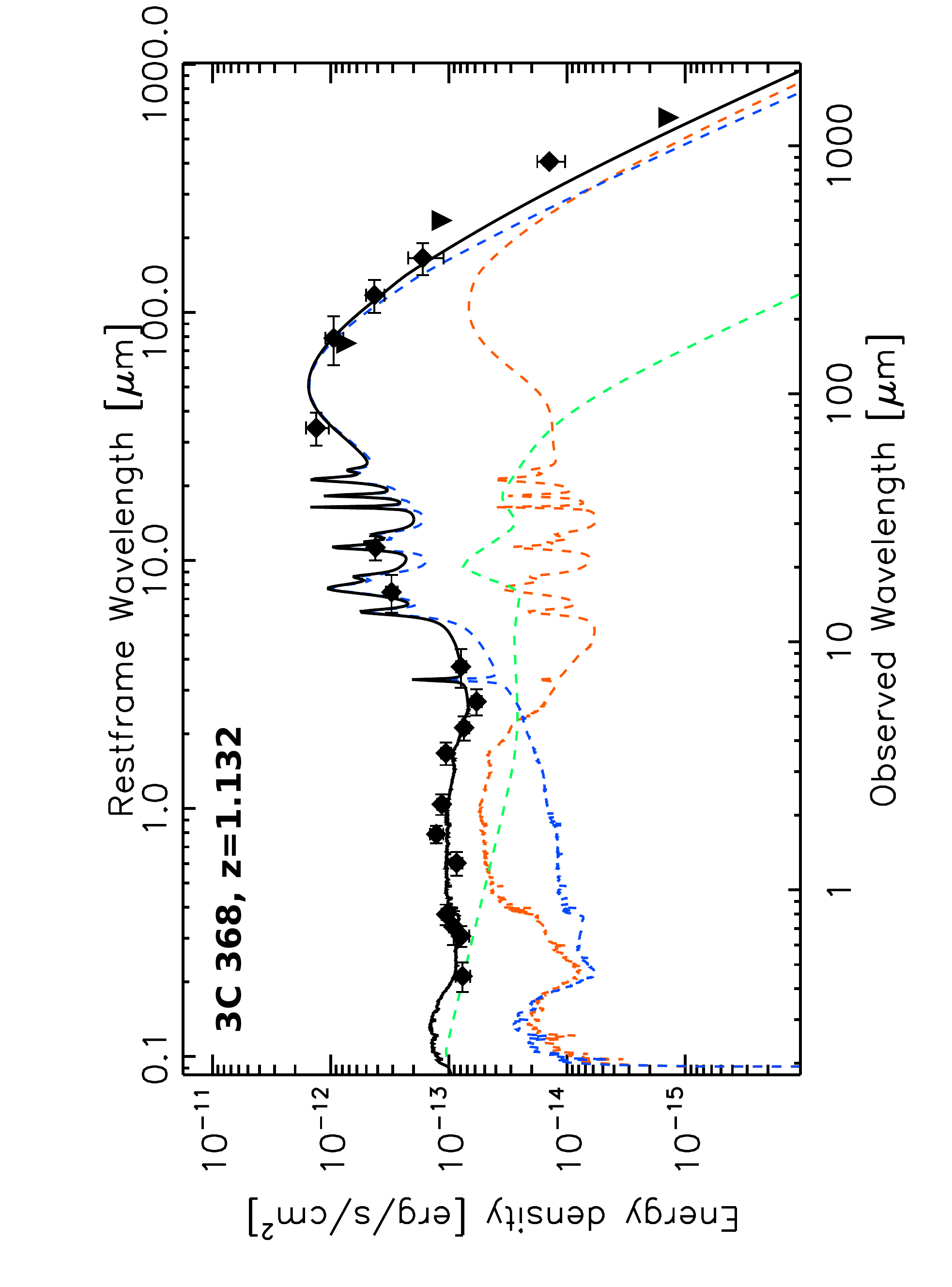}}
\end{overpic}
\caption{Best fit for 3C~368 ($z=1.132$). The colour code and the layout are the same as in Fig. \ref{fig:results_4C4117}.}
\label{fig:results_3C368}
\end{figure}

\clearpage
\begin{figure} \centering
\begin{overpic}[width=1.0\textwidth,angle=0,trim= 0 0 0 0,scale=1.2]{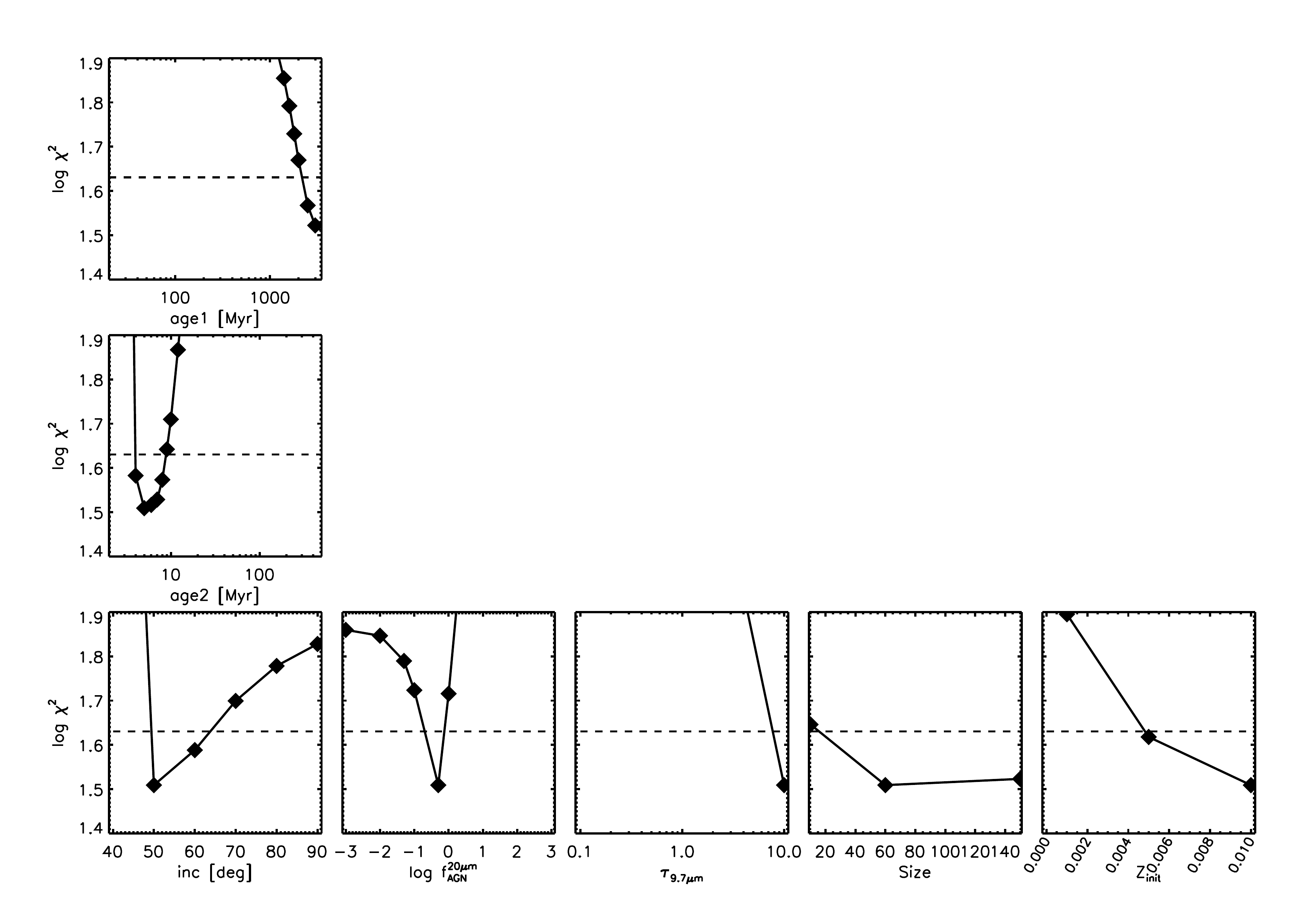}
\put(25,75){\includegraphics[height=0.7\textwidth,angle=270,trim= 0 0 0 0,scale=1.2]{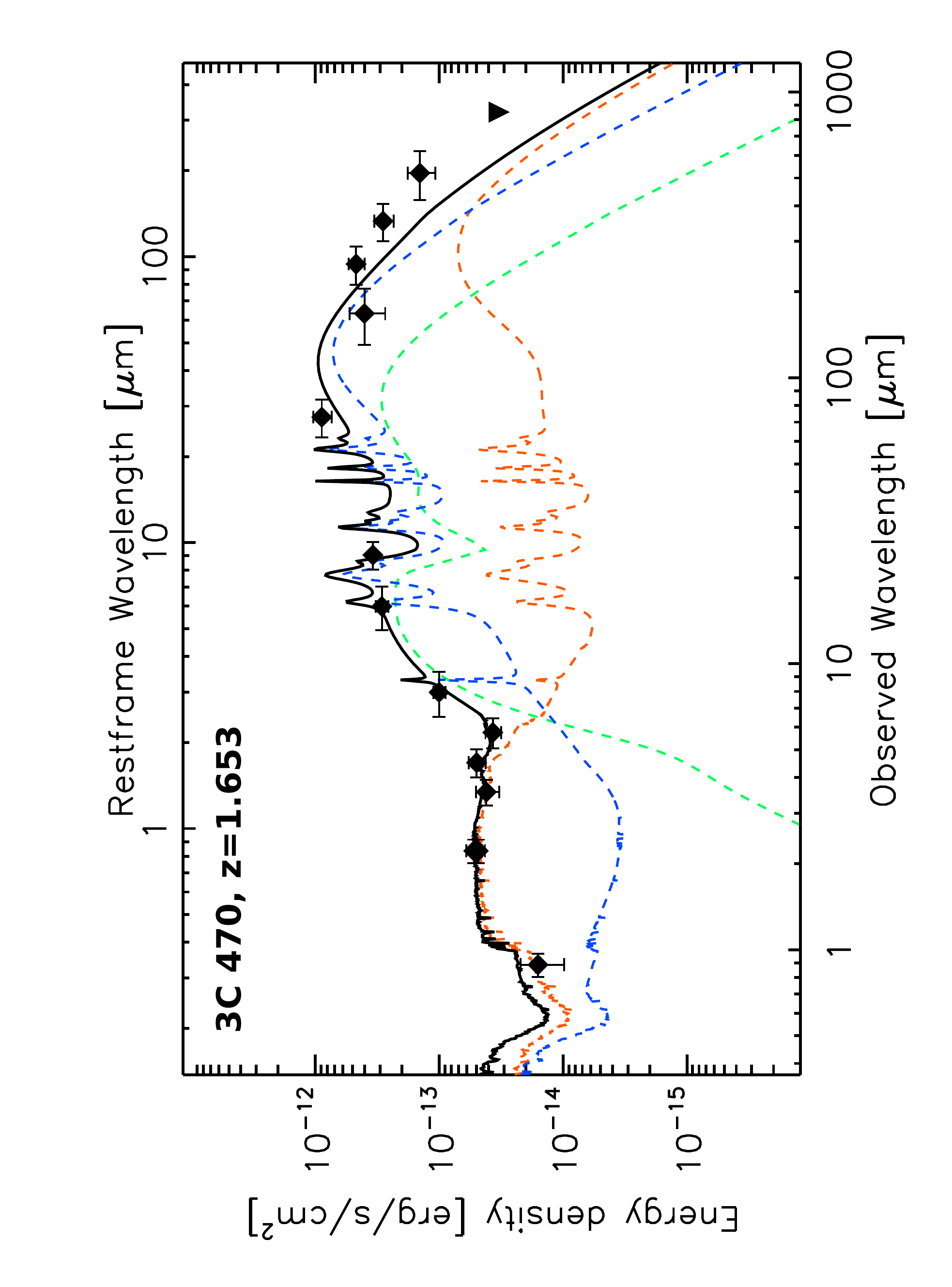}}
\end{overpic}
\caption{Best fit for 3C~470 ($z=1.653$). The colour code and the layout are the same as in Fig. \ref{fig:results_4C4117}. No polarisation data are available for this galaxy, therefore the insets present the results of the $without polarisation$ approach, reported in black.}
\label{fig:results_3C470}
\end{figure}

\clearpage
\begin{figure} \centering
\begin{overpic}[width=1.0\textwidth,angle=0,trim= 0 0 0 0,scale=1.2]{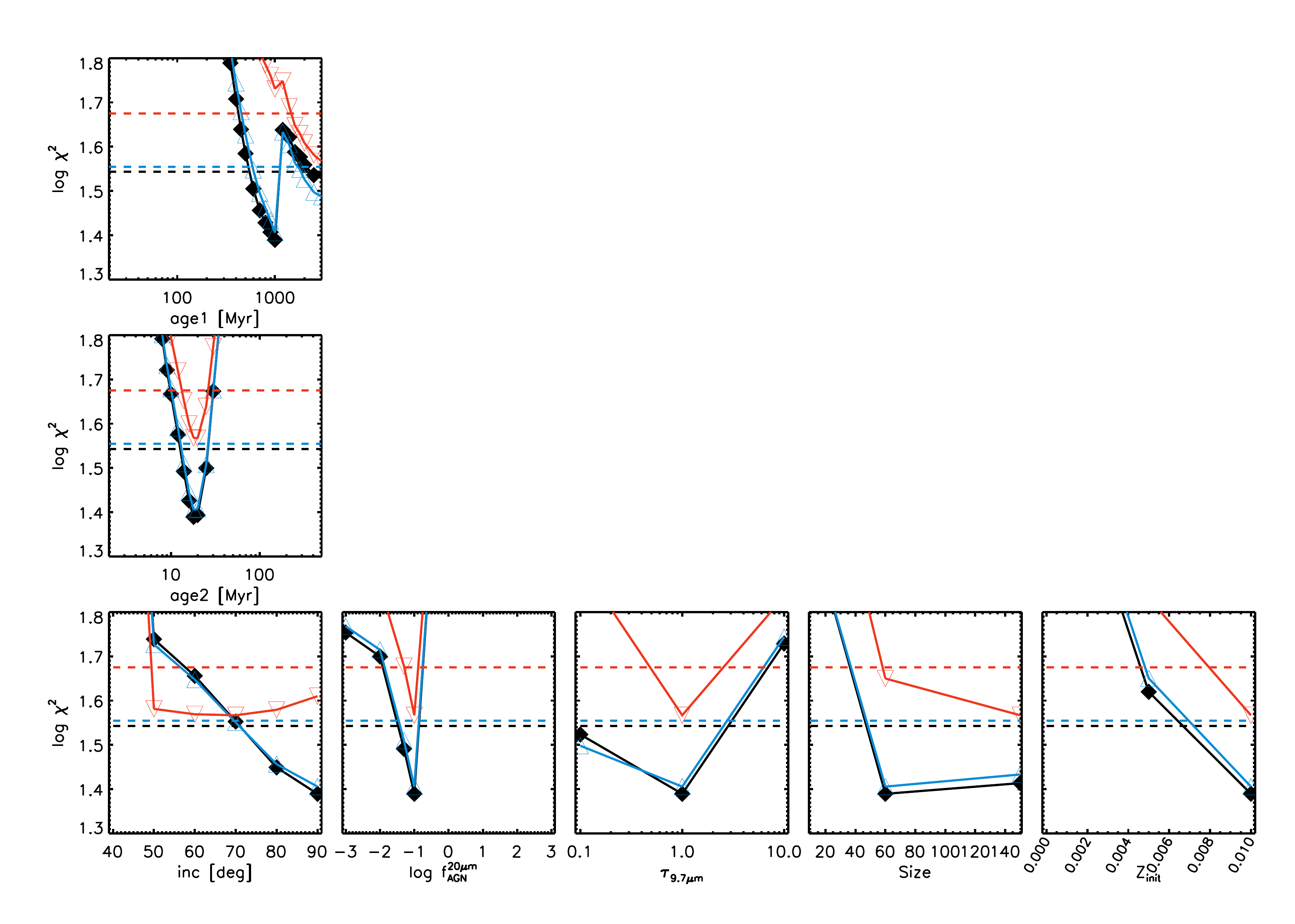}
\put(25,75){\includegraphics[height=0.7\textwidth,angle=270,trim= 0 0 0 0,scale=1.2]{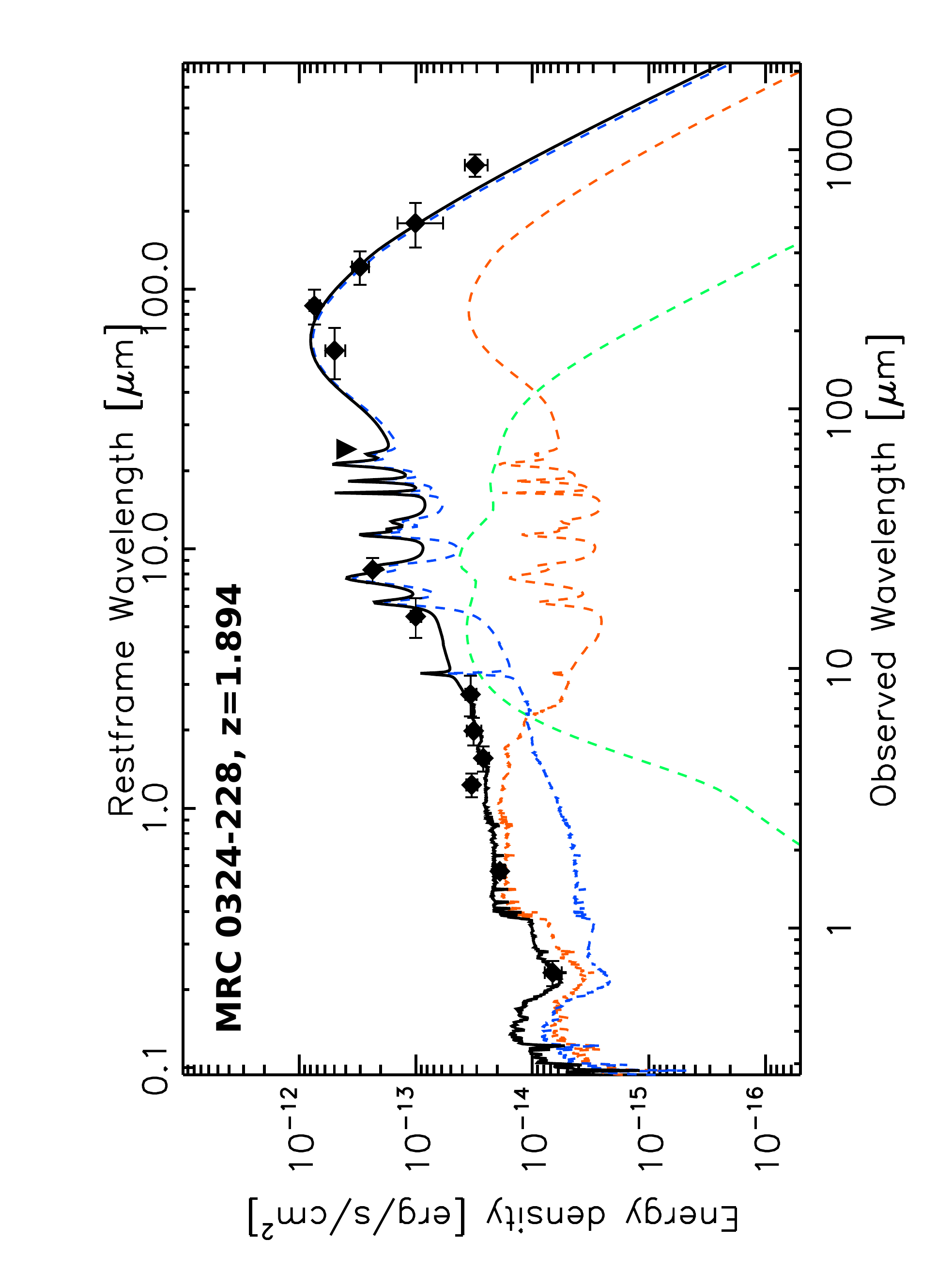}}
\end{overpic}
\caption{Best fit for MRC~0324-228 ($z=1.894$). The colour code and the layout are the same as in Fig. \ref{fig:results_4C4117}.}
\label{fig:results_MRC0324}
\end{figure}

\clearpage
\begin{figure} \centering
\begin{overpic}[width=1.0\textwidth,angle=0,trim= 0 0 0 0,scale=1.2]{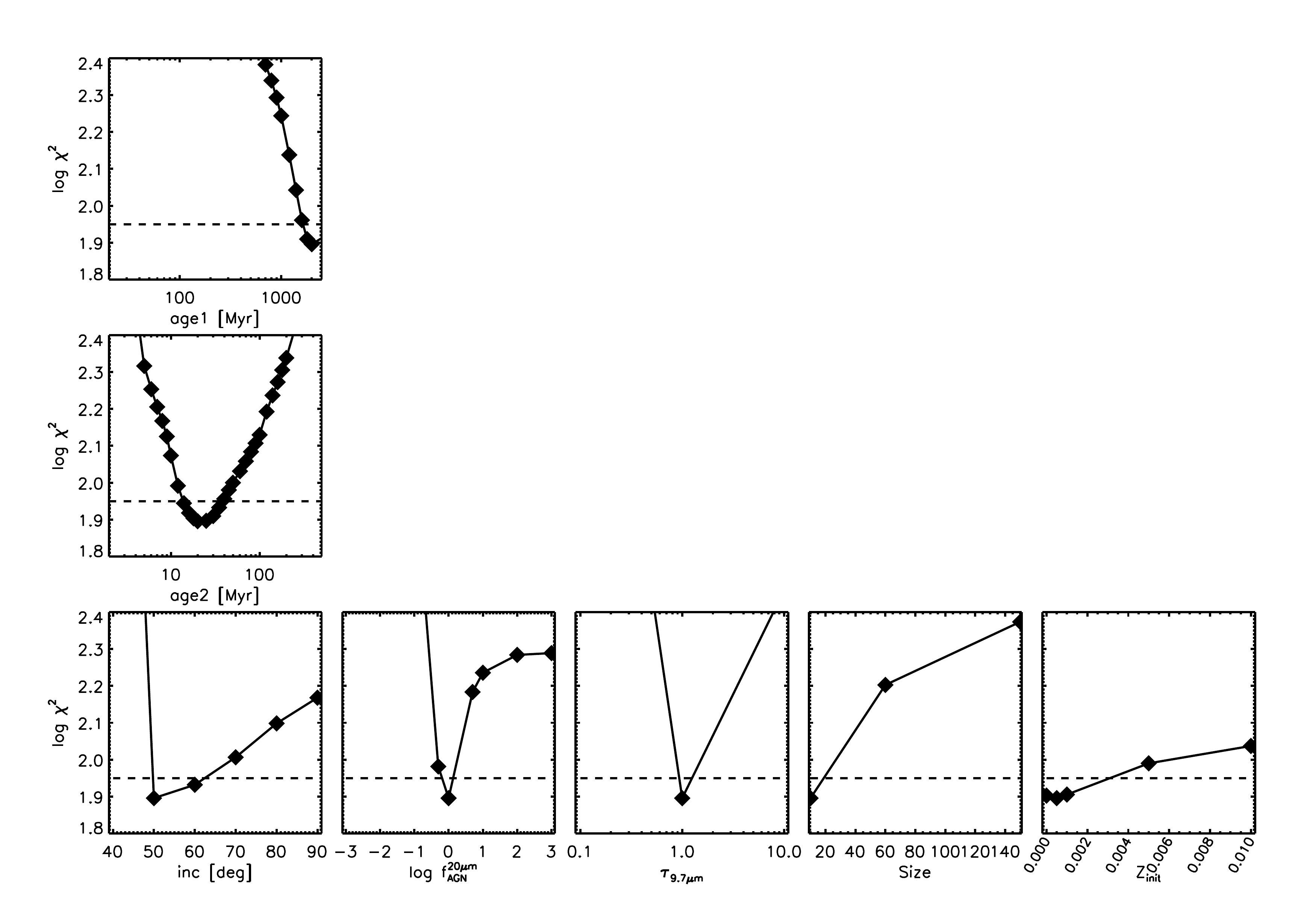}
\put(25,75){\includegraphics[height=0.7\textwidth,angle=270,trim= 0 0 0 0,scale=1.2]{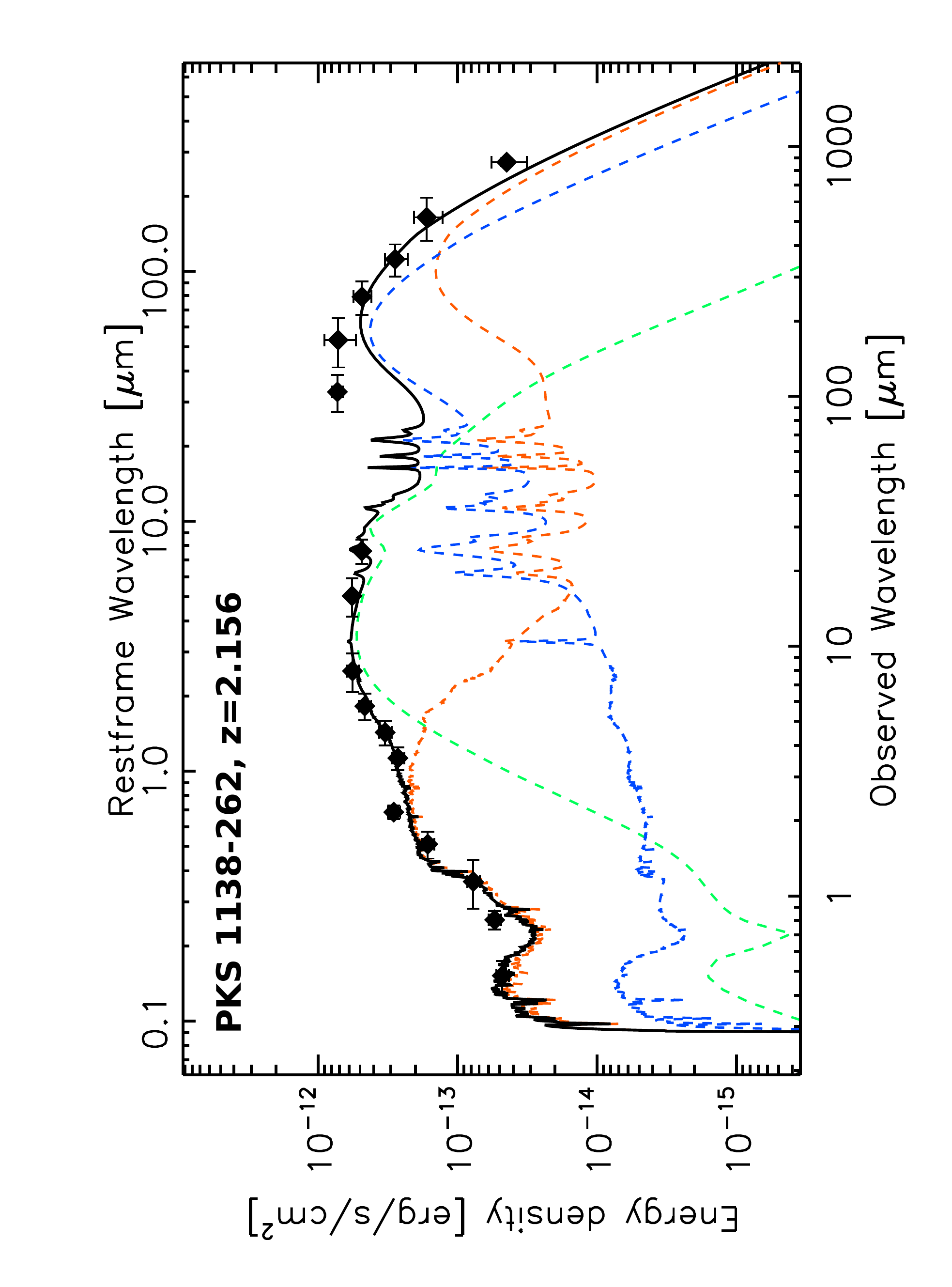}}
\end{overpic}
\caption{Best fit for PKS~1138-262 ($z=2.156$). The colour code and the layout are the same as in Fig. \ref{fig:results_4C4117}. No polarisation data are available for this galaxy, therefore the insets present the results of the $without polarisation$ approach, reported in black.}
\label{fig:results_PKS1138}
\end{figure}

\clearpage
\begin{figure} \centering
\begin{overpic}[width=1.0\textwidth,angle=0,trim= 0 0 0 0,scale=1.2]{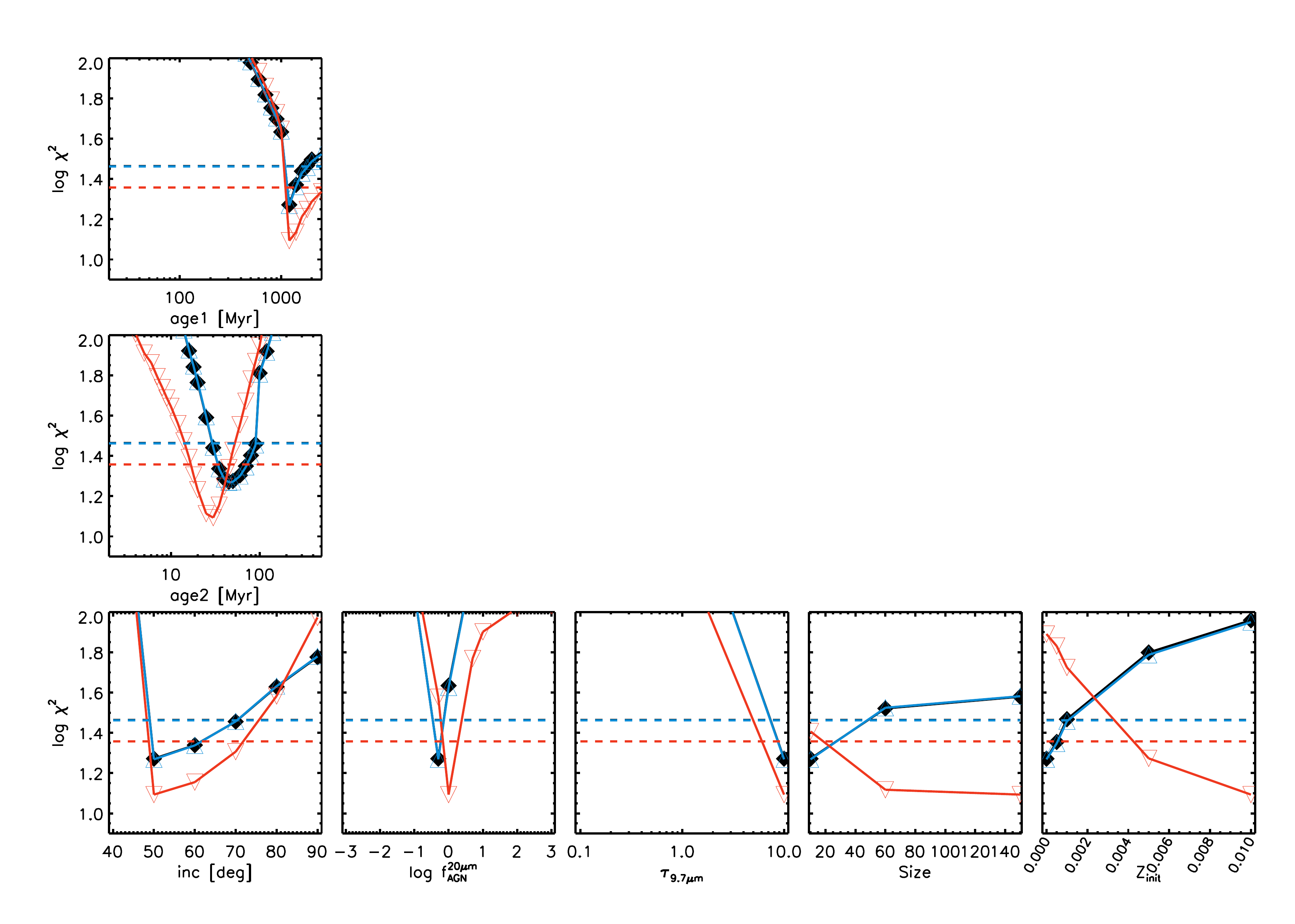}
\put(25,75){\includegraphics[height=0.7\textwidth,angle=270,trim= 0 0 0 0,scale=1.2]{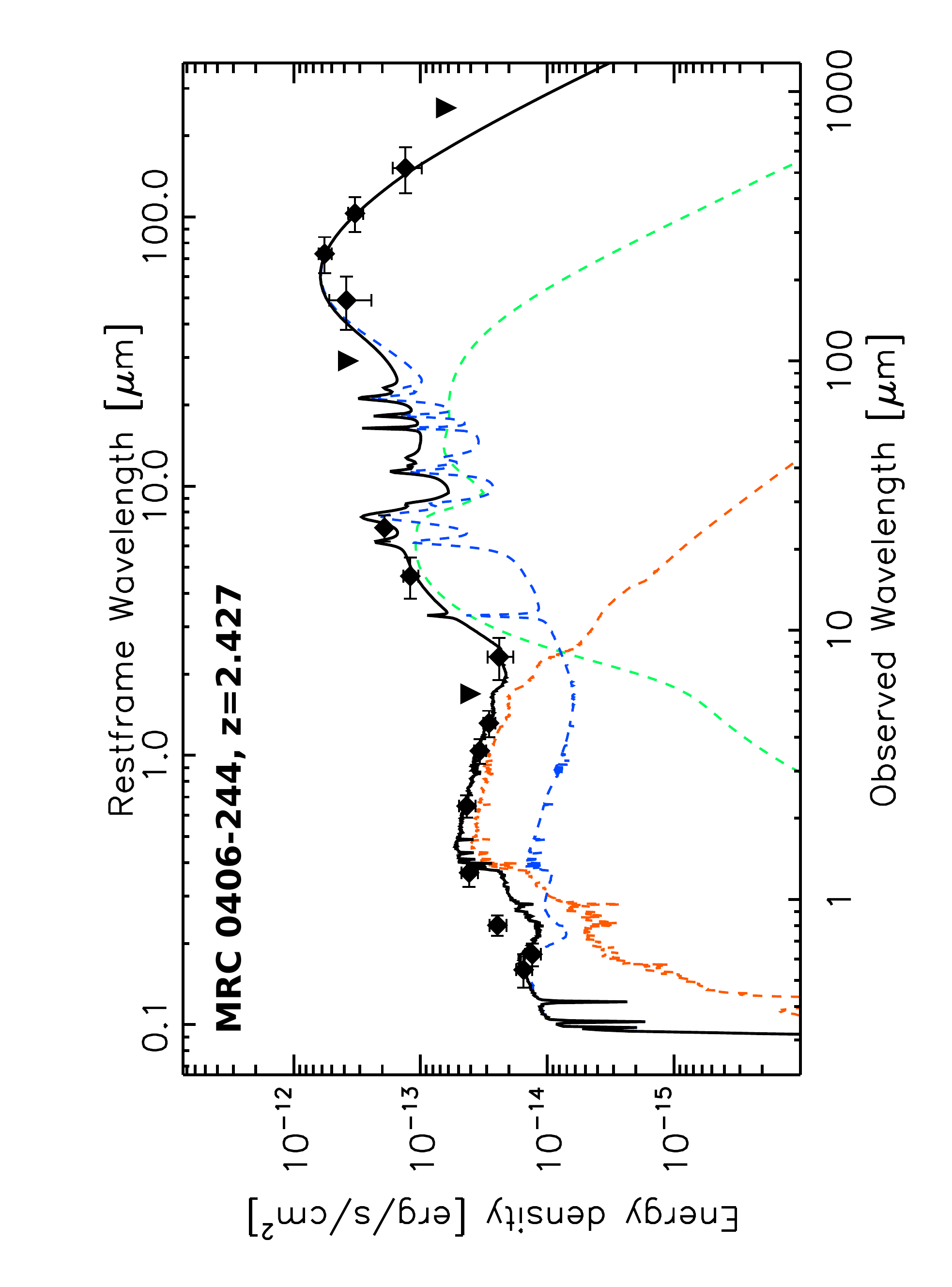}}
\end{overpic}
\caption{Best fit for MRC~0406-244 ($z=2.427$). The colour code and the layout are the same as in Fig. \ref{fig:results_4C4117}.}
\label{fig:results_MRC0406}
\end{figure}

\clearpage
\begin{figure} \centering
\begin{overpic}[width=1.0\textwidth,angle=0,trim= 0 0 0 0,scale=1.2]{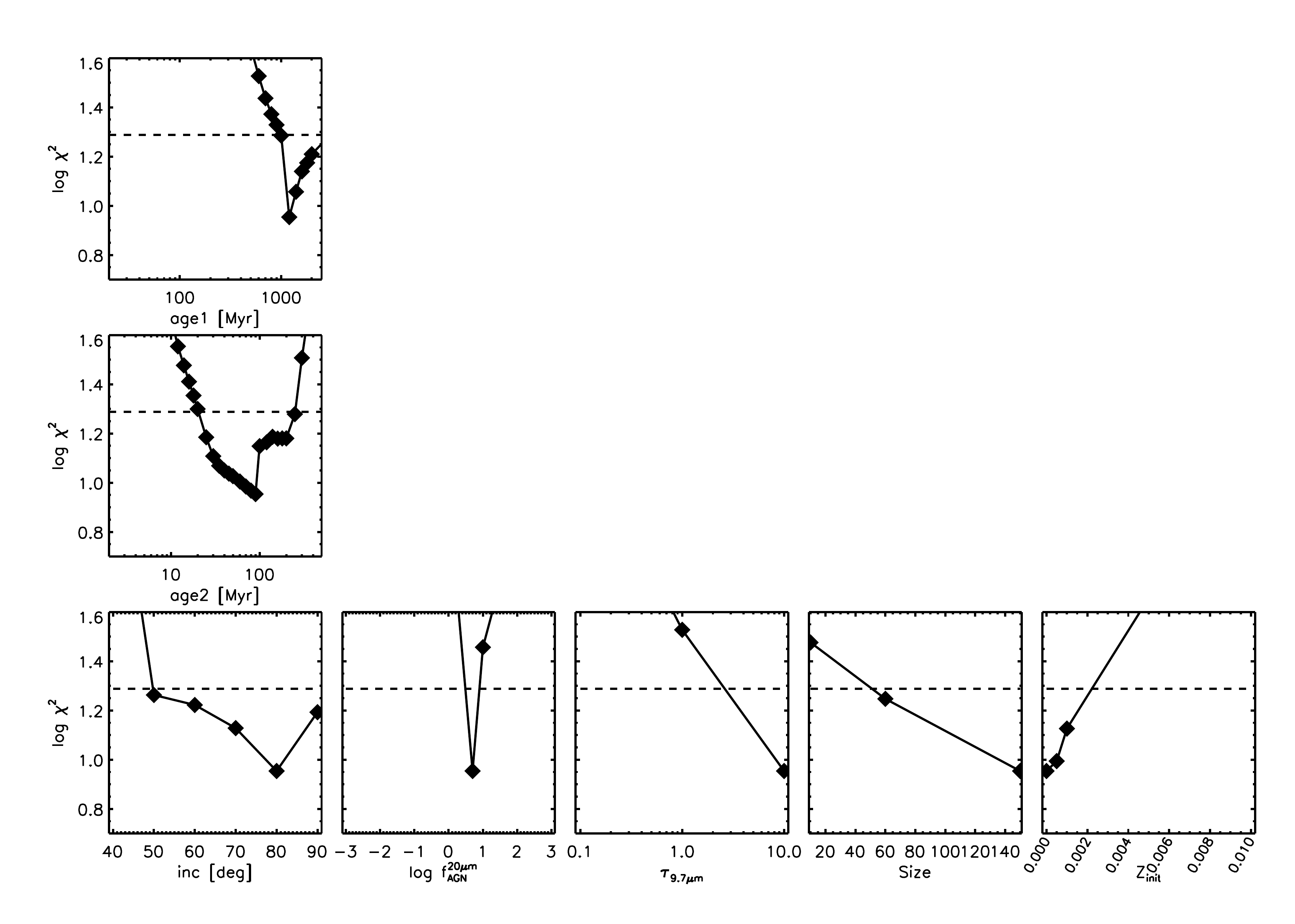}
\put(25,75){\includegraphics[height=0.7\textwidth,angle=270,trim= 0 0 0 0,scale=1.2]{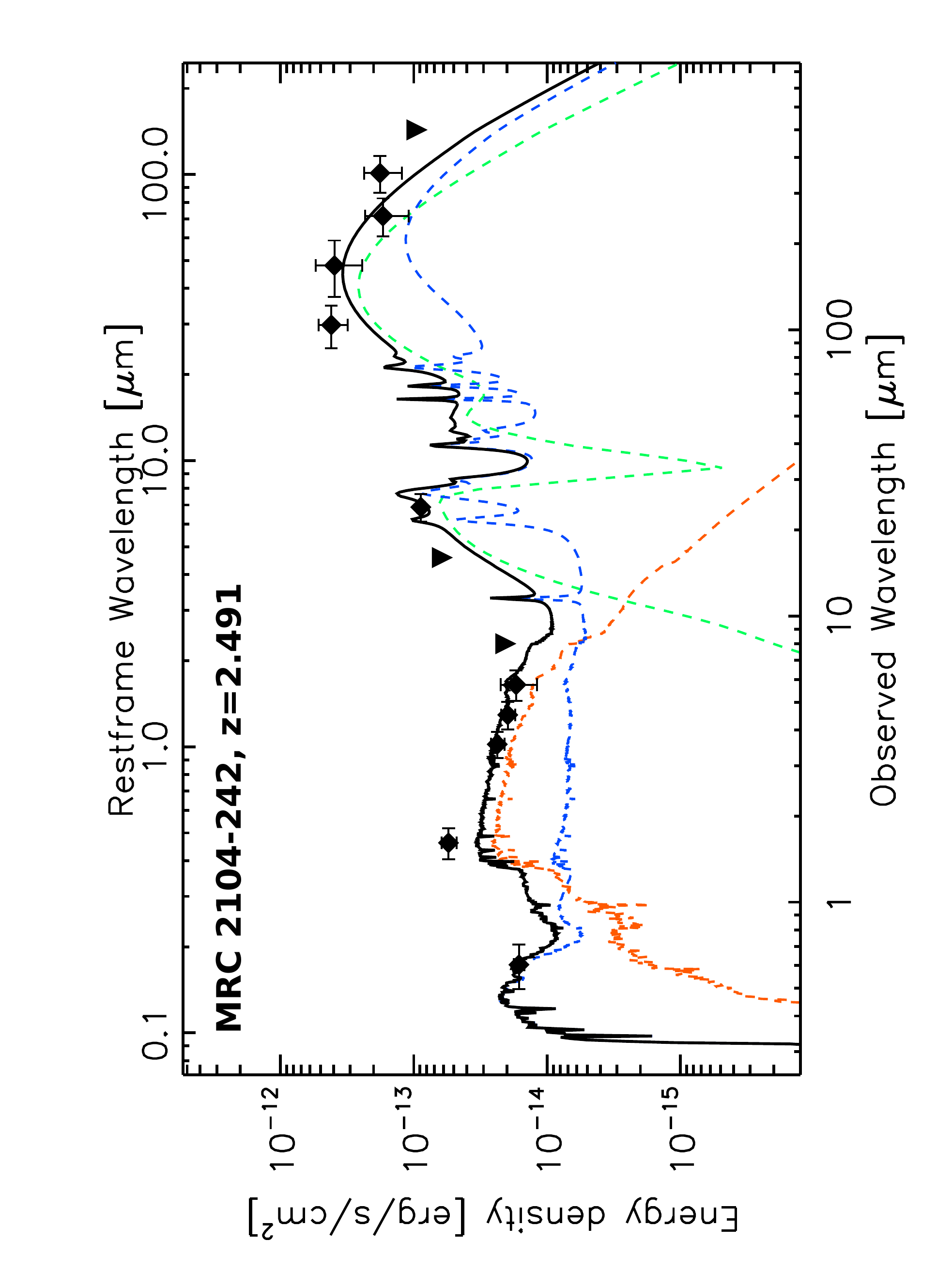}}
\end{overpic}
\caption{Best fit for MRC~2104-242 ($z=2.491$). The colour code and the layout are the same as in Fig. \ref{fig:results_4C4117}. No polarisation data are available for this galaxy, therefore the insets present the results of the $without polarisation$ approach, reported in black.}
\label{fig:results_MRC2104}
\end{figure}

\clearpage
\begin{figure} \centering
\begin{overpic}[width=1.0\textwidth,angle=0,trim= 0 0 0 0,scale=1.2]{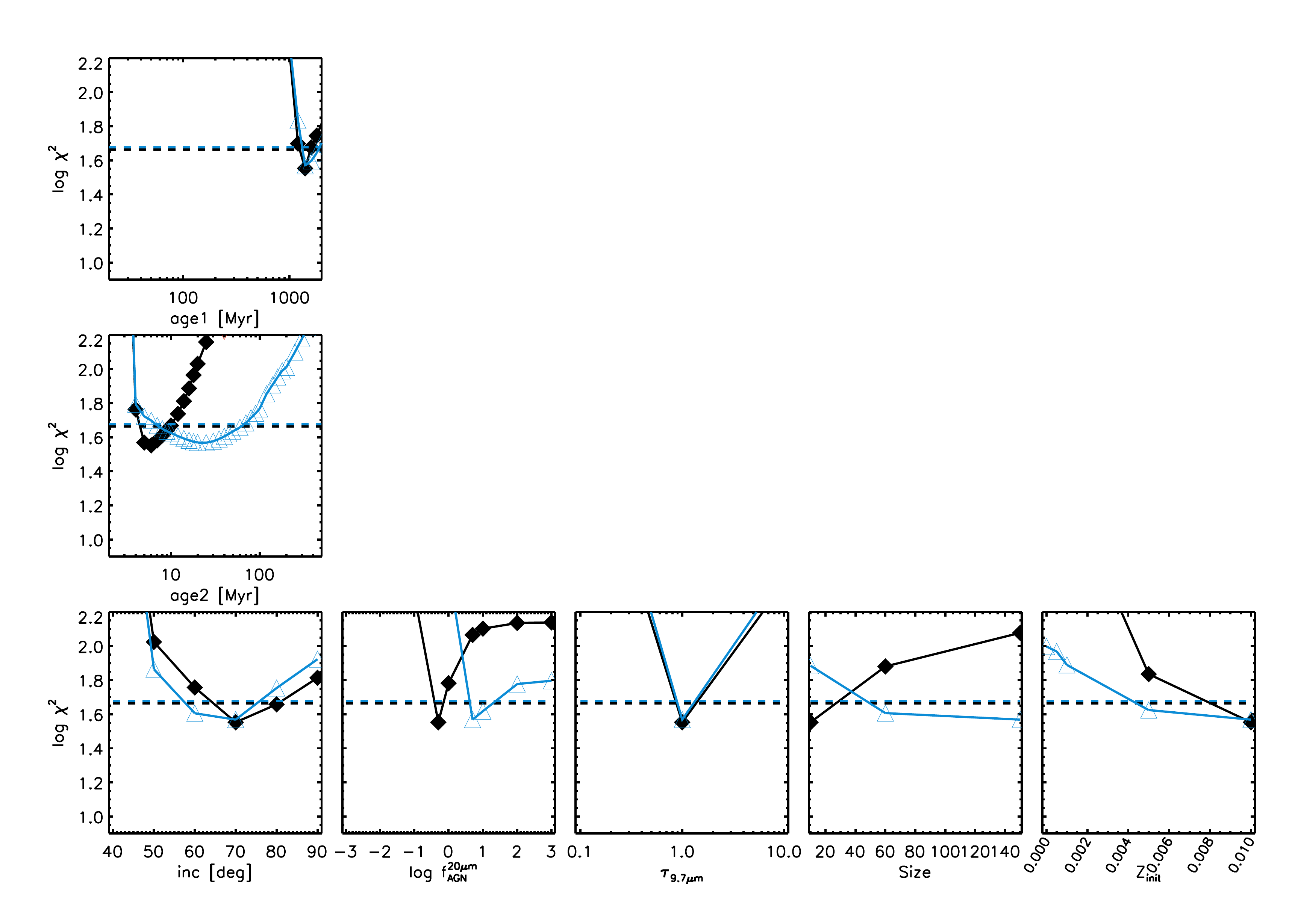}
\put(25,75){\includegraphics[height=0.7\textwidth,angle=270,trim= 0 0 0 0,scale=1.2]{./sed/USS_0828_bestfit_woAGN.eps}}
\end{overpic}
\caption{Best fit for USS~0828+193 ($z=2.572$). The colour code and the layout are the same as in Fig. \ref{fig:results_4C4117}. Only the minimal subtraction case (blue lines) is reported for this source (see Sect. \ref{sec:polarisation}).}
\label{fig:results_USS0828}
\end{figure}

\clearpage
\begin{figure} \centering
\begin{overpic}[width=1.0\textwidth,angle=0,trim= 0 0 0 0,scale=1.2]{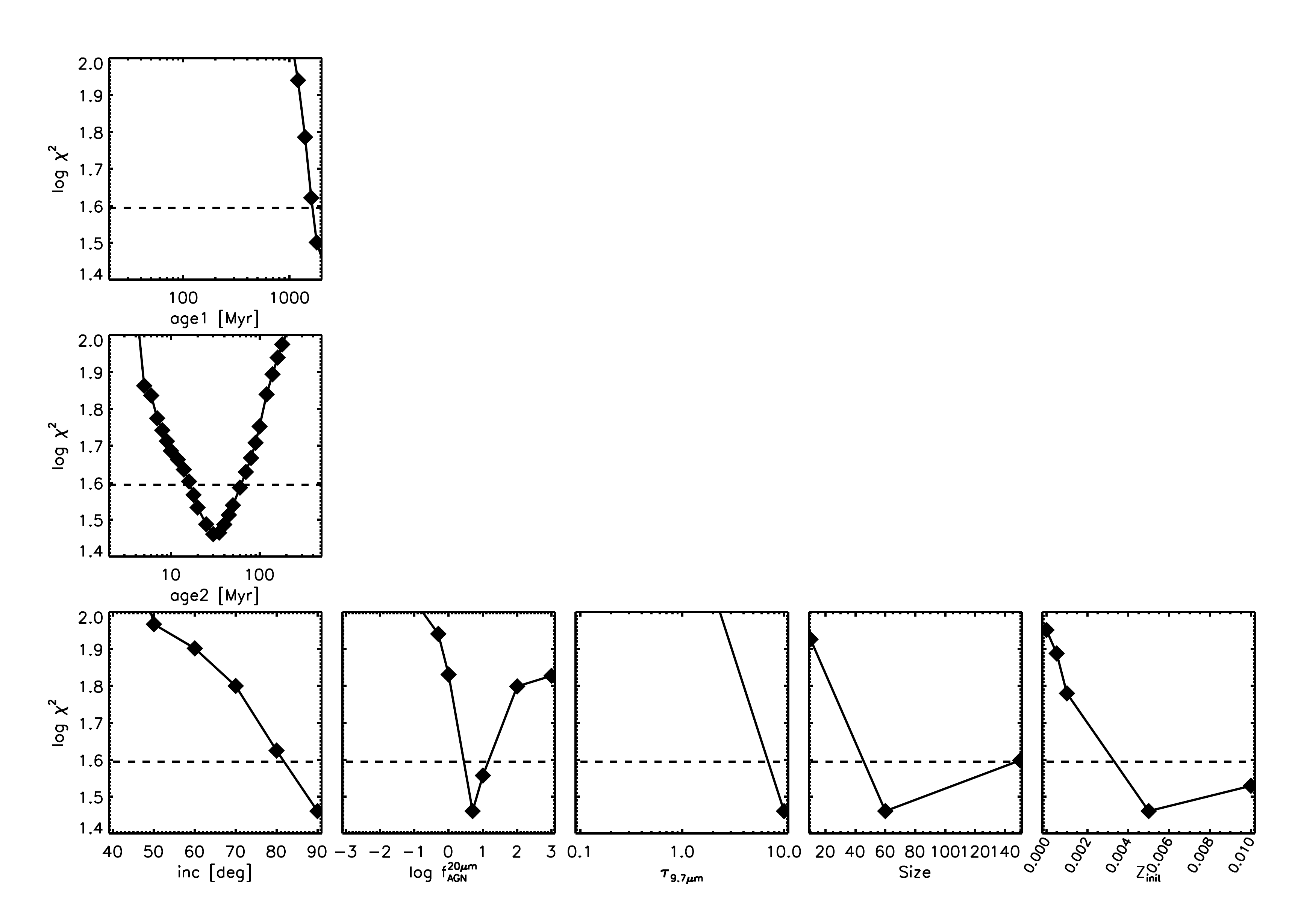}
\put(25,75){\includegraphics[height=0.7\textwidth,angle=270,trim= 0 0 0 0,scale=1.2]{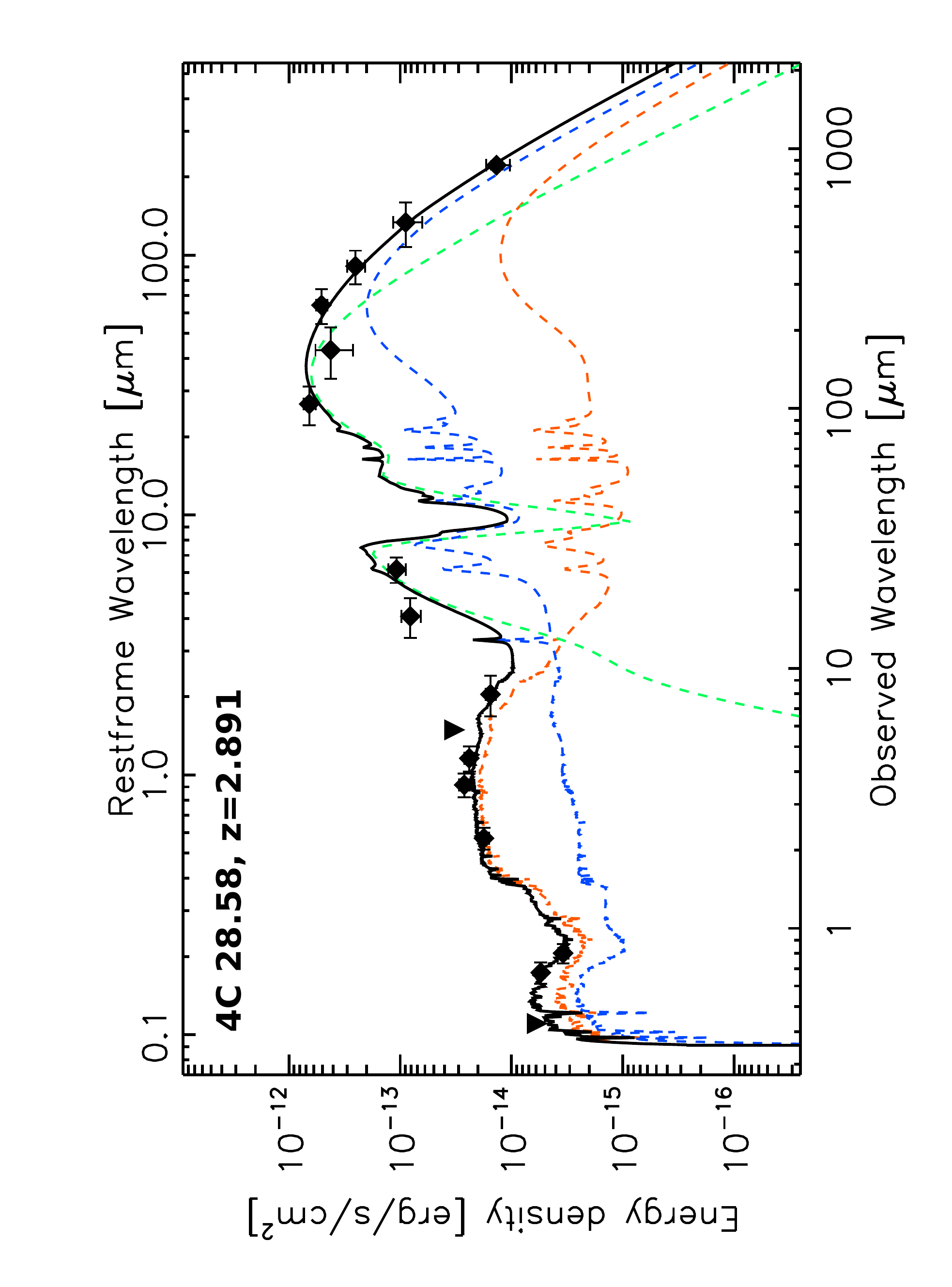}}
\end{overpic}
\caption{Best fit for 4C~28.58 ($z=2.891$). The colour code and the layout are the same as in Fig. \ref{fig:results_4C4117}. No polarisation data are available for this galaxy, therefore the insets present the results of the $without polarisation$ approach, reported in black.}
\label{fig:results_4C2858}
\end{figure}

\clearpage
\begin{figure} \centering
\begin{overpic}[width=1.0\textwidth,angle=0,trim= 0 0 0 0,scale=1.2]{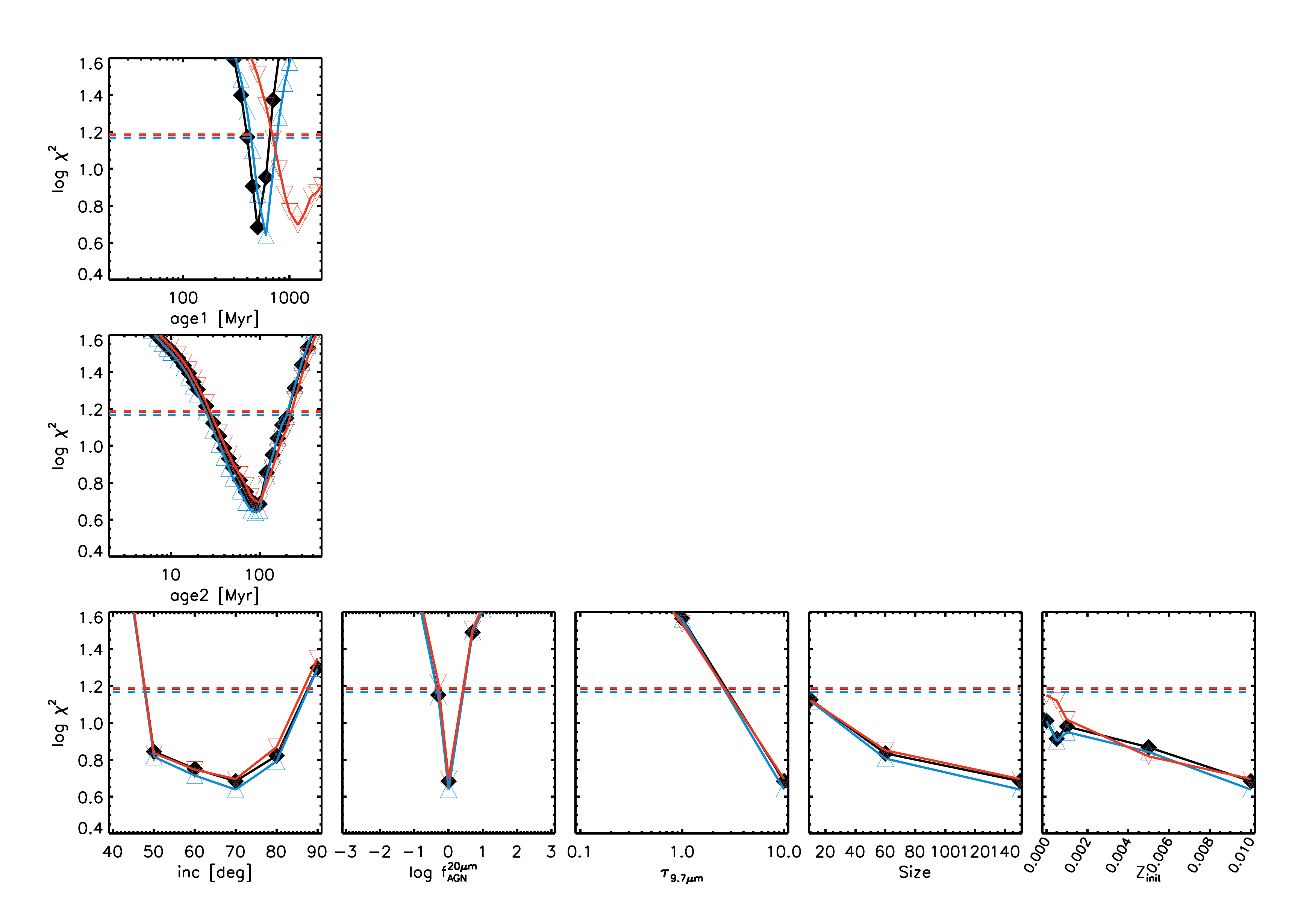}
\put(25,75){\includegraphics[height=0.7\textwidth,angle=270,trim= 0 0 0 0,scale=1.2]{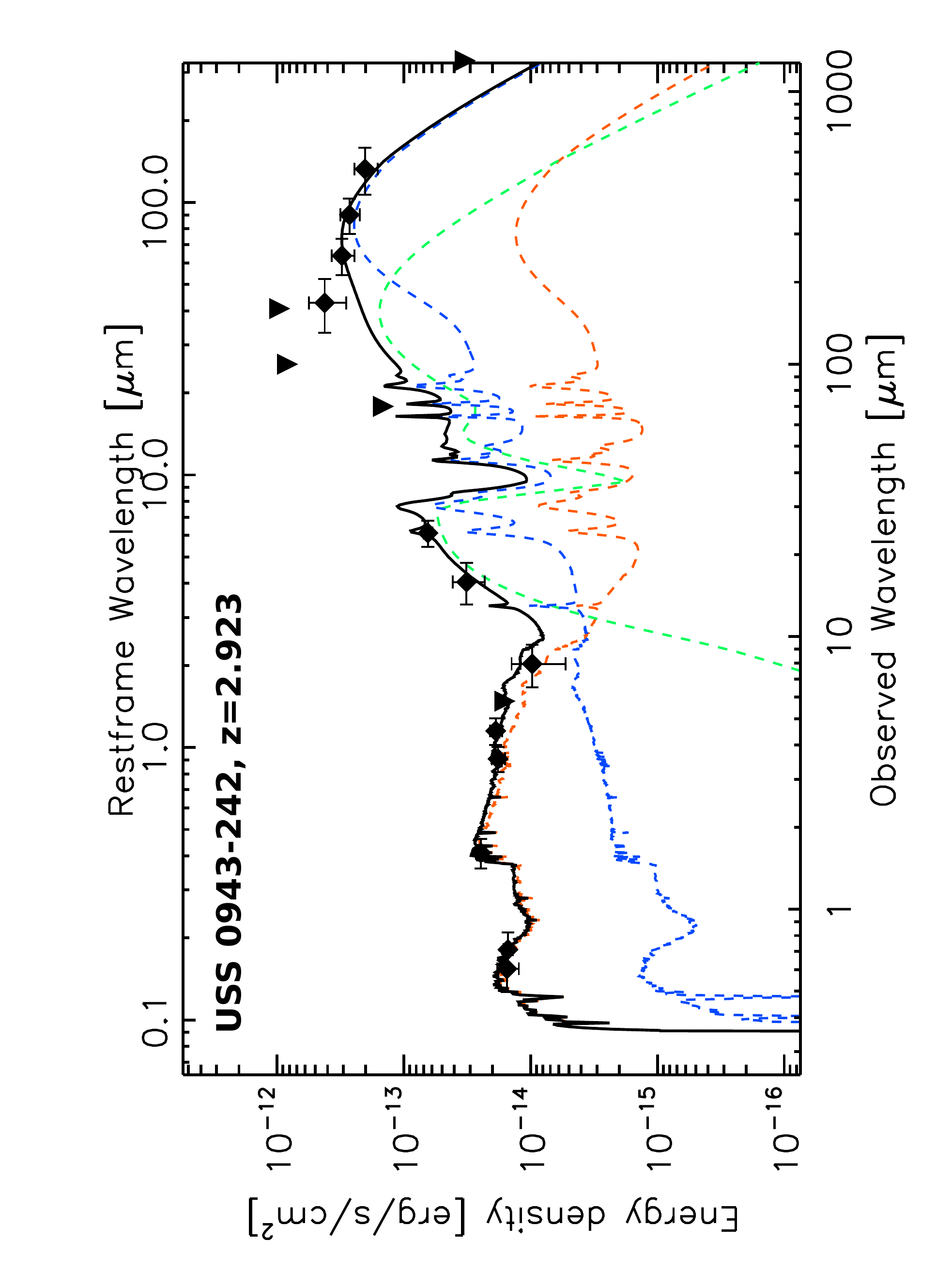}}
\end{overpic}
\caption{Best fit for USS~0943-242 ($z=2.923$). The colour code and the layout are the same as in Fig. \ref{fig:results_4C4117}.}
\label{fig:results_USS0943}
\end{figure}

\clearpage
\begin{figure} \centering
\begin{overpic}[width=1.0\textwidth,angle=0,trim= 0 0 0 0,scale=1.2]{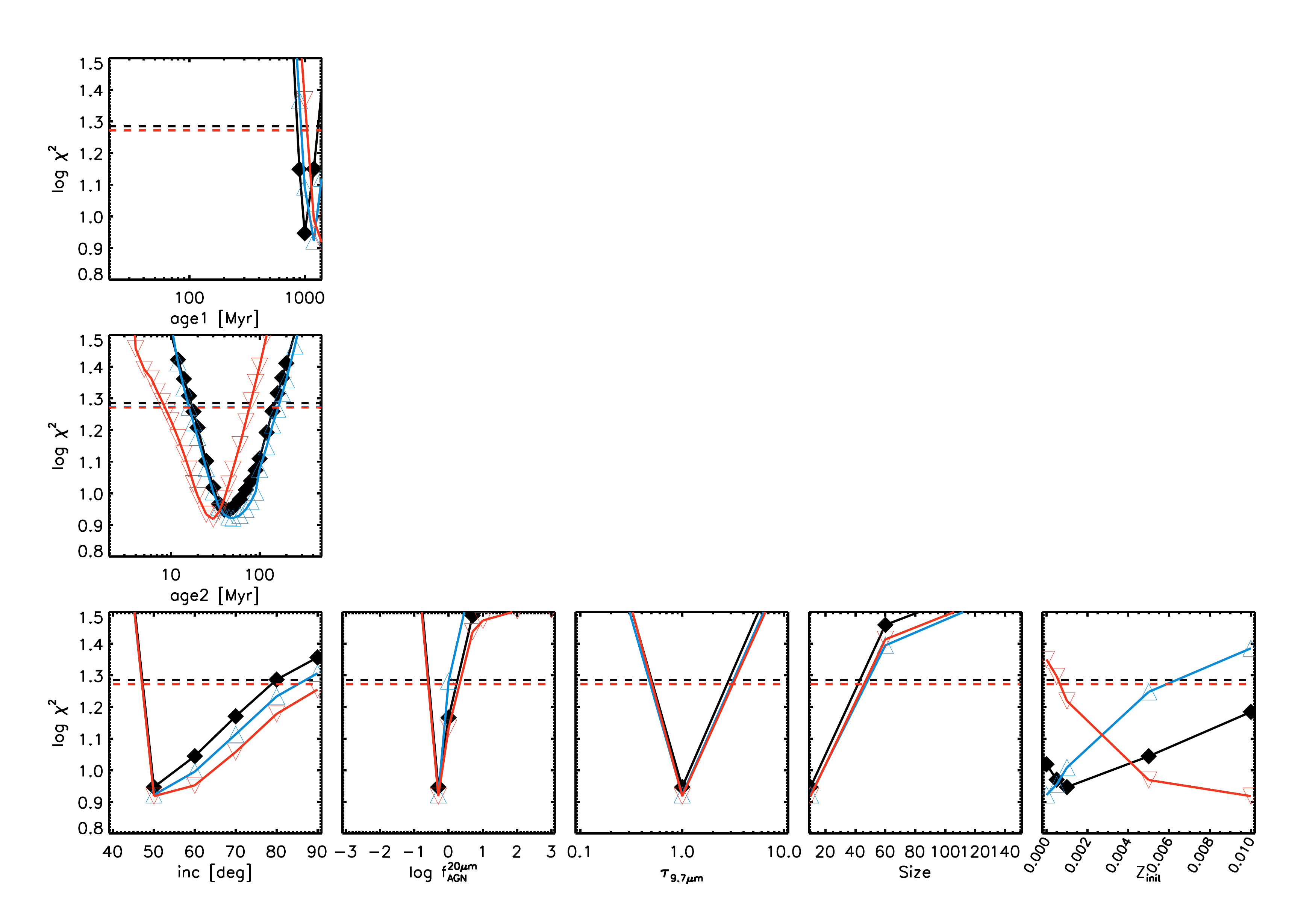}
\put(25,75){\includegraphics[height=0.7\textwidth,angle=270,trim= 0 0 0 0,scale=1.2]{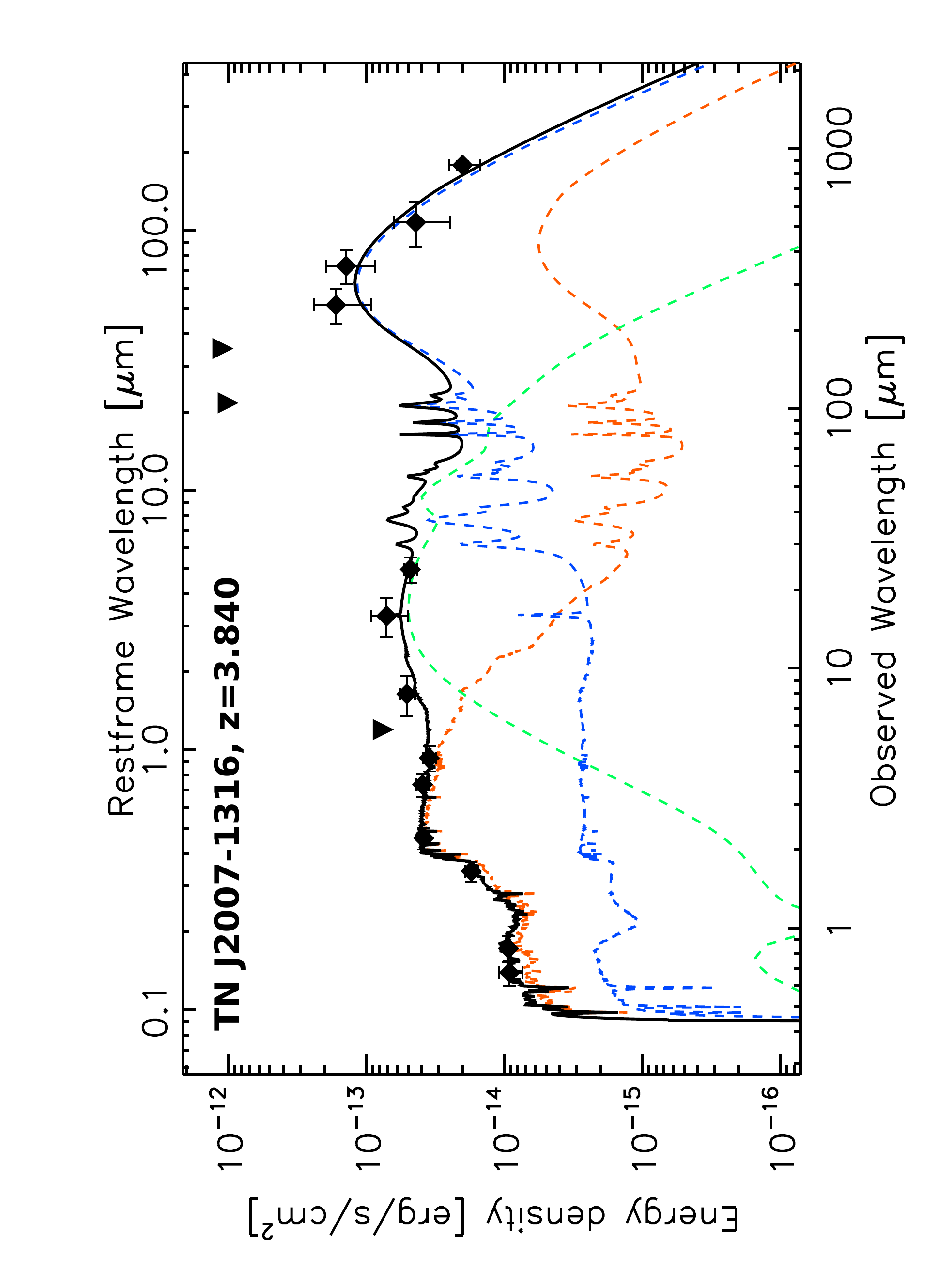}}
\end{overpic}
\caption{Best fit for TN~J2007-1316 ($z=3.840$). The colour code and the layout are the same as in Fig. \ref{fig:results_4C4117}.}
\label{fig:results_TNJ2007}
\end{figure}

\end{landscape}

\end{document}